\documentclass[twocolumn,prd,amsmath,superscriptaddress]{revtex4}

\usepackage{graphicx}
\usepackage{amsmath}
\usepackage{amssymb}
\usepackage{graphics}
\usepackage{hyperref}
\usepackage{ifthen}
\usepackage{bm}
\usepackage[usenames,dvipsnames]{color}

\newcommand{\aavgP}{\left< a_P^2 \right>_\varphi}
\newcommand{\aavgS}{\left< a_S^2 \right>_\varphi}
\newcommand{\afrag}{\alpha_\mathrm{frag}}
\newcommand{\ArefP}{A_{\mathrm{ref,}P}}
\newcommand{\ArefS}{A_{\mathrm{ref,}S}}
\newcommand{\Awin}{{\mathcal A}_\mathrm{win}}
\newcommand{\Bext}{B_0}
\newcommand{\bgam}{\beta_\gamma}
\newcommand{\bmat}{\beta_\mathrm{m}}
\newcommand{\Cgc}{{\mathcal C}}
\newcommand{\chiSqnull}{\chi^2_\mathrm{null}}
\newcommand{\Dzf}{\Delta z_\mathrm{f}}
\newcommand{\Faft}{F_\mathrm{aft}}
\newcommand{\Faftmean}{{\bar F}^\mathrm{(aft)}}
\newcommand{\Fdark}{F_\mathrm{dark}}
\newcommand{\fdiff}{f_\mathrm{diff}}
\newcommand{\Fgam}{F_\gamma}
\newcommand{\FigA}{({\em{Top}})}
\newcommand{\FigB}{({\em{Bottom}})}
\newcommand{\FigThreeA}{({\em{Left}})}
\newcommand{\FigThreeB}{({\em{Middle}})}
\newcommand{\FigThreeC}{({\em{Right}})}
\newcommand{\Fobs}{{\bar F}^\mathrm{(obs)}}
\newcommand{\Forange}{F_\mathrm{or,0}}
\newcommand{\Fpred}{{\bar F}^\mathrm{(pred)}}
\newcommand{\Fpump}{F_\mathrm{pump}}
\newcommand{\Fpumpmean}{F_\mathrm{pump,mean}}
\newcommand{\fref}{{\bar f}_\mathrm{ref}}
\newcommand{\Fsyst}{F_\mathrm{syst}}
\newcommand{\Ftran}{F_\mathrm{tran}}
\newcommand{\fvis}{f_\mathrm{vis}}
\newcommand{\fvol}{f_\mathrm{vol}}
\newcommand{\Gaft}{\Gamma_\mathrm{aft}}
\newcommand{\GammeV}{GammeV}
\newcommand{\GammeVCHASE}{CHASE}
\newcommand{\GammeVCHASEfull}{GammeV-CHASE}
\newcommand{\gcmc}{g$/$cm$^3$}
\newcommand{\Gdec}{\Gamma_{\mathrm{dec}}}
\newcommand{\gf}{g_\mathrm{F}}
\newcommand{\Gfrag}{\Gamma_\mathrm{frag}}
\newcommand{\Gorange}{\Gamma_\mathrm{or,0}}
\newcommand{\Gtot}{\Gamma_\mathrm{tot}}
\newcommand{\losc}{\ell_\mathrm{osc}}
\newcommand{\ltot}{\ell_\mathrm{tot}}
\newcommand{\meff}{m_\mathrm{eff}}
\newcommand{\meffprot}{m_\mathrm{eff,prot}}
\newcommand{\meffw}{m_\mathrm{eff,surf}}
\newcommand{\mel}{m_\mathrm{elec}}
\newcommand{\mf}{m_\mathrm{F}}
\newcommand{\mlatt}{m_\mathrm{lattice}}
\newcommand{\mmax}{m_\mathrm{max}}
\newcommand{\mpr}{m_\mathrm{prot}}
\newcommand{\Mpl}{M_\mathrm{Pl}}
\newcommand{\nB}{n_B}
\newcommand{\nL}{n_\mathrm{L}}
\newcommand{\Nphiw}{N_\phi^{(\omega)}}
\newcommand{\nR}{n_\mathrm{R}}
\newcommand{\Pabs}{{\mathcal P}_\mathrm{abs}}
\newcommand{\PabsBexit}{{\mathcal P}_\mathrm{abs}^{(B\,\mathrm{ exit})}}
\newcommand{\Paft}{P_\mathrm{aft}}
\newcommand{\Pdec}{P_{\mathrm{dec}}}
\newcommand{\Pdet}{P_\mathrm{det}}
\newcommand{\Pgc}{{\mathcal P}_{\gamma \leftrightarrow \phi}}
\newcommand{\phibulk}{\phi_\mathrm{bulk}}
\newcommand{\phiw}{\phi_\mathrm{surf}}
\newcommand{\psic}{\psi_\phi}
\newcommand{\Psie}{\Psi_\mathrm{elec}}
\newcommand{\psig}{{\vec \psi}_\gamma}
\newcommand{\psigBexit}{{\vec \psi}_\gamma^{(B\,\mathrm{ exit})}}
\newcommand{\Psip}{\Psi_\mathrm{prot}}
\newcommand{\rb}{r_\mathrm{Bohr}} 
\newcommand{\rhoe}{\rho_\mathrm{elec}}
\newcommand{\rhog}{\rho_{{\mathcal L}\mathrm{,EM}}}
\newcommand{\rhom}{\rho_\mathrm{mat}}
\newcommand{\rhop}{\rho_\mathrm{prot}}
\newcommand{\rhov}{\rho_\mathrm{vac}}
\newcommand{\rlens}{r_\mathrm{lens}}
\newcommand{\rp}{r_\mathrm{P}} 
\newcommand{\rpmt}{r_\mathrm{PMT}}
\newcommand{\sfrag}{s_\mathrm{frag}} 
\newcommand{\sigfrag}{\sigma_\mathrm{frag}} 
\newcommand{\sign}{\textrm{signum}}
\newcommand{\sigphiF}{\sigma_{\phi\psi\rightarrow\phi\psi}}
\newcommand{\sinc}{\mathrm{sinc}}
\newcommand{\sL}{s_\mathrm{L}}
\newcommand{\sM}{s_\mathrm{M}}
\newcommand{\sR}{s_\mathrm{R}}
\newcommand{\tL}{t_\mathrm{L}}
\newcommand{\tM}{t_\mathrm{M}}
\newcommand{\tprod}{t_\mathrm{prod}}
\newcommand{\tR}{t_\mathrm{R}}
\newcommand{\Veff}{V_\mathrm{eff}}
\newcommand{\Vvac}{{\mathcal V}_\mathrm{vac}} 
\newcommand{\xiL}{\xi_\mathrm{L}}
\newcommand{\xiM}{\xi_\mathrm{M}}
\newcommand{\xiR}{\xi_\mathrm{R}}

\begin{document}

\title{Designing dark energy afterglow experiments}

\author{Amol Upadhye}
\affiliation{Argonne National Laboratory, 9700 S. Cass Ave., Lemont, IL 60439}%

\author{Jason H. Steffen}
\author{Aaron S. Chou}
\affiliation{Fermi National Accelerator Laboratory, PO Box 500, Batavia, IL 60510}%

\date{\today}

\begin{abstract}
Chameleon fields, which are scalar field dark energy candidates, can evade fifth force constraints by becoming massive in high-density regions.  However, this property allows chameleon particles to be trapped inside a vacuum chamber with dense walls.  Afterglow experiments constrain photon-coupled chameleon fields by attempting to produce and trap chameleon particles inside such a vacuum chamber, from which they will emit an afterglow as they regenerate photons.  Here we discuss several theoretical and systematic effects underlying the design and analysis of the \GammeV~and \GammeVCHASE~afterglow experiments.  We consider chameleon particle interactions with photons, Fermions, and other chameleon particles, as well as with macroscopic magnetic fields and matter.  The afterglow signal in each experiment is predicted, and its sensitivity to various properties of the experimental apparatus is studied.  Finally, we use \GammeVCHASE~data to exclude a wide range of photon-coupled chameleon dark energy models.
\end{abstract}

\maketitle

\section{Introduction}
\label{sec:introduction}

Though the existence of the cosmic acceleration has been confirmed repeatedly, its cause remains a mystery.  The simplest explanation, a cosmological constant $\Lambda$, is completely consistent with the data~\cite{Komatsu_etal_2010,Larson_etal_2011,Suzuki_etal_2012,Sanchez_etal_2012}, but leads to more questions.  Why is $\Lambda$ some $120$ orders of magnitude below the Planck density?  If some new physics cancels this energy density, then why isn't the cancellation complete?  Several answers have been proposed~\cite{Abbott_1985,Brown_Teitelboim_1987,Bousso_Polchinski_2000,Steinhardt_Turok_2006,Dvali_Hofmann_Khoury_2007,deRham_etal_2008,deRham_Hofmann_Khoury_Tolley_2008,Agarwal_etal_2009}, with the simplest among them reducing at low energies to a single effective scalar field tunneling among the large number of local minima of its potential.  Beyond these specific models, it is worthwhile to ask whether generic low-energy effective theories, possibly responsible for the cosmic acceleration, would predict any effects detectable in laboratory experiments.

A single scalar field ``dark energy'' is the simplest dynamical generalization of $\Lambda$, but most ``natural'' models mediate unscreened fifth forces which have been excluded over a large range of scales~\cite{Adelberger_etal_2009}.  Thus, unless these models are prevented by symmetry~\cite{Frieman_Hill_Stebbins_Waga_1995,Carroll_1998} from coupling to matter, they must possess a mechanism for screening fifth forces locally.  Chameleon theories are scalar-tensor theories with potentials chosen to make their effective masses larger in higher-density regions of the universe, allowing them to ``hide'' from fifth force constraints~\cite{Khoury_Weltman_2004a,Khoury_Weltman_2004b,Brax_etal_2004,Gubser_Khoury_2004,Upadhye_Gubser_Khoury_2006,Mota_Shaw_2006,Mota_Shaw_2007}.  Symmetron theories~\cite{Hinterbichler_Khoury_2010,Olive_Pospelov_2007,Pietroni_2005} screen their fifth forces through a restoration of symmetry at high densities, while Galileons~\cite{Deffayet_etal_2001,Nicolis_Rattazzi_Trincherini_2008} have non-canonical kinetic terms which reduce their effective matter couplings.

Chameleons were the first of these screened theories to be discovered, and are likely the best-studied.  If they couple to photons as expected~\cite{Brax_etal_2010}, then the very effect which enables them to evade fifth force constraints also allows chameleons produced through photon oscillation to be trapped inside a vacuum chamber.  Photon regeneration from such chameleons could produce a detectable afterglow~\cite{Gies_Mota_Shaw_2008,Ahlers_etal_2008,Chou_etal_2009} which has been constrained by the \GammeV~and \GammeVCHASEfull~(hereafter \GammeVCHASE) experiments~\cite{Chou_etal_2009,Steffen_etal_2010,Steffen_Upadhye_2009,Upadhye_Steffen_Weltman_2010,Steffen_etal_2012}.  The goal of the current work is to study the behavior of chameleon particles in afterglow experiments.  Specific examples are based upon \GammeVCHASE, and constraints use \GammeVCHASE~data, but we aim to provide a general discussion of the design and analysis of afterglow experiments applicable to future experiments of this form.

Afterglow experiments rely on two effects: oscillation and reflection.  The rate at which a chameleon particle passing through a classical, macroscopic magnetic field oscillates into a photon, and vice versa, has been computed semiclassically~\cite{Raffelt_Stodolsky_1988,Upadhye_Steffen_Weltman_2010}.  Although the smooth variation of the magnetic field inside an afterglow experiment could lead to the adiabatic suppression of oscillation, we show that the quantum measurement of particle content made by glass windows inside the magnetic region almost completely mitigate this suppression.  Thus the chameleon production rate and the photon regeneration rate may be computed.  

Meanwhile, the reflection of a chameleon particle from a homogeneous region of high density is simply a matter of energy conservation; a particle with a given energy cannot enter a region of space where its effective mass exceeds its total energy.  We consider a real solid as a lattice of atomic nuclei surrounded by a nearly homogeneous electron cloud and show that such a solid may be approximated as homogeneous for the purpose of determining whether chameleons reflect.  Such reflection allows chameleons to be ``bottled'' in a vacuum chamber with dense walls, where they remain until regenerating photons.  Moreover, we show that averaging over photon polarizations washes out the dependence of the afterglow signal on the potential-dependent chameleon-photon phase, which was calculated by~\cite{Brax_etal_2007b}.  Thus the predicted afterglow signal is relatively robust with respect to the chameleon potential at high densities.

Finally, we predict the afterglow signal for \GammeVCHASE~using a Monte Carlo simulation which we cross-check against an analytic approximation improving upon~\cite{Upadhye_Steffen_Weltman_2010}.  This prediction is shown to be unaffected by chameleon scattering from atoms inside the laboratory vacuum, robust with respect to surface roughness in the chamber walls, and relatively insensitive to chamber properties such as the reflectivity of the walls and the geometry of the apparatus.  \GammeVCHASE~data are analyzed using the profile likelihood method~\cite{Rolke_Lopez_Conrad_2005}.  The model-independent constraints of~\cite{Steffen_etal_2010} are reproduced and elaborated upon, then extended to a wider variety of chameleon models such as dark energy models.  We then place \GammeVCHASE~constraints in context by comparing them to other chameleon constraints as well as to forecasts.

The paper proceeds as follows.  After introducing photon-coupled chameleon theories in Sec.~\ref{sec:basics_of_chameleon_physics}, we study chameleon particle interactions with Fermions, photons, and other chameleon particles in Sec.~\ref{sec:interactions_of_chameleon_particles}.  Section~\ref{sec:reflection_from_a_barrier} looks at chameleon particle reflection from barriers of matter.  The computation of \cite{Upadhye_Steffen_Weltman_2010} is corrected and improved upon in  Sec.~\ref{sec:chameleon-photon_oscillation_analytic_calculation}, which uses simple approximations for the chameleon initial conditions and the magnetic field. A more accurate model of the magnetic field, and the effects of windows inside the magnetic field region, are studied in Sec.~\ref{sec:through_the_magnetic_field}.  Sec.~\ref{sec:enhancements_in_gammev-chase} compares \GammeV~and \GammeVCHASE.  A Monte Carlo simulation of \GammeVCHASE~is used in Sec.~\ref{sec:chameleon-photon_oscillation_monte_carlo_simulation} to compute decay and afterglow rates as well as to study the sensitivity of the afterglow signal to properties of the vacuum chamber.  Sec.~\ref{sec:analysis_and_constraints} presents the data analysis, discusses systematic uncertainties, and uses \GammeVCHASE~data to constrain several chameleon field models.  We conclude in Section~\ref{sec:conclusion}.

\section{Basics of chameleon physics}
\label{sec:basics_of_chameleon_physics}

\subsection{Action and effective potential}

We study photon-coupled scalar chameleon theories with actions $S = S_\phi + S_\gamma + S_\mathrm{m}$ of the following form:
\begin{eqnarray}
S_\phi
&=&
\int d^4x \sqrt{-g} 
\left[
  \frac{1}{2} \Mpl^2 R
  - \frac{1}{2}\partial_\mu\phi \partial^\mu\phi
  - V(\phi)
  \right]
\label{e:S_phi}
\\
S_\gamma
&=&
\int d^4x \sqrt{-g} 
\left[
  - \frac{1}{4} \exp\left(\frac{\bgam\phi}{\Mpl}\right) F_{\mu\nu} F^{\mu\nu}
\right]
\label{e:S_gamma}
\\
S_\mathrm{m}
&=&
\int d^4x \sqrt{-g} 
\left[
  {\mathcal L}_\mathrm{m}\left(\exp\left(\frac{2\bmat\phi}{\Mpl}\right)g_{\mu\nu}
  , \psi_\mathrm{m}^i\right)
  \right].
\label{e:S_matter}
\end{eqnarray}
Here, ${\mathcal L}_\mathrm{m}$ is the Lagrangian density for matter fields $\psi^i$  moving along geodesics of the metric $\exp(2\bmat\phi/\Mpl)g_{\mu\nu}$.  This is equivalent to a coupling between $\phi$ and the trace of the matter stress tensor $T_{\mu\nu}$.  Since $\phi$ will vary by much less than $\Mpl/ \bmat$ and $\Mpl/\bgam$ in cases of interest, the precise functional forms of the couplings to matter and photons are not important.  Expanding $\exp(\bgam\phi/\Mpl)$ and $\exp(2\bmat\phi/\Mpl)$ to linear order in the expressions for $S_\gamma$ and $S_\mathrm{m}$ respectively, we find the usual photon and matter actions plus linear couplings to the chameleon field.  

In the presence of an electric field $\vec E(\vec x)$ and a magnetic field $\vec B(\vec x)$ as well as a nonrelativistic matter density $\rho(\vec x) = -T^\mu_\mu$, these couplings give the scalar field an effective potential
\begin{equation}
\Veff(\phi,\vec x)
=
V(\phi)
+
\frac{1}{2}\frac{\bgam \phi}{\Mpl}\left(|\vec B|^2 -|\vec E|^2\right)
+
\frac{\bmat\phi}{\Mpl} \rho
\label{e:Veff}
\end{equation}
excluding terms suppressed by higher powers of $\beta\phi/\Mpl$.  The scalar field equation of motion is then 
\begin{equation}
\partial^\mu\partial_\mu\phi 
= 
\frac{\partial \Veff}{\partial \phi}
=
V_{,\phi}
+ \frac{\bgam}{2\Mpl} \left(|\vec B|^2 -|\vec E|^2\right)
+ \frac{\bmat \rho}{\Mpl}.
\label{e:eom}
\end{equation}

\subsection{Chameleon and thin-shell effects}

Consider a static matter density $\rho(\vec x,t) = \rho(\vec x) \gg |\vec B|,\,|\vec E|$.  The scalar equation of motion (\ref{e:eom}) reduces to 
\begin{equation}
\nabla^2 \phi = V_{,\phi} + \frac{\bmat}{\Mpl}\rho(\vec x).
\label{e:eom_static}
\end{equation}

If $V_{,\phi}$ and its derivatives are negligible, then (\ref{e:eom_static}) is similar in form to the Poisson equation $\nabla^2 \Psi = \rho(\vec x) / (2\Mpl^2)$ for the gravitational potential $\Psi$.  The requirement that $\phi$ and $\Psi$ remain finite as $|\vec x|\rightarrow\infty$ implies that $\phi = 2\bmat \Mpl \Psi + $~constant.  As the size and density of the matter distribution sourcing $\phi$ and $\Psi$ increases, these fields also grow.  This regime of negligible $V,\phi$ is known as the linear regime of the chameleon since the equation of motion is linear.  

Now suppose that, as $\phi$ grows beyond a certain point, $V_{,\phi} < 0$ begins to increase in magnitude rapidly and nonlinearly with $\phi$.  Then $V_{,\phi}$ will partially cancel the matter source on the right hand side of (\ref{e:eom_static}), slowing the growth of the field $\phi$.  For a sufficiently large and dense object, $\phi$ will approach its bulk solution defined by $V_{,\phi}(\phibulk) + \bmat \rho / \Mpl = 0$, turning off the source altogether.  This is known as the nonlinear regime; nonlinearities in the equation of motion (\ref{e:eom_static}) are essential to determining the behavior of the field.  The nonlinear regime is characterized by a rapid growth in the effective mass $\meff = V_{,\phi\phi}^{1/2}$ of the field.  Since a large mass decreases the range of the force mediated by $\phi$, the field is able to ``hide'' from fifth force constraints, an effect known as the chameleon effect.  

Next, consider an object of constant density $\rho_0$.  The change in the field from its background value, $\phi(\vec x) - \phi_\infty$, will be approximately $2\bmat\Mpl(\Psi(\vec x) -\Psi_\infty)$ in the linear regime and will saturate at $\phibulk(\rho_0)-\phi_\infty$ in the nonlinear regime.  We know that the nonlinear regime has been reached at a point $\vec x$ inside the object when 
\begin{equation}
|\phibulk(\rho_0)-\phi_\infty| 
\ll
2\bmat\Mpl|\Psi(\vec x) -\Psi_\infty|.
\label{e:nonlinearity_condition}
\end{equation}
Since the source on the right hand side of (\ref{e:eom_static}) vanishes in the nonlinear regime, the field $\phi$ outside the object is effectively sourced only by the portion of the object that is in the linear regime.  That is, the field ``sees'' only a thin outer shell of a sufficiently large and dense object.  Such an object is said to be in the nonlinear regime or to have a thin shell.

\subsection{Power law and dark energy potentials}

In order to provide concrete examples, we study chameleon potentials of the power law form 
\begin{equation}
V(\phi) = g \left|\phi\right|^n,
\label{e:V_powerlaw}
\end{equation}
with $g>0$ and either $n<0$ or $n>2$.   Given our convention $\bmat > 0$, the sign $\sigma_n = \sign(\phi)$ is positive for $n<0$ and negative for $n>2$.  Eq.~(\ref{e:V_powerlaw}) is useful as a large-field approximation to the potential $M_\Lambda^4 \exp(\kappa M_\Lambda^N / \phi^N) \approx M_\Lambda^4 + \kappa M_\Lambda^{4+N} \phi^{-N}$ frequently used in the literature, in which $M_\Lambda = 2.4\times 10^{-3}$~eV is the dark energy scale and $\kappa$ is a dimensionless constant. We refer to the power-law-plus-constant model 
\begin{equation}
V_\mathrm{de}(\phi)
=
M_\Lambda^4 
+
g \left|\phi\right|^n,
\quad
g =
\left\{
\begin{array}{ll}
  \lambda / 4!           & \textrm{for }n=4,     \\
  \kappa M_\Lambda^{4-n} & \textrm{for }n\neq 4, \\
\end{array}
\right.
\label{e:V_chameleon_dark_energy}
\end{equation}
where $\kappa$ and $\lambda$ are dimensionless constants, as ``chameleon dark energy.''  We will assume $\kappa=1$ for $n\neq 4$ unless stated otherwise; such a model uses the same energy scale $M_\Lambda$ in the constant and $\phi$-dependent terms of the potential.

The field which minimizes the effective potential (\ref{e:Veff}) in a bulk of matter density $\rhom$ and electromagnetic field Lagrangian density $\rhog = \frac{1}{2} \bgam \Mpl^{-1} (|\vec B|^2-|\vec E|^2)$ is
\begin{equation}
\phibulk
=
\sigma_n
\left(\frac{\bmat\rhom + \bgam\rhog}{|n|g\Mpl}\right)^\frac{1}{n-1}.
\label{e:phibulk}
\end{equation}
Since $\rhom \gg \rhog$ in almost all cases, $\phibulk$ is typically a function of $\rhom$ alone.  
Increasing $\rhom$ causes the magnitude of $\phi$ to decrease for $n<0$ and to increase for $n>2$.  

Differentiating $V$, we obtain $V_{,\phi} = n \sigma_n g |\phi|^{n-1}$, which is always negative as appropriate to a chameleon theory.  The effective mass $\meff^2 = V_{,\phi\phi}(\phibulk)$ is found by differentiating once again:
\begin{equation}
\meff(\phibulk)
=
\sqrt{n(n-1)g} \left(\frac{\bmat\rhom + \bgam\rhog}{|n|g\Mpl}\right)^\frac{n-2}{2n-2}.
\label{e:meffbulk}
\end{equation}
This increases with density if $n<0$ or $n>2$.

\subsection{Chameleon-photon oscillation}
\label{subsec:chameleon-photon_oscillation}

Consider the passage of a chameleon particle through a region of constant matter density $\rhom$ and external magnetic field $\vec \Bext$, with no electric field.  Variation of the action (\ref{e:S_gamma}) with respect to the electromagnetic field leads to
\begin{equation}
\partial_\mu \left( \exp\left(\frac{\bgam\phi}{\Mpl}\right) F^{\mu\nu}\right),
\end{equation}
with the other two of Maxwell's equations unchanged.  Oscillation between chameleon particles and photons can be described by perturbing about the background fields.
Writing these perturbations in terms of a dimensionless chameleon amplitude $\psic$ and a dimensionless photon amplitude $\psig$ in the same direction as the magnetic field perturbation, we have $(-\partial^2/\partial t^2 - \vec k^2)\psig = k \bgam \Bext \Mpl^{-1} \hat k \times (\hat x \times \hat k) \psic$ for a plane wave with momentum $\vec k = k \hat k$.  We have assumed without loss of generality that $\vec \Bext = \Bext \hat x$.  Here, $\hat k$ and $\hat x$ are unit vectors.  In the relativistic, weak-mixing approximation, this is solved by~\cite{Raffelt_Stodolsky_1988,Upadhye_Steffen_Weltman_2010}
\begin{eqnarray}
\psig(t)
&=&
-i e^{-ikt-\frac{i\meff^2 t}{4k}} \frac{2 k \bgam \Bext}{\meff^2 \Mpl}
\sin\left(\frac{\meff^2 t}{4k}\right) \vec a(\hat k) \quad
\label{e:psig}
\\
\Pgc 
&=&
\left|\psig \right|^2
=
\Cgc^2 \sin^2\left(\frac{\meff^2 t}{4k}\right)
\left|\vec a(\hat k)\right|^2
\label{e:Pgc}
\\
\vec a(\hat k) 
&=&
 \hat k \times (\hat x \times \hat k)
\label{e:a_vec}
\\
\Cgc 
&=& 
2k\bgam\Bext\Mpl^{-1}\meff^{-2}.
\label{e:Cgc}
\end{eqnarray}
Here, $\Pgc(t)$ is the probability that a particle beginning in a pure chameleon state $\psig(0)=0$ at time $t=0$ will be a photon when measured at time $t$.  $\Cgc$ is sometimes written in terms of the ``mixing angle'' $\varpi$, with $\Cgc = \sin(2\varpi)$.  Note that $|\vec a|^2$ is zero for a plane wave travelling parallel to the magnetic field and one for a wave perpendicular to the field.  In the low-mass limit, for a wave with $\vec k \perp \vec \Bext$, the oscillation probability simplifies to $\Pgc \approx \bgam^2 \Bext^2 t^2 / (4\Mpl^2)$.  The probability for a particle beginning in a pure photon state to oscillate into a chameleon is also given by (\ref{e:Pgc}).

\subsection{An idealized afterglow experiment}
\label{subsec:an_idealized_afterglow_experiment}

A chameleon particle with energy $\omega$ will be excluded by energy conservation from a region in which its effective mass $\meff \gg \omega$.  (We will study this exclusion in greater detail in Sections~\ref{sec:interactions_of_chameleon_particles} and \ref{sec:reflection_from_a_barrier}.)  Consider an evacuated chamber whose internal mass density is $\rhov$ and whose walls have a density $\rhom$.  If $\meff(\phibulk(\rhov)) \ll \omega \ll \meff(\phibulk(\rhom))$, then the particle will propate freely inside the chamber, but will not be able to penetrate its walls.  That is, the chameleon particle will be trapped inside the chamber.  We refer to the condition $\omega \ll \meff(\phibulk(\rhom))$ as ``chameleon containment.''  

\begin{figure}[tb]
\begin{center}
\includegraphics[width=3.3in]{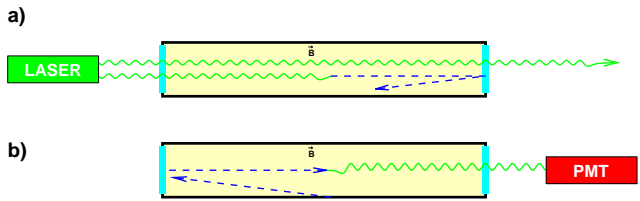}
\caption{An idealized afterglow experiment.  ({\em a}): Production phase.  Photons stream through chamber via entrance and exit windows, occaisionally oscillating into chameleon particles which are trapped inside.  ({\em b}):  Afterglow phase.  The photon source is turned off and a detector is uncovered.  Chameleon particles oscillate back into photons, which emerge from the chamber and reach the detector.  \label{f:gammev_apparatus_idealized}}
\end{center}
\end{figure}

Afterglow experiments rely on oscillation and containment to produce, trap, and detect chameleon scalar fields.  Figure~\ref{f:gammev_apparatus_idealized} shows a simple, idealized afterglow experiment.  An evacuated cylindrical chamber has metal walls and two glass windows.  Matter densities in the walls and windows are high enough that the chameleon containment condition is satisfied.  A large magnetic field $\vec \Bext$ inside the chamber points in a direction perpendicular to the cylinder axis.  

In the production phase of the experiment, shown in Fig.~\ref{f:gammev_apparatus_idealized}~({\em a}), photons are streamed through the chamber via the windows.  The background magnetic field allows some of them to oscillate into chameleons.  Since these chameleon particles are trapped, a population of chameleons builds up inside the chamber.  In the afterglow phase, Fig.~\ref{f:gammev_apparatus_idealized}~({\em b}), the photon source is turned off while the magnetic field is maintained.  Chameleon particles propagating in this magnetic field oscillate back into photons.  These regenerated photons may escape from the chamber through the windows, leading to an ``afterglow'' of photons from the chamber.

In order to predict the afterglow signal expected for a given chameleon model in such an experiment, we must compute the rate $\Gdec$ at which the chameleon population decays by photon regeneration, as well as the rate $\Gaft$ at which each chameleon particle produces detectable afterglow photons.  (We assume that photon regeneration is the dominant chameleon loss mode; we will consider another possibility in Sec.~\ref{subsec:chameleon_fragmentation}.)  Given these rates, the expected afterglow signal is shown in~\cite{Upadhye_Steffen_Weltman_2010} to be
\begin{equation}
F_\mathrm{aft}(t)
=
\frac{\Fgam \Pgc \Gaft}{\Gdec} \left(1-e^{-\Gdec \tprod}\right)e^{-\Gdec t}
\label{e:Faft}
\end{equation}
during the afterglow phase, $t>0$.  Here $\Fgam$ is the rate at which photons are streamed through the chamber during the production phase, and $\tprod$ is the duration of the production phase.

Sections~\ref{sec:chameleon-photon_oscillation_analytic_calculation} and \ref{sec:chameleon-photon_oscillation_monte_carlo_simulation} present accurate calculations of $\Gdec$ and $\Gaft$ for realistic afterglow experiments.  Here we can compute very rough estimates for our idealized experiment from Fig.~\ref{f:gammev_apparatus_idealized}.  Note that these estimates will not even be correct at the order-of-magnitude level; experimental constraints must be based on the accurate calculations of Sec.~\ref{sec:chameleon-photon_oscillation_monte_carlo_simulation}.  If the total chamber length is $\ltot$, then the total chameleon-photon conversion probability at low $\meff$, averaging over all angles, will be $\sim \Pgc = \bgam^2 \Bext^2 \ltot^2 / (4 \Mpl^2)$ to within a few orders of magnitude.  The time taken for a relativistic chameleon particle to travel between the windows, again averaging over angles, will be of order $\ltot$.  The decay rate will be the conversion probability per unit time,  $\sim \bgam^2 \Bext^2 \ltot / (4 \Mpl^2)$.  The fraction of these photons reaching a detector of size $r_\mathrm{det}$ outside the chamber will be of order $(r_\mathrm{det}/\ltot)^2$.  For $\ltot \sim 1$~m and $r_\mathrm{det} \sim 1$~cm, we would therefore expect the afterglow rate to be about four orders of magnitude less than the decay rate.

\section{Interactions of chameleon particles}
\label{sec:interactions_of_chameleon_particles}

The goal of this section is to compute the cross section for a chameleon particle to interact with other particles which it would encounter in the diffuse gas inside the vacuum chamber.  In particular, we are interested in the chameleon-atom scattering cross section.  The chameleon particle can interact directly with the proton and electron which make up a hydrogen atom.  It can also scatter from the static chameleon field sourced by the mass density of the atom.  We shall see that the latter effect is dominant, and that the cross section is approximately the square of the proton radius, so that chameleon-atom scattering is a negligible effect in an afterglow experiment.  We conclude with a discussion of interactions between two chameleon particles.

\subsection{Scattering from Fermionic point-particles}
\label{subsec:scattering_from_fermionic_point-particles}

Consider a scalar particle of fixed mass $m_\phi$ scattering from a Fermion of mass $\mf$.  The matter coupling from (\ref{e:S_matter}), with $-T_\mu^\mu = \mf {\bar \psi} \psi$, implies a Yukawa interaction between the scalar and the Fermion with coupling constant $\gf = \bmat \mf /\Mpl$.  Assuming that $\bmat \ll 10^{19}$, this will be small for nucleons and lighter Fermions.  Since the s- and t-channel Feynman diagrams have two vertices, the cross section $\sigphiF$ will be suppressed by four powers of $\gf$.  

Chameleon-photon oscillation is strongly suppressed for nonrelativistic chameleons, since the chameleon and photon do not remain in phase.  Thus we are interested in the case $m_\phi \ll p_\phi$, where $p_\phi$ is the chameleon momentum.  Furthermore, chameleons in afterglow experiments are produced by lasers and detected in photomultiplier tubes, so we expect $p_\phi \sim 1$~eV, much less than the masses of electrons and nucleons.  The limit applicable to afterglow experiments is $m_\phi \ll p_\mathrm{CM} \ll \mf$, where $p_\mathrm{CM} \sim p_\phi$ is the particle momentum in the center-of-mass frame.  To lowest order the cross section is 
\begin{equation}
\sigphiF
=
\frac{\gf^4}{24\pi \mf^2}.
\label{e:sigma_phi_fermion}
\end{equation}

For a proton or neutron, treated as a point particle, this is $6 \gf^4 \times 10^{-34}$~m$^2$ with $\gf = 4 \times 10^{-19} \bmat \ll 1$.  At fixed $\bmat$, $\sigphiF$ scales as $\mf^2$, so the cross section for an electron is smaller by six orders of magnitude.  Since electrons and nucleons are present in roughly equal numbers, we may neglect the scattering between chameleons and electrons treated as point particles.

\subsection{Chameleon scattering from background chameleon field}
\label{subsec:chameleon_scattering_from_background_chameleon_field}

Given a static matter density $\rho_0(\vec x)$ (assuming negligible $|\vec B|$ and $|\vec E|$) with corresponding static solution $\phi_0(\vec x)$, we can linearize the equation of motion (\ref{e:eom}) about $\phi_0$.  With $\phi(\vec x, t) = \phi_0(\vec x) + \delta\phi(\vec x,t)$ we find
\begin{equation}
\left(\Box - \meff(\vec x)^2\right)\delta\phi = 0.
\label{e:eom_linear}
\end{equation}
This tells us that, as the effective mass $\meff(\vec x) = \meff(\phi_0(\vec x)) = V_{,\phi\phi}(\phi_0(\vec x))^{1/2}$ varies with position, the total energy $E = (p^2 + \meff^2)^{1/2}$ of a chameleon particle remains constant.  

As a chameleon particle approaches an object with a thin shell, nonlinearity in $V_{,\phi}$ causes $\meff$ to rise sharply.  Energy conservation prevents a chameleon particle of energy $E$ from entering a region in which $\meff > E$; such a particle will bounce off of the object.  Thus an object of density $\rho_0$ with a thin shell will scatter chameleon particles with energies less than $\meff(\phibulk(\rho_0))$.  Although tunneling is possible, it is negligible for macroscopic objects such as the glass windows used in afterglow experiments.

The scattering of chameleon particles from the background field $\phi_0$ of a massive object is used by afterglow experiments to trap chameleon particles.  An evacuated chamber with a ``vacuum'' of density $\rhov$ allows chameleon particles of energy $E > \meff(\phibulk(\rhov))$.  These same particles will bounce from the chamber walls, of density $\rhom$, if the containment condition 
\begin{equation}
E < \meff(\phibulk(\rhom))
\label{e:containment}
\end{equation}
is satisfied.  In \GammeVCHASEfull~$\rhov \sim 10^{-14}$~g/cm$^3$ and $\rhom \sim 1$~g/cm$^3$.  This difference of fourteen orders of magnitude means that chameleon particles with a large range of potentials can be trapped inside the vacuum chamber.

\subsection{Scattering from atoms}

We have studied chameleon scattering from pointlike Fermions as well as extended matter distributions.  An atom is both.  Here, we model the proton and the electron cloud as uniform-density spheres of radius $\rp = 0.83$~fm and $\rb = 0.529$~\AA, respectively, in a laboratory vacuum of density $\rhov = 10^{-14}$~g/cm$^3$.  
We compare the resulting cross sections to those obtained in Sec.~\ref{subsec:scattering_from_fermionic_point-particles}.

In our approximation, the electron cloud, with mass $\mel = 511$~keV/c$^2$, has a density $\rhoe = 3 \mel / (4\pi \rb^3) = 1.5\times 10^{-3}$g/cm$^3$.  Defining the gravitational potential to be zero at infinity, we have $\Psie(0) = -\rhoe \rb^2 / (4\Mpl^2) = -1.9\times 10^{-47}$ at the center of the cloud.  The proton has mass $\mpr = 938$~MeV/c$^2$, mean density $\rhop = 7.0\times 10^{14}$g/cm$^3$, and gravitational potential $\Psip(0) = -2.2\times 10^{-39}$.  The electron cloud and the proton, respectively, will have thin shells only if 
\begin{eqnarray}
|\phibulk(\rhoe) - \phi_\infty| < 2\bmat\Mpl|\Psie(0)|.
\label{e:electron_thinshell}
\\
|\phibulk(\rhop) - \phi_\infty| < 2\bmat\Mpl|\Psip(0)|.
\label{e:proton_thinshell}
\end{eqnarray}
Recall that an object with a thin shell represents a large perturbation to the background chameleon field, from which incident chameleon particles may scatter.

\begin{figure}[tb]
\begin{center}
\includegraphics[angle=270,width=3.3in]{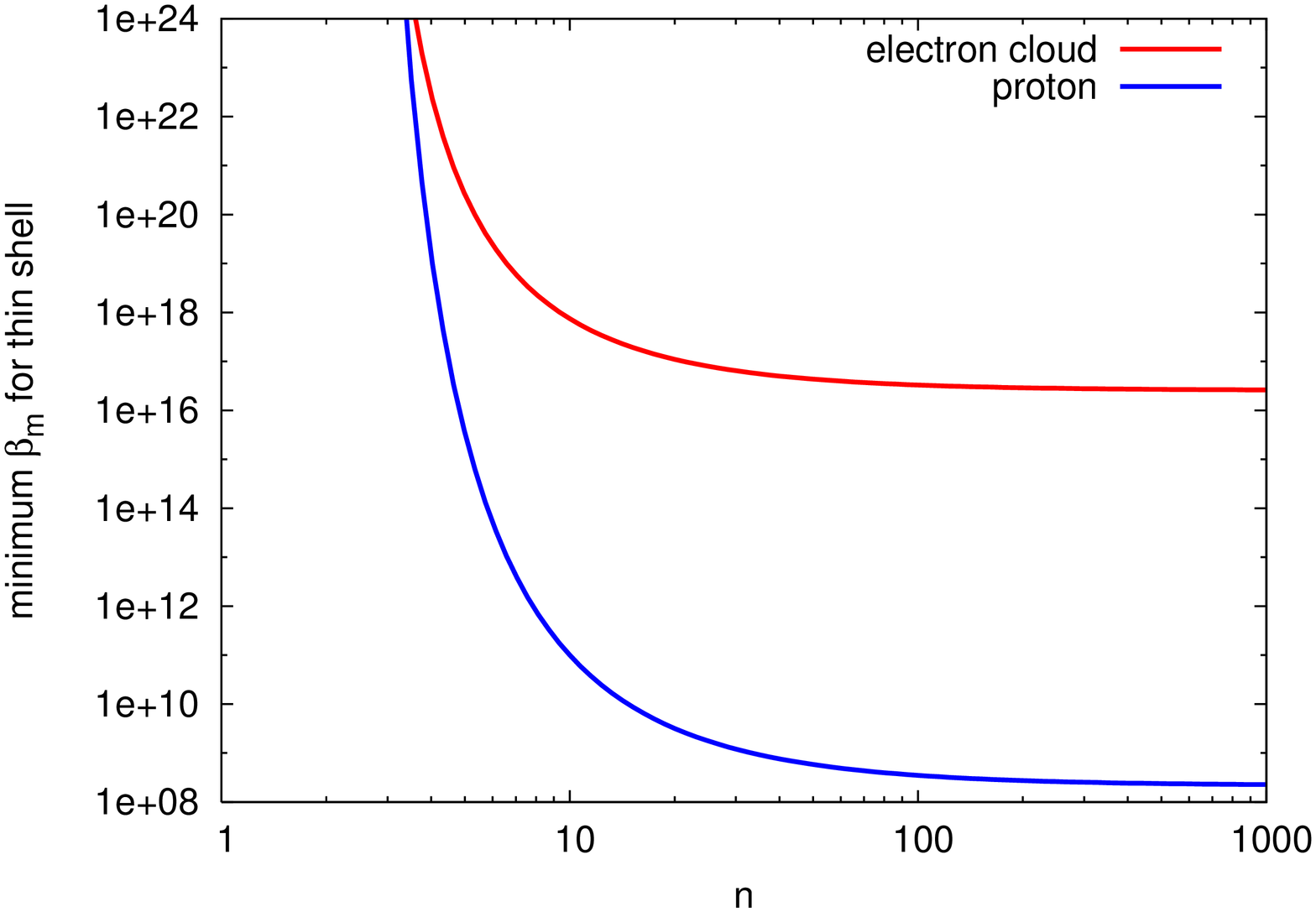}
\includegraphics[angle=270,width=3.3in]{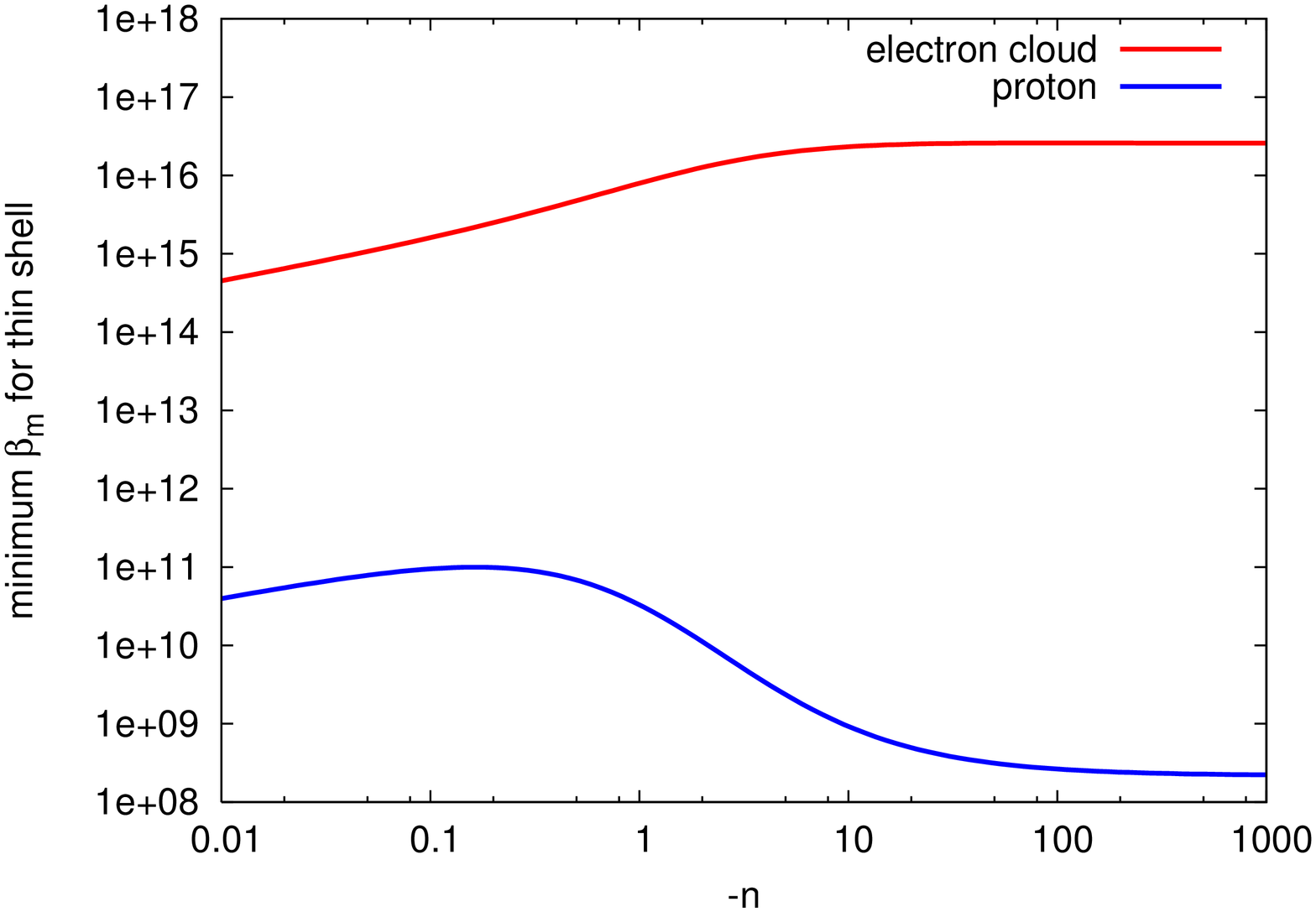}
\caption{Minimum value of the matter coupling $\bmat$ required for the electron cloud and the proton, respectively, to have thin shells, in the case of a chameleon dark energy $V(\phi) = M_\Lambda^4 + M_\Lambda^{4-n}|\phi|^n$.  \FigA: $n>2$.  \FigB: $n<0$.  \label{f:beta_min}}
\end{center}
\end{figure}

Consider power law potentials of the form (\ref{e:V_powerlaw}).  If $n>2$, then the left hand sides of (\ref{e:electron_thinshell},~\ref{e:proton_thinshell}) will be dominated by $|\phibulk|$.  For chameleon dark energy, $V(\phi) = M_\Lambda^4 + M_\Lambda^{4-n}|\phi|^n$, the thin shell condition becomes 
\begin{equation}
\bmat
>
\left(\frac{\rho}{|n|M_\Lambda^3 \Mpl}\right)^\frac{1}{n-2}
\left(\frac{M_\Lambda}{2\Mpl |\Psi|}\right)^\frac{n-1}{n-2}
\label{e:beta_min}
\end{equation}
with the density $\rho$ and gravitational potential $\Psi$ appropriate to each object.
Figure~\ref{f:beta_min}~\FigA~shows the minimum values of $\bmat$ necessary for each object to have a thin shell.  
If $n<0$, then the left hand sides of (\ref{e:electron_thinshell},~\ref{e:proton_thinshell}) will be dominated by the background field value $|\phi_\infty|$.  Assuming that the vacuum chamber containing the atom is much larger than the chameleon Compton wavelength $\meff(\phibulk(\rhov))^{-1}$ at the vacuum density $\rhov$, the thin shell condition is just (\ref{e:beta_min}) with $\rho = \rhov$.  Fig.~\ref{f:beta_min}~\FigB~shows the minimum $\bmat$ required for a thin shell in the $n<0$ case.

Colliders already exclude $\bmat \gtrsim 10^{15}$ for similar chameleon models~\cite{Brax_etal_2010}.  In the remaining parameter space, the electron cloud will only have a thin shell when $n$ is tuned to be very close to zero, $-10^{-2} \lesssim n < 0$, and we do not consider this case further.  The proton does have a thin shell in a substantial fraction of the allowed parameter space.  Thus a chameleon particle incident upon a hydrogen atom will pass right through the electron cloud, which represents only a small perturbation to the background mass, and will interact solely with the proton.

In order to find the background chameleon field $\phi_0(r)$ due to the proton, we solve~(\ref{e:eom_static}) for the spherical tophat density distribution by which we approximate the proton.  Since the boundary conditions $\phi_0'(0) = 0$ and $\lim_{r\rightarrow\infty} \phi(r) = \phi_\infty$ are defined at different $r$, we solve~(\ref{e:eom_static}) numerically using the shooting method: we guess a value of $\phi_0(0)$, solve to find the field at large $r$, and refine our guess.  Since scattering will be important only when the proton has a thin shell, $\meff(\phibulk(\rhop)) r \gg 1$, we focus on this regime.  We immediately run into numerical difficulties due to the exponential sensitivity of the large-$r$ field to the central value.  This sensitivity can be seen by linearizing $\phi_0(r) = \phibulk(\rhop) + \delta\phi(r)$ inside the proton, resulting in $\nabla^2 \delta\phi = \meffprot^2 \delta\phi$ with $\meffprot = \meff(\phibulk(\rhop))$.  The linearized equation has the one-parameter family of solutions $\delta\phi(r) = F \sinh(\meffprot r) / (\meffprot r)$, valid as long as $|\delta \phi| \ll |\phibulk(\rhop)|$.  Although this linear approximation will be inapplicable at the proton's surface, $r=\rp$, it will be valid for $r < \rp - \Delta r$ for some $\Delta r$.  In the thin shell case, we can choose  $\meffprot^{-1} \ll \Delta r \ll \rp$, begin our numerical solution at $\rp - \Delta r$ using the linear approximation, and apply the shooting method to $F$ rather than $\phi(0)$.  

\begin{figure}[tb]
\begin{center}
\includegraphics[angle=270,width=3.3in]{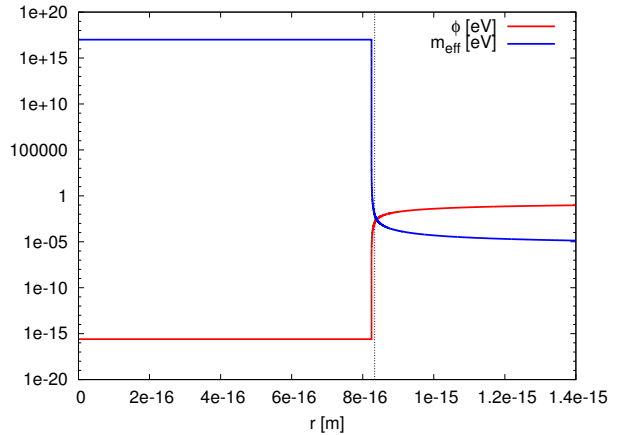}
\caption{Chameleon field and effective mass for a proton, approximated as a uniform-density ball of mass $\mpr = 938$~MeV/c$^2$ and radius $\rp = 0.83$~fm.  The potential $V(\phi)$ is that of a chameleon dark energy with $n=-1$ and a matter coupling $\bmat = 10^{12}$.  A vertical dotted line shows the surface of the proton.  \label{f:phi_meff_proton}}
\end{center}
\end{figure}

Figure~\ref{f:phi_meff_proton} shows the result of this calculation for a chameleon dark energy with $n=-1$ and $\bmat = 10^{12}$.  For this model, $\phibulk(\rhop) = 2.5\times 10^{-16}$~eV and $\meffprot = 9.9\times 10^{16}$~eV, implying a Compton wavelength of $\meffprot^{-1} = 2.4\times 10^{-9}\rp$.  Choosing $\Delta r \approx 10^{-2} \rp$, we find using the shooting method that $\log(F/\phibulk(\rhop)) = -8.58\times 10^{9}$.  At $r = \rp - \Delta r$, $\log|\delta\phi/\phibulk(\rhop)| = -8.17\times 10^9$, well within the regime of validity of the linearized equation of motion.

This numerical result is consistent with theoretical expectations.  Since the effective mass inside the proton deviates from $\meffprot$ for $r > \rp - \Delta r$, the proton has a thin shell of thickness $\lesssim \Delta r$, which includes a fraction $\approx 3 \Delta r / \rp$ of its mass.  The surface gravitational potential of this mass shell is $\Psi_\mathrm{shell} = -\frac{3 \mpr \Delta r}{8\pi \Mpl^2 \rp^2} = -4.5\times 10^{-41}$, or of order $1\%$ of the total gravitational potential.  Then $2\bmat |\Psi_\mathrm{shell}| = 0.2$~eV, of the order of $\phi_\infty = \phibulk(\rhov) = 0.07$~eV.  Therefore this  shell alone is enough to saturate the chameleon field inside the proton.

Suppose that a chameleon particle approaches the proton from large $r$ with an energy $E_\phi \sim 1$~eV characteristic of a laser oscillation experiment.  Evidently from Fig.~\ref{f:phi_meff_proton}, the effective mass will rise to equal $E_\phi$ at $r \approx \rp - \Delta r \approx \rp$.  Thus the incoming chameleon particle will scatter off of a background chameleon ``ball'' of radius approximately $\rp$, implying a cross section of $4\pi \rp^2$.  In our coupling constant regime of interest, $\bmat \lesssim 10^{15}$, for which $\gf \ll 1$, this will be much larger than $\sigphiF \sim \gf^4 / \mpr^2 \sim \gf^4 \rp^2$ from (\ref{e:sigma_phi_fermion}).  Therefore, the cross section for chameleon-hydrogen scattering is dominated by semiclassical scattering of chameleon particles by the background chameleon potential of the proton, which is well-approximated by a hard sphere of radius $\rp$.  This is true for any chameleon model for which the proton, but not the electron cloud, has a thin shell.

For laser oscillation experiments such as \GammeVCHASEfull, chameleon-atom scattering has a negligible effect on the chameleon afterglow signal.  The chameleon-atom scattering rate for a vacuum density of $\rhov = 10^{-14}$~g/cm$^3$ made up almost entirely of hydrogen atoms is $\sigma n = 4\pi \rp^2 \rhov / \mpr = 1.6\times 10^{-5}$Hz, or of order one scattering event per chameleon per $10^{12}$ passes through a $10$~meter chamber.  Thus we expect atom scattering to correct our predicted afterglow signal at the $10^{-12}$ level.  Incidentally, even if the electron cloud also had a thin shell, and we replaced $\rp$ in the above expression by the Bohr radius, the rate would go up by $(\rb/\rp)^2 \sim 10^{10}$.  Thus atom scattering would still only be a percent-level effect.

\subsection{Chameleon fragmentation}
\label{subsec:chameleon_fragmentation}

Fragmentation is the process by which one chameleon particle interacts with another to produce more than two chameleon particles.  Since photomultiplier tubes are sensitive only to energies of order $1$~eV, repeated fragmentation would result in a population of chameleon particles with too little energy to produce detectable photons. Thus fragmentation lowers the signal expected in afterglow experiments.

Let the cross section for chameleon fragmentation be $\sigfrag$.  The fragmentation rate of chameleons at an inital energy of $\omega$, $\Gfrag = \Nphiw \sigfrag / \Vvac \equiv \Nphiw \sfrag$, depends on the number $\Nphiw$ of chameleons at that energy as well as on the volume $\Vvac$ of the  vacuum chamber; $\sfrag$ is the cross section per unit volume.  

Next we compute $\Nphiw$ as a function of time for a typical afterglow experiment.  At initial time $-\tprod$, the number of chameleon particles is zero.  During the production phase, $-\tprod < t \leq 0$, photons of energy $\omega \sim 1$~eV are streamed through the vacuum chamber at a rate $\Fgam$, which is about $8.8\times 10^{18}$~Hz in \GammeVCHASE.  Each of these has a probability $\Pgc$ of oscillating into a chamelelon particle.  At time $t=0$, the photon source is turned off, and the population of detectable chameleons decreases from its peak of $\Nphiw(0)$ due to photon regeneration (with a rate $\Gdec$) as well as fragmentation.  At $t>0$, the afterglow phase of the experiment, a detector outside the chamber can look for evidence of chameleon-photon oscillation.

The evolution of $\Nphiw$ is described by
\begin{equation}
\frac{d\Nphiw}{dt}
=
\Fgam \Pgc \Theta(-t)
- \Gdec \Nphiw
- \sfrag {\Nphiw}^2
\end{equation}
where $\Theta$ is the step function.  In the production phase $t \leq 0$,
\begin{equation}
\Nphiw
=
\frac{2\Fgam\Pgc \sinh\left(\frac{\Gtot(t+\tprod)}{2}\right)}
{\Gtot \cosh\left(\frac{\Gtot(t+\tprod)}{2}\right) + \Gdec \sinh\left(\frac{\Gtot(t+\tprod)}{2}\right)}
\end{equation}
where $\Gtot^2 \equiv {\Gdec^2 + 4 \Fgam \Pgc\sfrag}$.  In the afterglow phase $t > 0$,
\begin{equation}
\Nphiw
=
\frac{\Nphiw(0)}
{\left(1 + \frac{\sfrag \Nphiw(0)}{\Gdec}\right)e^{\Gdec t}
  - \frac{\sfrag \Nphiw(0)}{\Gdec}}.
\end{equation}
At times much smaller than the decay time, $0 < t \ll \Gdec^{-1}$, the chameleon population is approximately $\Nphiw(0) / (1 + (\sfrag\Nphiw(0) + \Gdec)t)$; fragmentation will dominate over decay if $\sfrag \Nphiw(0) \gg \Gdec$.  Since the fragmentation rate decreases with $\Nphiw$, decays via photon regeneration will eventually come to dominate, leading to a chameleon population of $\Nphiw(0) e^{-\Gdec t} / (1 + \sfrag \Nphiw(0)/\Gdec)$.

\begin{figure}[tb]
\begin{center}
\includegraphics[width=3.3in]{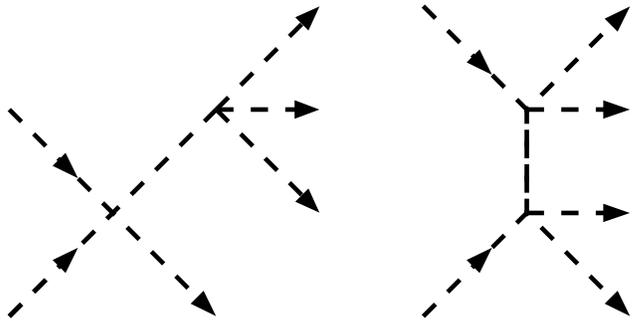}
\caption{Fragmentation processes in $\phi^4$ chameleon models.  \label{f:phi4frag}}
\end{center}
\end{figure}

Thus far our discussion has used the purely phenomenological parameter $\sigfrag$.  For the particular case of $\phi^4$ theory, we can estimate the fragmentation cross secton from processes such as those in Figure~\ref{f:phi4frag} using dimensional analysis, $\sigfrag  = \afrag \lambda^4 / \omega$.  Here $\afrag \ll 1$ is a dimensionless numerical factor resulting from an integral over the four-body phase space of outgoing particles.  We assume $\afrag = 1$ in order to be conservative; as we shall see in Section~\ref{sec:analysis_and_constraints}, \GammeVCHASE~constraints are not competitive with Casimir force constraints for this potential, so a lengthy numerical calculation of $\afrag$ is unwarranted.  This order-of-magnitude calculation shows that fragmentation cannot be neglected in \GammeVCHASE~for $\lambda \gtrsim 0.001$.

In principle, any potential of the forms (\ref{e:V_powerlaw},\ref{e:V_chameleon_dark_energy}) can be expanded in Taylor series about the expectation value $\phibulk$ of the field in a given matter density.  Consider chameleon dark energy with $n=-1$.  The $\ell$-th order term in the series for $\ell > 4$, $(-1)^\ell M_\Lambda^5 \phibulk^{-(\ell+1)}  (\phi-\phibulk)^\ell \equiv v_\ell (\phi-\phibulk)^\ell$, which describes a process in which two chameleon particles fragment into $\ell-2$, should contribute a quantity $\sigma_\ell \sim |v_\ell|^2 \omega^{2\ell-10} = (M_\Lambda/\omega)^{10} (\omega/\phibulk)^{2\ell} / \phibulk^2 $ to $\sigfrag$, by dimensional analysis.  The prefactor $(M_\Lambda / \omega)^{10} \sim 10^{-30}$ in \GammeVCHASE.  However, for typical parameter values, $\phibulk$ can be less than $\omega$, implying that $\sigma_\ell \rightarrow \infty$ as $\ell \rightarrow \infty$.  The same is true for other $n<0$; $\phibulk$ is typically within several orders of magnitude of the energy scale $M_\Lambda$ in the potential.  

The underlying problem is that chameleon theories are low-energy effective field theories whose cutoff energies can be below the particle energy $\omega \sim 1$~eV in an afterglow experiment.  A proper calculation of the fragmentation cross section would require a more fundamental theory, but UV completions of chameleon theories are not yet well-understood~\cite{Hinterbichler_Khoury_Nastase_2011}, so such a calculation is beyond the scope of this paper.  Henceforth we neglect chameleon fragmentation for all potentials other than $\frac{\lambda}{4!}\phi^4$.

\section{Reflection from a barrier}
\label{sec:reflection_from_a_barrier}


\subsection{A barrier as a lattice of atoms}
\label{subsec:barrier_as_lattice_of_atoms}

Section~\ref{subsec:chameleon_scattering_from_background_chameleon_field} showed that a chameleon particle of energy $E$ will be excluded by, and will therefore reflect from, a region in which the effective chameleon mass is greater than $E$.  For a homogeneous object of density $\rho_0$ which satisfies the nonlinearity condition (\ref{e:nonlinearity_condition}), this reflection condition becomes $\meff(\phibulk(\rho_0)) > E$.  

However, ordinary matter is not homogeneous, and homogeneity is not necessarily a good approximation in the case of a nonlinear field such as the chameleon.  
Mota and Shaw, in references~\cite{Mota_Shaw_2006,Mota_Shaw_2007}, considered matter as a cubic lattice of homogeneous spheres of radius $r_0$, with a lattice spacing $d_0 \gg r_0$.  
Since the spheres approximate atomic nuclei, we expect $r_0 \sim 1$~fm and $d_0 \sim 1$~\AA.  
The mean density $\rho_0 = M_0 d_0^{-3}$ of such matter depends on the mass $M_0$ of each sphere as well as the spacing $d_0$, but is independent of $r_0$.  Suppose that at some large value of $r_0$ one sphere considered individually is in the linear regime of a chameleon theory.  
As $r_0$ is decreased beyond a certain point, the sphere will acquire a thin shell, and the chameleon will effectively ``see'' only a fraction of it. 
Thus the nonlinear chameleon theory will be sourced by a density smaller than the mean density $\rho_0$, and its mass will be smaller than $\meff(\phibulk(\rho_0))$.  Refs.~\cite{Mota_Shaw_2006,Mota_Shaw_2007} approximate the actual chameleon mass inside such matter as the lesser of $\meff(\phibulk(\rho_0))$ and 
\begin{equation}
m_\mathrm{crit} 
\approx
\sqrt{3|n-1|} d_0^{-1} (r_0/d_0)^{q(n)/2},
\end{equation}
with $q(n) = \min(1,\frac{n-4}{n-1})$, for a power law or chameleon dark energy potential.

\begin{figure}[tb]
\begin{center}
\includegraphics[angle=270,width=3.3in]{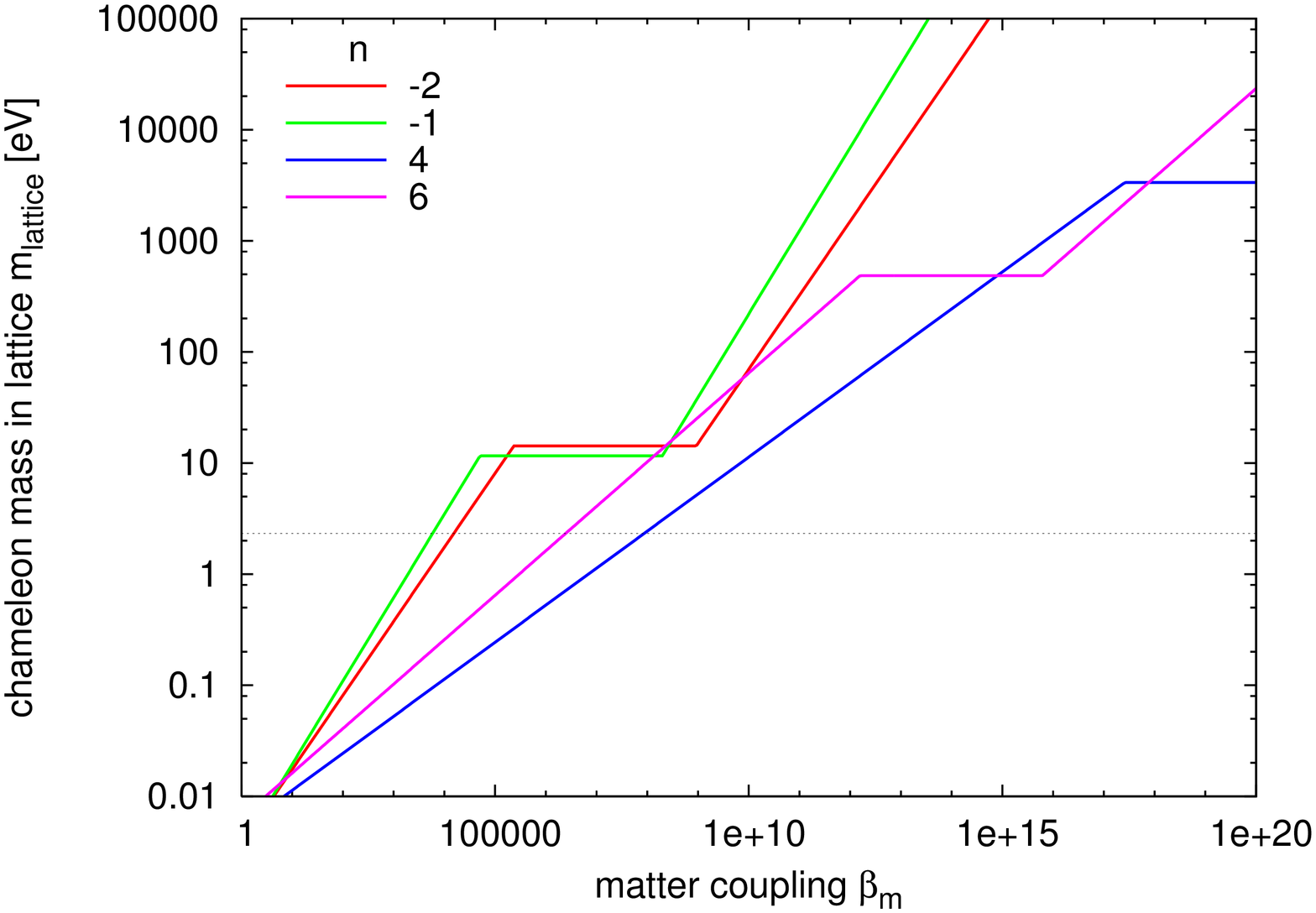}
\includegraphics[angle=270,width=3.3in]{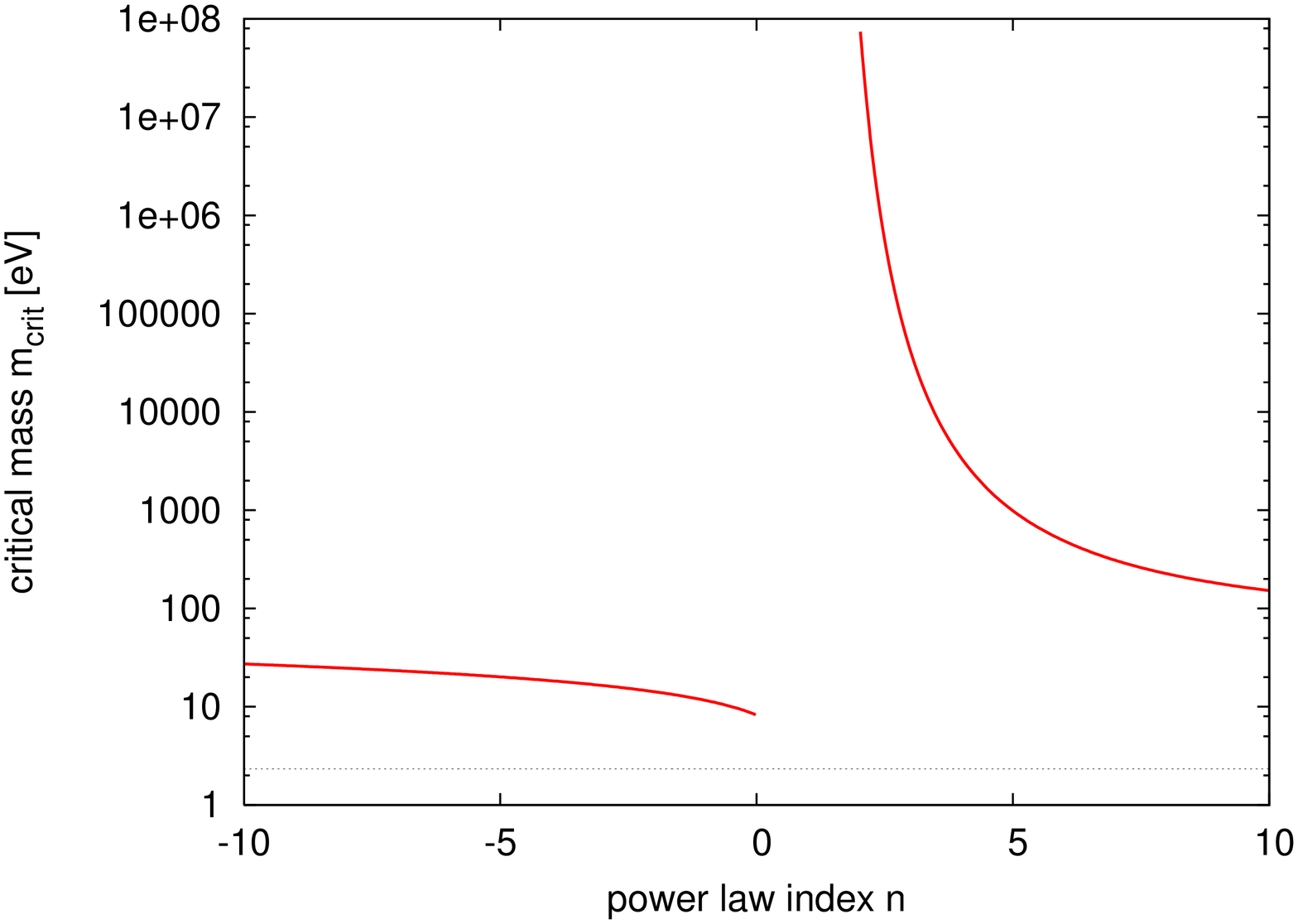}
\caption{Chameleon mass in matter, modelled as a lattice of spherical nuclei in a homogeneous gas of electrons.  In each plot, the horizontal dotted line shows $m = \omega = 2.33$~eV, the chameleon energy in \GammeVCHASE.  \FigA~Approximate chameleon mass.  \FigB~Critical mass of refs.~\cite{Mota_Shaw_2006,Mota_Shaw_2007}, the mass above which the inhomogeneous nature of the matter becomes apparent to the chameleon field.    \label{f:m_lattice}}
\end{center}
\end{figure}

Well inside the nonlinear regime, when the fraction of each nucleus seen by the chameleon is much less than $\mel/\mpr \sim 10^{-3}$, the chameleon will effectively ignore the nuclei and be sourced by the approximately homogeneous gas of electrons making up the solid.  This gas has a density $\rho_{0\mathrm{,e}} = \rho_0 (\mel/\mpr) (\mathcal{Z}_0/\mathcal{A}_0)$ where $\mathcal{Z}_0$ and $\mathcal{A}_0$ are, respectively, the atomic number and mass number of the atoms making up the lattice.  Even a highly nonlinear chameleon field should have a mass no lower than $\meff(\phibulk(\rho_{0\mathrm{,e}}))$.  Thus we approximate the chameleon mass in such a lattice by
\begin{eqnarray}
\mlatt
&\approx&
\max\Big[ \meff(\phibulk(\rho_{0\mathrm{,e}})), 
\nonumber\\
&&\qquad\quad
\min(m_\mathrm{crit}, \: \meff(\phibulk(\rho_0)))
\Big].
\end{eqnarray}
Figure~\ref{f:m_lattice}~\FigA~shows $\mlatt$ as a function of the matter coupling $\bmat$ for several power law dark energies.  Three regimes are seen for each model.  At low $\bmat$ the chameleon effectively sees a homogeneous solid, and the mass corresponds to the bulk field value expected for such a solid.  At intermediate $\bmat$ the mass flattens out as a greater portion of each atomic nucleus is ``hidden'' behind a thin shell.  At the largest $\bmat$, the homogeneous electron gas dominates the density seen by the chameleon field, causing the chameleon mass to rise with $\bmat$ once again.  

In each case, the mass required for chameleon containment, $\meff = \omega = 2.33$~eV in the case of \GammeVCHASE, lies in the low-coupling regime.  Figure~\ref{f:m_lattice}~\FigB~generalizes this conclusion to chameleon dark energy models of arbitrary $n$.  Thus we are justified in treating the chamber wall as a homogeneous solid of density $\rho_0$ for the purpose of testing the containment condition. 

\subsection{Phase change due to reflection}
\label{subsec:phase_change_due_to_reflection}

Let there be a vacuum $\rho=0$ in the region $x \geq 0$ and a constant density $\rhom$ at $x<0$.  This is a simple model for a planar slab thick enough that the chameleon reaches its bulk value.  Given this density and a power law potential, (\ref{e:eom_static}) is solved in the region $x \geq 0$ by
\begin{eqnarray}
\phi_0(x)
&=&
\phiw \left(1 + \sqrt{g/2}|n-2| \phiw^\frac{n-2}{2} x\right)^{-\frac{2}{n-2}}\quad
\label{e:phi0_planar}
\\
\phiw
&=&
(1-1/n)\phibulk(\rhom)
\label{e:phiw_planar}
\\
\meff(x)
&=&
\meffw \left(1 + \frac{|n-2| \meffw \; x }{\sqrt{2n(n-1)}}\right)^{-1}
\label{e:meff0_planar}
\\
\meffw
&=&
\sqrt{\frac{g(n-1)^{n-1}}{n^{n-3}}} \left(-\frac{\bmat \rhom}{n g \Mpl}\right)^\frac{n-2}{2n-2},
\label{e:meffw_planar}
\end{eqnarray}
where a subscript ``surf'' denotes the value of a function at the surface $x=0$.

We wish to find solutions to (\ref{e:eom}) representing plane wave perturbations to $\phi_0(x)$ far away from the wall, $\phi-\phi_0 \propto e^{-i\omega t}$ when $x \gg 0$.  With the definition $\phi(x,t) = \phi_0(x) + e^{-i \omega t}\delta\phi(x)$, (\ref{e:eom_linear}) reduces to 
\begin{equation}
\delta\phi''(x)
=
\left(\frac{\ell^2}{(x+\ell \omega/\meffw)^2} - 1\right) \omega^2 \delta\phi,
\label{e:eom_linear_planar}
\end{equation}
with the length $\ell = \omega^{-1}\sqrt{2n(n-1)}/|n-2|$ of order the wavelength of the plane wave.  This is solved by
\begin{eqnarray}
\delta\phi
&=&
\tilde x^{1/2}
 \left(c_1 J_\alpha(\tilde x) + c_2 Y_\alpha(\tilde x)\right)
\\
\tilde x
&=&
\omega(x + \ell\omega/\meffw)
\\
\alpha^2
&=&
\omega^2\ell^2 + 1/4
\end{eqnarray}
where $J_\alpha$ and $Y_\alpha$ are, respectively, Bessel functions of the first and second kind, of order $\alpha$.
At the surface of the wall, $x=0$, so $\tilde x = \ell\omega^2/\meffw$.  We expect $\omega\ell \sim 1$ while $\omega/\meffw \ll 1$, so $\tilde x \approx 0$ at the wall.  Thus the coefficient of the irregular Bessel function must be small, $c_2 \ll c_1$.  Far from the wall, $\delta\phi$ will therefore be dominated by the regular Bessel function:
\begin{equation}
\delta\phi_\mathrm{far}(x)
\approx
c_1 \sqrt{\omega x} J_\alpha(\omega x)
\propto
c_1 \sin(\omega x - \pi\alpha/2 + \pi/4).
\label{e:delta_phi_far}
\end{equation}
Writing this perturbation as the difference of an incoming wave and an outgoing wave with a $V(\phi)$-dependent phase shift $\xi_V$, in the relativistic limit $\omega = k$, we have
\begin{eqnarray}
\delta\phi(x)
&\propto&
e^{-i\omega x} - e^{+i\omega x - i\xi_V}
\nonumber\\
&=&
-2ie^{-i\xi/2} \sin(\omega x - \xi_V/2).
\label{e:delta_phi_phaseshift}
\end{eqnarray}

Equating the above to (\ref{e:delta_phi_far}) gives the phase shift for potentials (\ref{e:V_powerlaw},\ref{e:V_chameleon_dark_energy}),
\begin{equation}
\xi_{\phi^n} 
=
\pi\left(\alpha - \frac{1}{2}\right)
=
\pi\left(\left|\frac{3n-2}{2n-4}\right| - \frac{1}{2}\right),
\label{e:xi_pwrlaw}
\end{equation}
which we note is independent of the energy $\omega$ of the incoming particle.  In a real experiment particles will approach the wall with a range of incident angles.  Suppose now that the incoming chameleon wave has a nonzero angle of incidence, and let the $xy$ plane be the plane of reflection.  Since the $y$ momentum $k_y$ is unaffected by the reflection, we may factor out $e^{i k_y y}$ as well as the time dependence: $\phi(x,t) = \phi_0(x) + e^{i k_y y - i\omega t} \delta\phi(x)$.  Defining $k_x = (\omega^2 - k_y^2)^{1/2}$, we see that $\omega$ in (\ref{e:eom_linear_planar}) is replaced by $k_x$.  Since (\ref{e:xi_pwrlaw}) is independent of $\omega$, we have shown that the phase shift for power law  and chameleon dark energy potentials is independent of incident angle as well as energy.

More generally, if $\meff(x)^2 = \meff(\phi_0(x))^2 = A (x + D)^{-2}$ for some constants $A$ and $D$, as in (\ref{e:meff0_planar}), then the chameleon reflection phase $\xi_V = \pi ((A+1/4)^{1/2}-1/2)$ will be independent of incident angle and energy. Since $\frac{d^2}{dx^2}\phi_0(x) = \frac{d}{d\phi} V(\phi)$ in the vacuum, we have $\meff^2 = \frac{d^2}{d\phi^2} V(\phi) = \frac{d}{d\phi} \frac{d^2}{dx^2}\phi_0 = \frac{d^3 \phi_0}{dx^3} / \frac{d\phi_0}{dx} = A (x+D)^{-2}$.  This will be satisfied if $d\phi_0/dx \propto (x+D)^\zeta$ for $\zeta$ satisfying $\zeta(\zeta-1) = A$, that is, if $\phi_0(x) \propto (x+D)^{\zeta+1}$ or $\phi_0(x) \propto \log(x+D)$ up to an additive constant.  Substitution into $\frac{d^2}{dx^2}\phi_0(x) = \frac{d}{d\phi} V(\phi)$ then implies $V(\phi) = g\phi^n$ or $V(\phi) = g \exp(\phi/M)$, up to additive constants, for constants $g$, $n$, and $M$.  Thus we have extended our result about the independence of $\xi_V$ on incident angle and energy to exponential potentials.
Incidentally, choosing $\zeta = -1$, for an exponential potential 
\begin{equation}
V(\phi) = g \exp(\phi/M) + \textrm{const.},
\label{e:V_exp}
\end{equation} 
results in the phase shift
\begin{equation}
\xi_{\mathrm{exp}(\phi)}
=
\pi.
\end{equation}

\begin{figure}[tb]
\begin{center}
\includegraphics[angle=270,width=3.3in]{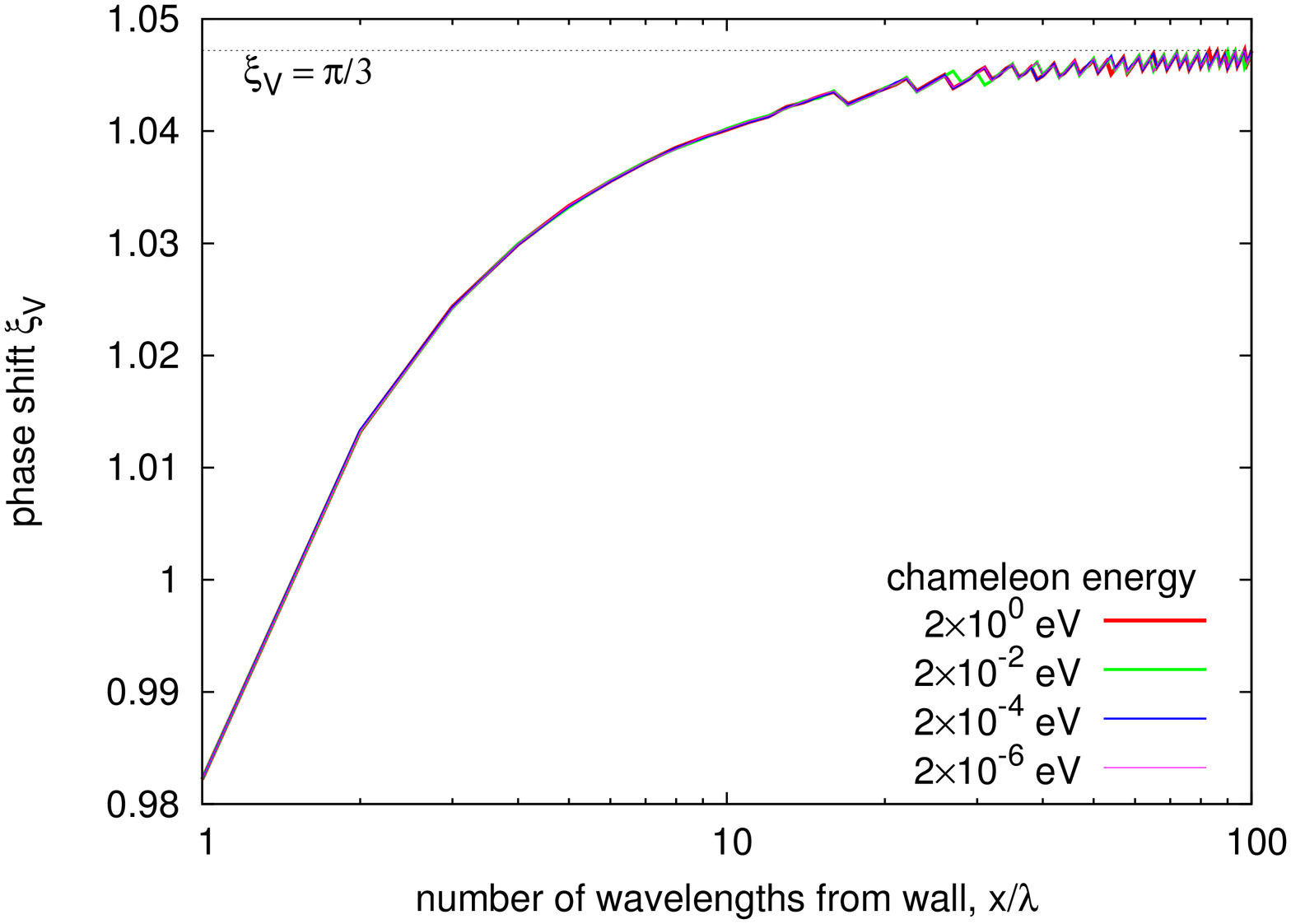}
\includegraphics[angle=270,width=3.3in]{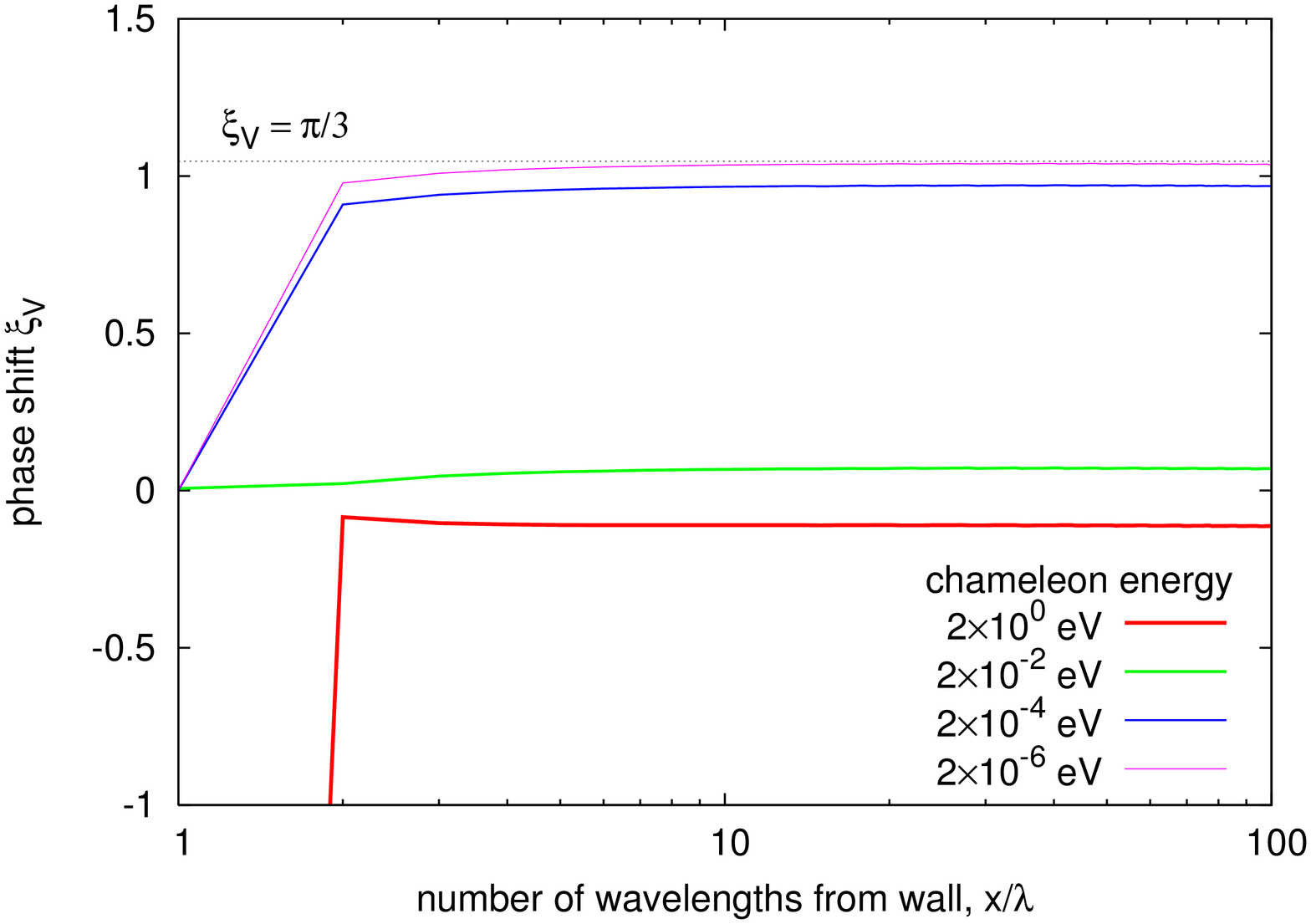}
\caption{Chameleon reflection phase $\xi_V$ vs. distance from wall, with $\bmat = 10^{12}$ and $\rhom  = 1$~g/cm$^3$.  
\FigA~$n=-1$ chameleon dark energy (\ref{e:V_chameleon_dark_energy}), $V(\phi) = M_\Lambda^4(1 + M_\Lambda/\phi)$.  
\FigB~potential $V(\phi) = M_\Lambda^4 \exp(M_\Lambda/\phi)$. 
In both cases, the dotted line shows $\xi_V=\pi/3$, the expected phase in the $x \rightarrow \infty$ limit for $n=-1$ chameleon dark energy.
\label{f:xi_exp_n-1}}
\end{center}
\end{figure}

However, for more general potentials, the phase $\xi_V$ will depend on the incident angle and energy.  A chameleon particle with greater momentum $k_x$ perpendicular to the wall will approach closer to the wall, allowing it to probe a different region of the potential $V(\phi)$ than a lower-$k_x$ particle.  Although exact solutions do not exist for general potentials, we can compute $\phi_0(x)$ numerically using  $\frac{d^2}{dx^2}\phi_0(x) = \frac{d}{d\phi} V(\phi)$.  Given $\phi_0(x)$ we may find $\meff(x)$ and then solve numerically for $\delta\phi(x)$ using $\frac{d^2}{dx^2} \delta\phi(x) = (\meff(x)^2 - k_x^2)\delta\phi(x)$.  Comparison of the zeros of this numerical $\delta\phi(x)$ to those of (\ref{e:delta_phi_phaseshift}) is used to find the phase shift $\xi_V$; far from the wall, it will converge to a constant.  Fig.~\ref{f:xi_exp_n-1}~\FigA~applies this technique to the $n=-1$ power law model, verifying the energy-independence of $\xi_V$ shown in (\ref{e:xi_pwrlaw}).  Furthermore, the phase values converge to $\xi_{\phi^{-1}} = \pi/3$ from (\ref{e:xi_pwrlaw}), shown as a dotted line.

An interesting example is the ``exponential-inverse'' potential 
$V(\phi) = M_\Lambda^4 e^\frac{M_\Lambda}{\phi}$,
which is frequently used in the literature.  At low energies, $\phi \gg M_\Lambda$ and the potential will approximate the $n=-1$ chameleon dark energy, eq.~(\ref{e:V_chameleon_dark_energy}); at high energies, it will differ from (\ref{e:V_chameleon_dark_energy}).  Fig.~\ref{f:xi_exp_n-1}~\FigB~shows the phase shift for several different chameleon energies.  As expected, for $\omega < M_\Lambda = 2.4\times 10^{-3}$~eV, the potential approaches the $n=-1$ chameleon dark energy, and the resulting chameleon phase approaches $\xi_{\phi^{-1}} = \pi/3$.  At larger energies $\xi_V$ differs substantially from $\pi/3$.

In the limit that the wall in an actual afterglow experiment is smooth, the afterglow rate will be dominated by chameleon particles bouncing with grazing incidence, $k_x / \omega \lesssim \textrm{radius / length} \sim 10^{-3} \Rightarrow k_x \lesssim 10^{-3}$~eV.  Thus the exponential-inverse model from Fig.~\ref{f:xi_exp_n-1} can reasonably be approximated as an $n=-1$ chameleon dark energy (\ref{e:V_chameleon_dark_energy}) for the purpose of computing the phase, although this approximation should worsen at larger $\bmat$.  

For potentials inconsistent with this power law approximation, it is necessary to compute numerically the chameleon reflection phase $\xi_V$ as a function of the incident angle, the coupling $\bmat$, and the parameters of the potential.  Although simple, this calculation is time-consuming and must be repeated for each new potential.  In the absence of a compelling reason for choosing a different potential, we restrict ourselves henceforth to the potentials (\ref{e:V_powerlaw},\ref{e:V_chameleon_dark_energy},\ref{e:V_exp}).

\subsection{Reflection and absorption of photons}
\label{subsec:reflection_and_absorption_of_photons}

The standard treatment of photon reflection from a conductive medium adds an imaginary component to the index of refraction of the medium, $n_1 - i \tilde n_1$.  Assume that the incident wave propagates through a medium whose index of refraction $n_0 \approx 1$ is purely real.  For stainless steel like that in the \GammeVCHASE~chamber walls, $n_1 = 1.6$ and $\tilde n_1 = 3.2$.  These quantities imply a mean reflection probability of $\fref = 0.65$.  The actual measured value is $\fref = 0.53$, or $18\%$ lower; the unpolished chamber walls are less reflective than polished stainless steel.  We model this discrepancy by assuming that $18\%$ of the chamber wall is obscured and perfectly absorbing.  Thus we multiply the computed reflection probabilities by a visibility correction factor $\fvis = 0.82$.  

\begin{figure}[tb]
\begin{center}
\includegraphics[angle=270,width=3.3in]{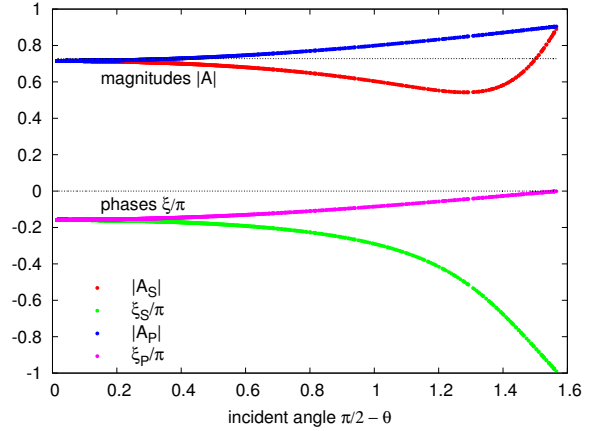}
\caption{Norms and arguments of the complex amplitude reflectivities $\ArefS$ and $\ArefP$.   Colored lines assume the parameters $n_1 = 1.6$, $\tilde n_1 = 3.2$, and $\fvis = 0.82$ appropriate to the \GammeVCHASE~chamber walls.  Dotted lines show the reflectivities in the thin-skin case; the two amplitudes are equal and real. See the text for a discussion of the conventions used here.  In brief, our $S$ and $P$ directions differ from the standard definition, and our phases are normalized so that $\xi_S = \xi_P$  at normal incidence.  \label{f:reflectivity}}
\end{center}
\end{figure}

The complex index of refraction implies that the photon reflection occurs over a skin depth $\delta_1 = \omega^{-1}\tilde n_1^{-1}$.  This introduces a phase shift which depends on the polarization direction as well as the incident angle.  Fig.~\ref{f:reflectivity} plots the amplitude reflectivities $\ArefP \equiv |\ArefP| e^{i\xi_P}$ and $\ArefS \equiv |\ArefS| e^{i\xi_S}$ for the P and S polarizations, respectively, as a function of incident angle.  (Recall that we keep track of the magnetic field component of the photon, $\psig \propto \delta \vec B$, so that our P and S polarizations are switched with respect to standard conventions.)  The norm of the amplitude reflectivity is the square root of the reflection probability, and the complex phase is the phase shift of the photon due to reflection.  In our convention, this phase shift is measured relative to that of a chameleon particle with Dirichlet boundary conditions and no additional phase shift; that is, our conventions for the photon phase shift differ from standard conventions by $\pi$.  

Reflection becomes perfect in the thin-skin limit $\tilde n_1 \rightarrow \infty$.  $\ArefP$ and $\ArefS$ both approach unity in this case.  Ref.~\cite{Upadhye_Steffen_Weltman_2010} implicitly used this approximation, setting $\fvis = 0.53$ to match the observed reflection probability.  Fig.~\ref{f:reflectivity} shows that, with this value of $\fvis$ and $\tilde n_1 = 10^6$, the thin-skin limit of that reference is reproduced.  Henceforth we use the values $n_1 = 1.6$, $\tilde n_1 = 3.2$, and $\fvis = 0.82$.

We will see that this polarization-dependence of the photon phase makes $\Gaft$ nearly independent of $\xi_V$.  The afterglow rate is dominated by particles bouncing at grazing incidence, for which $\xi_S$ and $\xi_P$ differ by $\approx \pi$.  Averaging over polarizations washes out the dependence on additional phases such as $\xi_V$.

\section{Chameleon-photon oscillation: Analytic calculation}
\label{sec:chameleon-photon_oscillation_analytic_calculation}

\subsection{Decay and afterglow rates}
\label{subsec:decay_and_afterglow_rates_1}

Scalar-photon oscillation was studied in \cite{Raffelt_Stodolsky_1988} and applied to afterglow experiments in \cite{Upadhye_Steffen_Weltman_2010}.  Although~\cite{Upadhye_Steffen_Weltman_2010} did not include the photon polarization-dependent reflectivities just described,
the data analysis in~\cite{Steffen_etal_2010} did incorporate this effect.  Here we summarize the calculation of \cite{Upadhye_Steffen_Weltman_2010} and add a photon polarization-dependence to the reflectivity and phase shift.  

\begin{figure}[tb]
\begin{center}
\includegraphics[width=3.3in]{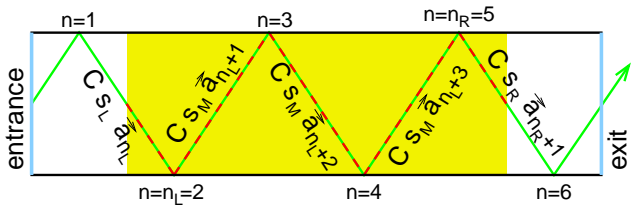}
\caption{Particle in an afterglow experiment.  The magnetic field region is shaded.  Inside this region, where chameleon-photon oscillation takes place, the particle trajectory is drawn as a dashed path.  Next to each dashed segment is shown the contribution of that segment to the total photon amplitude.  From~\cite{Upadhye_Steffen_Weltman_2010}. \label{f:psi_gamma}}
\end{center}
\end{figure}

Figure~\ref{f:psi_gamma} shows a cylindrical oscillation chamber with windows at the ends and a magnetic field region (shaded) offset from the windows.  The coordinate system used is identical to that of~\cite{Upadhye_Steffen_Weltman_2010}.  The origin lies at the center of the entrance window.  The $z$ axis is the cylinder axis of the vacuum chamber, with the unit vector $\hat z$ chosen to point in the direction from the entrance window to the exit.  $\hat x$ is a unit vector in the direction of the external magnetic field $\vec \Bext$, orthogonal to the cylinder axis, and $\hat y = \hat z \times \hat x$.

Consider a particle beginning in a pure chameleon state at the center of the entrance window.  This simplifies the probem considerably by maintaining cylindrical symmetry.  Ref.~\cite{Upadhye_Steffen_Weltman_2010} shows that, for $\meff^2 \ll 4\pi \omega / R$, where $\omega$ is the particle energy and $R$ the chamber radius, averaging over initial positions on the entrance window may be neglected for the purpose of computing the afterglow rate.  Although a particle beginning at the center of the entrance window is more likely to reach the detector outside the chamber, this is a purely geometrical effect for which we can correct quite easily.  

Since the total momentum of the chameleon particle is $k$, there is a two-parameter family of initial conditions beginning at the center of the entrance window.  We parameterize the initial conditions by the angle $\theta$ between $\hat k$ and the $z$ axis, $\cos(\theta) = \hat k \cdot \hat z$, and the angle $\varphi$ between the $x$ axis and the projection of $\hat k$ onto the $xy$ plane, $\cos(\varphi) = \hat k \cdot \hat x / |\hat k - (\hat k \cdot \hat z)\hat z|$.  (Recall that the $x$ and $z$ axes are specified by the background magnetic field and the cylinder axis, respectively.)  Let $\theta$ and $\varphi$ be specified for a particle.  Then the total number $N(\theta)$ of bounces inside the chamber, the first bounce $\nL(\theta)$ inside the region with magnetic field $\vec \Bext$, and the final bounce $\nR(\theta)$ in the $\Bext$ region depend on $\theta$ but not $\varphi$ by symmetry.  For the path shown in Fig.~\ref{f:psi_gamma}, $\nL=2$, $\nR = 5$, and $N = 6$.  As shown in ref.~\cite{Upadhye_Steffen_Weltman_2010}, the photon amplitude at the exit of the $\Bext$ region is found by summing up the contributions from each segment inside the $\Bext$ region shown in Fig.~\ref{f:psi_gamma}, properly accounting for photon absorption and chameleon-photon phase shifting at each bounce.  

Eq.~(\ref{e:psig}) gives the contribution of each segment to the photon amplitude.  Up to a phase factor, this is $\Cgc \sin(\meff^2 t / (4k)) \vec a(\hat k)$, with $t$ the time spent inside the $\Bext$ region and $\vec a$ expressed in the basis of S and P polarization states; recall that $\vec a$ and $\Cgc$ are defined by (\ref{e:a_vec},~\ref{e:Cgc}).  Since each segment in Fig.~\ref{f:psi_gamma} beginning and ending inside the $\Bext$ region has the same value of $t$ by symmetry, only three values of $t$ are necessary: the time $\tL$ taken by the particle between $\Bext$ region entry and bounce $\nL$; the time $\tM$ between two bounces inside the $\Bext$ region; the time $\tR$ between bounce $\nR$ and the $\Bext$ region exit.  We define $\sL = \sin(\tL)$, $\sM = \sin(\tM)$, and $\sR = \sin(\tR)$, so that
\begin{eqnarray}
\sL
&=&
\sin\left(\frac{\meff^2(z_{\nL}-\ell_1)}{4k\cos\theta}\right)
\\
\sM
&=&
\sin\left(\frac{\meff^2 R}{2k\sin\theta}\right)
\\
\sR
&=&
\sin\left(\frac{\meff^2 (\ell_1 + L - z_{\nR})}{4k\cos\theta}\right)
\end{eqnarray}
where $z_n$ is the $z$ value of the $n$th bounce, $\ell_1$ is the distance from the entrance window to the beginning of the $\Bext$ region, $\ell_2$ is the distance from the end of the $\Bext$ region to the exit window, and $L$ is the length of the $\Bext$ region.

Chameleon-photon phase shifts come from three different sources: a potential-dependent chameleon phase shift $\xi_V$ due to reflection from the walls, computed in Sec.~\ref{subsec:phase_change_due_to_reflection}; a photon polarization-dependent phase shift due to a nonzero skin depth during reflection, computed in Sec.~\ref{subsec:reflection_and_absorption_of_photons}; and a phase shift during propagation $\xi_\mathrm{propagation}(t)$ through the $\Bext$ region due to a chameleon-photon mass difference, computed in \cite{Upadhye_Steffen_Weltman_2010} and included in the time-dependent phase factor in (\ref{e:psig}).  $\xi_\mathrm{propagation}(t)$ must be computed over the time intervals $\tL$, $\tM$, and $\tR$ defined earlier, leading to 
\begin{eqnarray}
\xiL
&=&
\frac{\meff^2}{4k\cos\theta}\left(\left(\nL - \frac{3}{2}\right)\Delta z - \ell_1\right)
\\
\xiM
&=&
\frac{\meff^2 R}{k\sin\theta}
\\
\xiR
&=&
\frac{\meff^2}{4k\cos\theta}
\left(\left(\nR  + \frac{1}{2}\right)\Delta z - \ell_1-L\right)
\end{eqnarray}
where $\Delta z = 2R\cot\theta$ is the $z$ distance between bounces.  Following the convention of \cite{Upadhye_Steffen_Weltman_2010}, we redefine the complex reflectivities to include $\xiM$ and the chameleon phase shift $\xi_V$:
\begin{eqnarray}
A_P
&=&
\ArefP \exp(i \xiM + i \xi_V)
\\
A_S
&=&
\ArefS \exp(i \xiM + i \xi_V).
\end{eqnarray}

Next, we compute the components of $\vec a(\hat k)$ in the basis of S and P polarization states.  $\vec a$ and $\hat k$ are constant between bounces.  Let $\hat k_n$ be the particle direction {\em before} the $n$th bounce, and let $\vec a_n = \vec a(\hat k_n)$; thus a subscript $1$ denotes initial values.  We have $\vec a_1 = (1 - \sin^2\theta \, \cos^2\varphi)\hat x - (\sin^2\theta \, \sin\varphi\,\cos\varphi)\hat y - (\sin\theta\,\cos\theta\,\cos\varphi) \hat z$.  A bounce from the chamber wall switches the sign of the $z$ component of $\vec a$ while leaving the $x$ and $y$ components unchanged: $\vec a_2 = \vec a_1 - 2(\vec a_1 \cdot \hat z)\hat z$.  Thus $\vec a_n$ will equal $\vec a_1$ for odd $n$ and $\vec a_2$ for even $n$.  Since the plane of incidence for each bounce is spanned by $\hat z$ and $\hat k_1$, we define the S polarization direction by $\hat S = (\hat z \times \hat k_1) / |\hat z \times \hat k_1|$; this remains the same during all bounces.  For each $n$, the set ($\hat S$, $\hat P_n$, $\hat k$) must be orthonormal, with $\hat P_n$ the P polarization direction before the $n$th bounce.  We choose the sign of $\hat P_n$ such that $\hat S \times \hat P_n = \hat k_n$.  With these definitions we find $\hat S = -(\sin\varphi)\hat x + (\cos\varphi)\hat y$ and $\hat P_1 = -(\cos\theta\,\cos\varphi)\hat x - (\cos\theta\,\sin\varphi)\hat y + (\sin\theta)\hat z$; $\hat P_2$ has the same $x$ and $y$ components as $\hat P_1$, but its $z$ component has the opposite sign.  Finally, since $\hat S$ has no $z$ component, and since the $z$ components of $\vec a$ and $\hat P$ both change after a bounce, we have $a_S = \vec a_n \cdot \hat S = -\sin\varphi$ and $a_P = \vec a_n \cdot \hat P_n = -\cos\theta\,\cos\varphi$ independent of $n$.  Thus $\vec a_n = a_S \hat S + a_P \hat P_n$.

We proceed to compute the photon amplitude at the exit of the $\Bext$ region step by step.  Immediately before the first bounce in the $\Bext$ region, $\nL$, the photon amplitude is due entirely to the oscillation which took place between the entrance of the $\Bext$ region and the first bounce.  Up to a constant phase factor, it is $\Cgc \sL (a_S \hat S + a_P \hat P_n)$.  The bounce from the chamber wall rotates the amplitude to the new basis ($\hat S$, $\hat P_2$) and multiplies the individual components by the reflection factors $A_S$ and $A_P$.  Immediately after the bounce, the amplitude is $\psig^{(\nL^+)} = \Cgc \sL e^{i\xiL} (a_S A_S \hat S + a_P A_P \hat P_{\nL+1})$.  As the particle approaches bounce $\nL + 1$, further oscillation adds $\Cgc \sM \vec a_{\nL+1}$ to the amplitude.  Immediately after bounce $\nL + 1$ the amplitude is $\psig^{(\nL+1^+)} = \Cgc a_S (\sL e^{i\xiL} A_S^2 + \sM A_S) \hat S + \Cgc a_P (\sL e^{i\xiL} A_P^2 + \sM A_P) \hat P_{\nL+2}$.  Repeating this procedure for each of the $\nB = \nR - \nL + 1$ bounces inside the $\Bext$ region, we find
\begin{eqnarray}
\psig^{(\nR^+)}
&=&
\Cgc a_S 
\left( 
\sL e^{i\xiL} A_S^{\nB} 
+
\sM \sum_{n=1}^{\nB-1} A_S^n
\right) \hat S
\nonumber\\
& &
+
\Cgc a_P
\left(
\sL e^{i\xiL} A_P^{\nB}
+
\sM \sum_{n=1}^{\nB-1} A_P^n
\right) \hat P_{\nR+1}.\qquad 
\end{eqnarray}
Finally, we add the contribution $\Cgc \sR e^{i\xiR} \vec a_{\nR + 1}$ due to oscillation between bounce $\nR$ and the exit of the $\Bext$ region, leading to
\begin{eqnarray}
&&
\psigBexit
=
\label{e:psigBexit}
\\
& &
\quad
\Cgc a_S
\left(
\sL e^{i\xiL} A_S^{\nB} + \sM \frac{A_S - A_S^{\nB}}{1-A_S} + \sR e^{i\xiR}
\right) \hat S
\nonumber\\
& &
\quad
+
\Cgc a_P
\left(
\sL e^{i\xiL} A_P^{\nB} + \sM \frac{A_P - A_P^{\nB}}{1-A_P} + \sR e^{i\xiR}
\right) \hat P_{\nR+1}
\nonumber\\
&&
\left| \psigBexit \right|^2
=
\label{e:psig2Bexit}
\\
& &
\quad
\Cgc^2 a_S^2 
\left|
\sL e^{i\xiL}A_S^{\nB} + \sM \frac{A_S - A_S^{\nB}}{1-A_S} + \sR e^{i\xiR}
\right|^2
\nonumber\\
& &
\quad
+
(S \rightarrow P) 
\nonumber
\end{eqnarray}
where $(S \rightarrow P)$ denotes the preceding terms with S replaced by P.

After exiting the magnetic field region, the particle bounces from the walls $N-\nR$ more times.  This multiplies the S and P components of $\psig$ by $A_S^{N-\nR}$ and $A_P^{N-\nR}$, respectively.  No further oscillation occurs.  Once the particle reaches the end of the chamber, the exit window performs a quantum-mechanical measurement of particle content.  Since photons pass through the window while chameleons reflect, the chameleon-photon superposition is collapsed into one of those two states.  The probability that this measurement results in a photon is $\psig^* \cdot \psig$ evaluated at the exit window.  Assuming that a photon produced by a chameleon with initial conditions ($\theta$, $\varphi$) will reach the detector outside the chamber as an afterglow signal, the expected number of afterglow photons generated by this particle is $\psig^* \cdot \psig$ in the small-mixing-angle limit.  The time taken by the particle to reach the exit window from the entrance is $\ltot \sec\theta$, where $\ltot = \ell_1 + L + \ell_2$ is the total chamber length.  The contribution of this particle to the detectable afterglow rate $\Gaft$ is the expected photon number over the time, $|(\psigBexit\cdot\hat S) A_S^{N-\nR} |^2 \cos(\theta) / \ltot + (S \rightarrow P)$.

Only a small fraction of the photons produced emerge from the chamber and reach the detector.  Most are either absorbed in the walls or exit the chamber but miss the detector.  In order to find the total decay rate $\Gdec$ of the population of chameleon particles in the chamber, we must account for all of these photons.  Since oscillation stops after the $\Bext$ region exit, and since we already have $|\psig|^2$ at that point, we need only compute the probability of photon absorption inside the $\Bext$ region.  At bounce $n$, the absorption probability is given by the difference in photon probabilities before and after the bounce, $\Pabs^{(n)} = |\psig^{(n^-)}|^2 - |\psig^{(n^+)}|^2$.  Computing the photon amplitudes as above, we find
\begin{eqnarray}
\Pabs^{(n)}
&=&
\Cgc^2 a_S^2 (1-|A_S|^2)
\left|
\sL e^{i\xiL} A_S^{n-\nL} 
+
\sM \frac{1 - A_S^{n-\nL}}{1-A_S}
\right|^2
\nonumber\\
& &
+
(S \rightarrow P).
\end{eqnarray}
The total absorption probability is the sum over all $n$ from $\nL$ to $\nR$:
\begin{eqnarray}
\PabsBexit
&=&
\Cgc^2 a_S^2 (1-|A_S|^2)
\Bigg[
  \sL^2 \frac{1-|A_S|^{2\nB}}{1-|A_S|^2}
  \nonumber
  \\
  & & + 
  \frac{\sM^2 
  \left(
  \nB - \frac{1-A_S^{\nB}}{1-A_S} 
  - \frac{1-(A_S^*)^{\nB}}{1-A_S^*} 
  + \frac{1-|A_S|^{2\nB}}{1-|A_S|^2}  
  \right)}
       {(1-A_S)(1-A_S^*)}
  \nonumber
  \\
  & & +
  \sL \sM
  \frac{(1 - A_S^{\nB})e^{i\xiL} + (1-(A_S^*)^{\nB})e^{-i\xiL}}
       {(1-A_S)(1-A_S^*)}
  \nonumber
  \\
  & & -
  \sL \sM
  \frac{1-|A_S|^{2\nB}}{1-|A_S|^2}
  \left(
  \frac{e^{-i\xiL}}{1-A_S}
  + \frac{e^{i\xiL}}{1-A_S^*}
  \right)
\Bigg]
\nonumber\\
& &
+
(S \rightarrow P).
\label{e:PabsBexit}
\end{eqnarray}
The contribution to $\Gdec$ of this particle with initial ($\theta$, $\varphi$) is found by dividing the total conversion probability by the time, $(|\psigBexit|^2 + \PabsBexit) \cos(\theta) / \ltot$.

Finally, we average over angles in order to obtain the decay rate $\Gdec$ and the afterglow rate $\Gaft$:
\begin{eqnarray}
\Gdec
&=&
\frac{1}{2\pi}
\int_\frac{\Omega}{2} d^2\Omega
\left(
\PabsBexit
+ \left|\psigBexit\right|^2
\right)
\frac{\cos\theta}{\ltot}
\label{e:Gdec}
\\
\Gaft
&=&
\frac{1}{4\pi}
\int_\frac{\Omega}{2} d^2\Omega
\left| (\psigBexit\cdot\hat S) A_S^{N-\nR} \right|^2
\Pdet(\theta)
\frac{\cos\theta}{\ltot}
\nonumber\\
& &
+
(S \rightarrow P).\qquad 
\label{e:Gaft}
\end{eqnarray}
where $\Omega/2$ is the half-sphere with $0\leq \theta < \pi/2$ and $0 \leq \varphi < 2\pi$.
Here, $\Pdet(\theta)$, the probability that the particle will reach the detector, is a geometric factor which will be computed in Sec.~\ref{subsec:geometric_factor_Pdet}.  The extra factor of $2$ in the decay rate is due to the fact that particles travelling in the negative $z$ direction, away from the exit window, contribute to the decay rate but not to the afterglow rate.  Examining the integrands, we see that the only quantities which depend on $\varphi$ are $a_S^2$ and $a_P^2$; each term in the integrand contains one of these multiplying  a $\varphi$-independent quantity.  Thus the integrals over $\varphi$ may easily be performed using
\begin{eqnarray}
\aavgS
&=&
\frac{1}{2\pi} \int_0^{2\pi} a_S^2 d\varphi
=
\frac{1}{2}
\label{e:aavgS}
\\
\aavgP
&=&
\frac{1}{2\pi} \int_0^{2\pi} a_P^2 d\varphi
=
\frac{1}{2} \cos^2\theta.
\label{e:aavgP}
\end{eqnarray}

\subsection{Geometric factor $\Pdet(\theta)$}
\label{subsec:geometric_factor_Pdet}

\begin{figure}[tb]
\begin{center}
\includegraphics[width=3.3in]{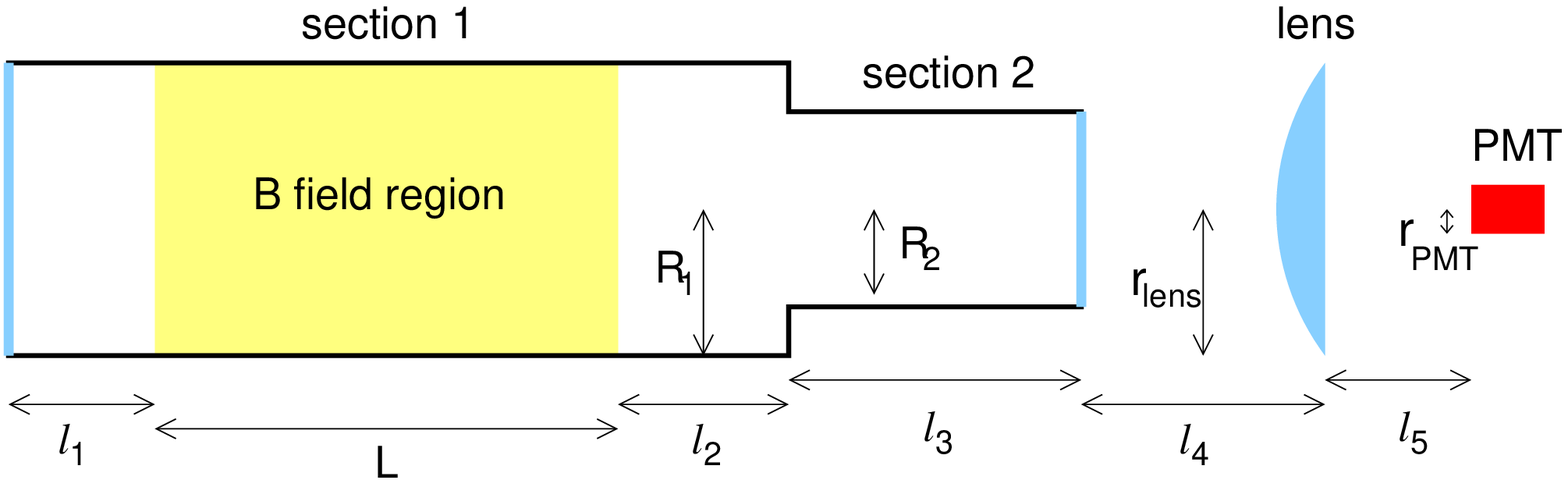}
\vskip0.1in
\includegraphics[width=3.3in]{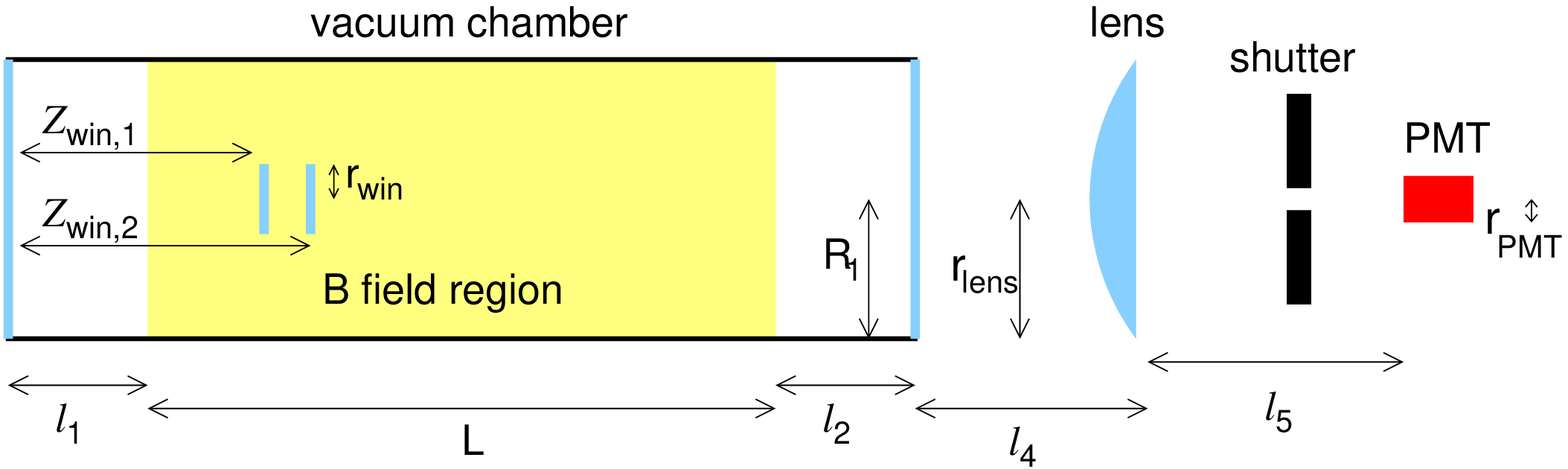}
\caption{\FigA~\GammeV~apparatus and \FigB~\GammeVCHASE~apparatus (side views, not to scale).  Note that \GammeVCHASE~has no section 2 ($\ell_3 = 0$) and has glass windows dividing the $\Bext$ region into three partitions. Values for these parameters are given in Table~\ref{t:gammev_apparatus}. \label{f:gammev_apparatus}}
\end{center}
\end{figure}

\begin{table}[tb]
\begin{center}
\begin{tabular}{|c||c|c|}
\hline
$\quad$quantity $\quad$& $\quad$GammeV$\quad$   & $\quad$\GammeVCHASE$\quad$ \\
\hline
\hline
$\ell_1$               & $2.36$~m               & $1.61$~m      \\
$L$                    & $6.0$~m                & $6.0$~m       \\
$\ell_2$               & $1.16$~m               & $1.74$~m      \\
$\ell_3$               & $2.51$~m               & N/A           \\
$\ell_4$               & $2.03$~m               & $0.64$~m      \\
$\ell_5$               & $0.10$~m               & $0.10$~m      \\
$Z_{\mathrm{win},1}$   & N/A                    & $2.61$~       \\
$Z_{\mathrm{win},2}$   & N/A                    & $2.91$~       \\
$R_1$                  & $2.38$~cm              & $3.175$~cm    \\
$R_2$                  & $1.75$~cm              & N/A           \\
$r_\mathrm{win}$       & N/A                    & $1.27$~cm     \\
$r_\mathrm{lens}$      & $2.54$~cm              & $2.54$~cm     \\
$r_\mathrm{PMT}$       & $2.5$~mm               & $2.5$~mm      \\
$t_0$                  & $1006$~sec             & $120$~sec     \\
$\Delta t$             & $3616$~sec             & $600-3600$~sec\\
$\fref$                & $0.53$                 & $0.53$        \\
$k\approx \omega$      & $2.33$~eV              & $2.33$~eV     \\
$B$                    & $5.0$~Tesla            & $5.0$~Tesla   \\
$V_\mathrm{pump}$      & $0.026$~m$^3$          & $0.014$~m$^3$ \\         
\hline
\end{tabular}
\end{center}
\caption{Properties of the GammeV and \GammeVCHASE~experiments, including the dimensions shown in Fig.~\ref{f:gammev_apparatus}.  \label{t:gammev_apparatus}}
 \end{table}

References~\cite{Upadhye_Steffen_Weltman_2010,Steffen_etal_2010} describe the geometry of the \GammeVCHASE~experiment.  Diagrams of both experiments are shown in Fig.~\ref{f:gammev_apparatus} with numerical values given in Table~\ref{t:gammev_apparatus}.  Note that, in \GammeVCHASE, $\ell_3 = 0$; there is no second chamber section.  Henceforth we set $\ell_3 = 0$ and $R_1 = R_2 = R$.  Outside of the oscillation chamber, a distance $\ell_4$ beyond the exit window, is a lens with radius $\rlens$ and focal length $\ell_5$.  At the focal point of the lens is positioned a photomultiplier tube (PMT) whose accepting area has radius $\rpmt$.  

We will see in Sec.~\ref{sec:enhancements_in_gammev-chase} that the \GammeVCHASE~$\Bext$ region is divided by glass windows into three partitions.  This smoothes out sharp features in the oscillation rate during the production phase of the experiment.  However, we will postpone a discussion of these partitions until that section, neglecting them for now. 

The previous discussion assumed that the initial chameleon particle started out at the center of the entrance window.  In the limit that the oscillation length $4\pi \omega / \meff^2$ is small compared with the cylinder radius, or $\meff \lesssim 0.01$~eV in \GammeV, the  dominant errors introduced by this approximation are purely geometrical.  Thus we only need to compute them once, independent of the chameleon model.  

The simplest geometric effect arises from the fact that the lens and detector are centered on the $z$ axis, which is where particles start out in our approximation.  This artificially enhances the probability that a regenerated photon will reach the detector.  

A more subtle geometric effect gives particles in our approximation a different radial probability distribution from the average particle.  The volume of cylindrical shell at a distance $r = (x^2 + y^2)^{1/2}$ from the $z$ axis, $\Delta {\mathcal V} = 2 \pi \ltot r \Delta r$, increases with $r$.  Thus a homogeneous, isotropic gas of chameleon particles will have a radial probability distribution $P(r) \propto r$.  However, consider a particle beginning at the center of the entrance window; without loss of generality, assume $\theta > 0$ and $\varphi = 0$.  At the first bounce, and all subsequent odd-numbered bounces, $x = R$.  At the second bounce, and all subsequent even-numbered bounces, $x = -R$.  The path taken by this particle, when projected onto the $xy$ plane, will bounce back and forth between the same two points.  Following \cite{Upadhye_Steffen_Weltman_2010} we call these ``2-point paths.''   All $r$ values are equally likely along such a path, implying that the radial probability distribution $P(r) = \textrm{constant}$.  Therefore, compared with the average particle, the particles which we consider spend more time at low $r$ and are more likely to reach the detector.  Incidentally, in the case of \GammeV, particles beginning at $r=0$ are $R_1/R_2$ times more likely to enter the second section of the chamber because of this effect.

We can account for both of these geometric effects at once by properly normalizing our afterglow rate integrand.  The fraction of 2-point paths reaching the detector is found by integrating over solid angles, 
\begin{eqnarray}
f_\mathrm{2-point}
&=&
\frac{1}{4\pi}
\int_\frac{\Omega}{2} d^2\Omega
\Theta_\mathrm{det}(\theta)
\nonumber\\
&=&
\frac{1}{2}
\int_0^\frac{\pi}{2} \sin\theta \, d\theta
\Theta_\mathrm{det}(\theta),
\end{eqnarray}
where $\Theta_\mathrm{det}$ is one if the path reaches the detector and zero otherwise.  For the \GammeVCHASE~geometry we find $f_\mathrm{2-point} = 1.57 \times 10^{-4}$.  We do a similar calculation for the average particle, whose radial position $r_\mathrm{ex}$ and direction ($\theta_\mathrm{ex}$, $\varphi_\mathrm{ex}$) at the exit window are appropriate to a homogeneous, isotropic distribution:
\begin{eqnarray}
f_\mathrm{avg}
&=&
\frac{1}{\pi R^2} \frac{1}{4\pi}
\int_0^R 2\pi r_\mathrm{ex} \,dr_\mathrm{ex} 
\int_0^\frac{\pi}{2} \sin\theta_\mathrm{ex} \, d\theta_\mathrm{ex}
\nonumber\\
& &
\times
\int_0^{2\pi} d\varphi_\mathrm{ex}
\Theta_\mathrm{det}(r_\mathrm{ex},\theta_\mathrm{ex},\varphi_\mathrm{ex}).
\end{eqnarray}
For the \GammeVCHASE~geometry $f_\mathrm{avg} = 9.84\times 10^{-5}$.  Thus, when we restrict our calculation to chameleon particles beginning at the center of the exit window, we increase our afterglow rate by a factor of $f_\mathrm{2-point} / f_\mathrm{avg} = 1.59$.  We correct for this by normalizing $\Pdet(\theta)$ in (\ref{e:Gaft}):
\begin{equation}
\Pdet(\theta)
=
\frac{f_\mathrm{avg}}{f_\mathrm{2-point}}
\Theta_\mathrm{det}(\theta).
\end{equation}
In the case of \GammeV, we include an extra factor of $R_2/R_1$ to account for the increased probability of reaching the second chamber section.  This simple, model-independent normalization turns out to be reasonably accurate.  The properly normalized afterglow rate calculated for 2-point paths differs by $6\%$ from that calculated for the ``3-point paths'' of \cite{Upadhye_Steffen_Weltman_2010} and by $8\%$ from a Monte Carlo calculation with arbitrary initial conditions.  This applies at low masses $\meff \ll (4\pi \omega / L)^{1/2}$, or about $0.001$~eV in \GammeV.  

Unfortunately there is no equivalent procedure for correcting the decay rate.  We will see that the 2-point decay rate disagrees with the Monte Carlo calculation of Sec.~\ref{sec:chameleon-photon_oscillation_monte_carlo_simulation} by about $40\%$.  However, the experimental upper bounds on the photon coupling $\bgam$ are extremely insensitive to the decay rate; at low $\bgam$, the decay time $\Gdec^{-1}$ is much larger than the duration of the experiment.  Furthermore, to the extent that $\Gdec$ matters, using the 2-point calculation is a conservative approximation; (\ref{e:Gdec}) computed using 2-point paths overestimates the decay rate, hence underpredicts the signal.

\begin{figure}[tb]
\begin{center}
\includegraphics[angle=270,width=3.3in]{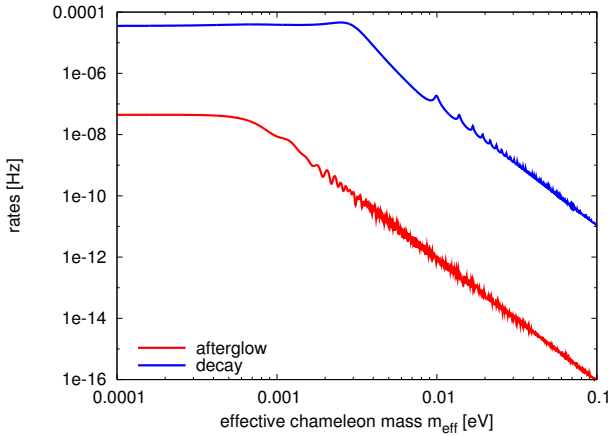}
\caption{Two-point afterglow and decay rates for $\Bext = 5$~Tesla, $\bgam = 10^{12}$, and $\xi_V = 0$, for the \GammeVCHASE~afterglow experiment (unpartitioned).  \label{f:Gaft_Gdec_2pt}}
\end{center}
\end{figure}

Both the decay rate (\ref{e:Gdec}) and the afterglow rate (\ref{e:Gaft}) depend on $\bgam$ and $\Bext$ only through the prefactor $\Cgc^2$.  Thus $\Gdec$, $\Gaft \propto \bgam^2 \Bext^2$.  The rates need only be computed at one value of the photon coupling and magnetic field; a simple rescaling can be used to find the rates at other values of these two parameters.  Figure~\ref{f:Gaft_Gdec_2pt} plots these rates as a function of chameleon mass in the 2-point computation.  Both rates are approximately independent of mass when $\meff \ll 10^{-3}$~eV and scale as $\meff^{-4}$ at $\meff \gg 10^{-3}$~eV, consistent with the mass scaling of (\ref{e:Pgc}).

\section{Through the magnetic field}
\label{sec:through_the_magnetic_field}

\subsection{The real magnetic field}
\label{subsec:the_real_magnetic_field}

Thus far we have treated the magnetic field as a ``tophat'' function suddenly increasing from zero to its peak value and then dropping again.  We have assumed a magnetic field $B_\mathrm{tophat}(z) = \Bext \Theta(z-\ell_1) \Theta(\ell_1 + L - z)$, where $\Theta(z)$ is the step function.  The real \GammeV~and \GammeVCHASE~magnetic field falls off from its peak value over a distance $\Dzf \approx 5$~cm in a way that is well approximated by $B_{\mathrm{tanh}} (z) = \Bext T_{\Dzf}(z-\ell_1) T_{\Dzf}(\ell_1 + L - z)$ with $T_{\Dzf}(z) = (1 + \tanh(z/\Dzf + 1))/2$.  For models with oscillation length $\losc = 4\pi k / \meff^2 \gg \Dzf$, or $\meff \ll 0.01$~eV in \GammeV~and \GammeVCHASE, this difference is negligible; the field is effectively a tophat function.  

Larger-mass chameleons, however, will see a slowly-varying magnetic field in these falloff regions. Their oscillation probabilities will differ from tophat function approximations.  In the extreme case that $\losc \ll \Dzf$ the spatial variation of the magnetic field will represent an adiabatic perturbation.  A particle which begins in a chameleon state and experiences an adiabatically evolving magnetic field will return to a pure chameleon state once the magnetic field evolves back to zero.  Thus $\psig = 0$ in the adiabatic limit.

\begin{figure}[tb]
\begin{center}
\includegraphics[angle=270,width=3.3in]{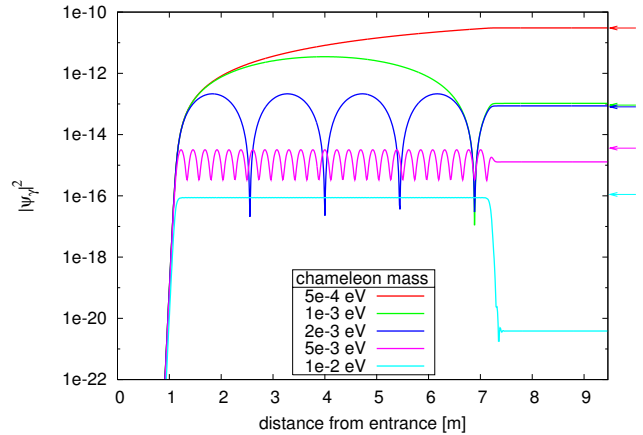}
\caption{Photon amplitude vs. $z$ for a magnetic field $\Bext = 5$~Tesla with falloff length $\Dzf = 5$~cm.  A coupling $\bgam = 10^{12}$ is assumed.  Arrows at the right show expected oscillation probabilities for a tophat magnetic field.  Low-mass chameleons effectively see a tophat field while chameleons with $\meff \gtrsim 0.01$~eV see a nearly adiabatic evolution of the field.  \label{f:realB_nowin}}
\end{center}
\end{figure}

\begin{figure}[tb]
\begin{center}
\includegraphics[angle=270,width=3.3in]{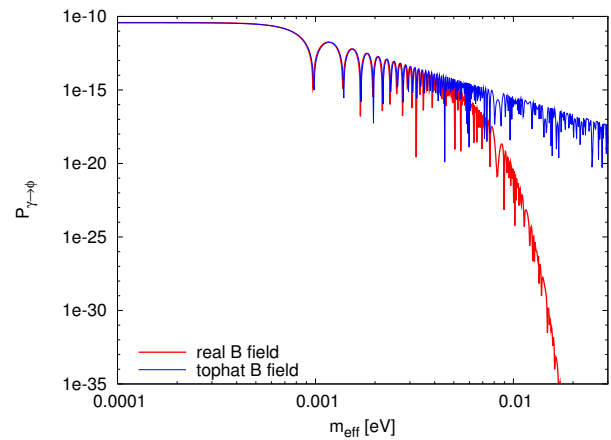}
\caption{Oscillation probability vs. chameleon mass for the real magnetic field $B_\mathrm{tanh}(z)$ and the tophat approximation.  $\bgam = 10^{12}$ is assumed.  \label{f:Pgc_realB_nowin}}
\end{center}
\end{figure}

Since our integrated solution (\ref{e:psig}) is not valid for a varying field $B_\mathrm{tanh}$, the preceding equation of motion must be integrated numerically.  Figure~\ref{f:realB_nowin} shows the results for several chameleon masses, assuming a particle trajectory along the $\hat z$ axis.  For comparison, arrows to the right of the plot show the final probabilities expected from the tophat approximation (\ref{e:Pgc}).  Note that the tophat approximation is excellent for masses up to the dark energy scale of $\approx 2\times 10^{-3}$~eV, and is valid up to a factor of order unity at twice that scale.  However, for $\meff \gtrsim 0.01$~eV, $\Bext(z)$ rises and falls slowly enough to represent an adiabatic evolution of the background, and the oscillation probability is less than the tophat prediction by some four orders of magnitude.  Figure~\ref{f:Pgc_realB_nowin} shows the oscillation probability for $B_\mathrm{tanh}(z)$ and $B_\mathrm{tophat}(z)$ over a range of chameleon masses.  As expected, chameleons with $\meff \gtrsim 0.01$~eV see a nearly adiabatically varying magnetic field.  

\subsection{Window in magnetic field region}
\label{subsec:window_in_magnetic_field_region}

Adiabatic suppression of oscillation can be substantially reduced by causing a sudden change in the background chameleon field inside the $\Bext$ region.  This is accomplished by placing a glass window inside that region.  Close to the surface of this window, the background chameleon field (\ref{e:phi0_planar}) and mass (\ref{e:meff0_planar}) change on distance scales of order $\meff(\phibulk(\rhom))^{-1}$.  By the containment condition this must be less than the wavelength $\sim k^{-1}$, which is in turn much smaller than $\losc$.  

\begin{figure}[tb]
\begin{center}
\includegraphics[angle=270,width=3.3in]{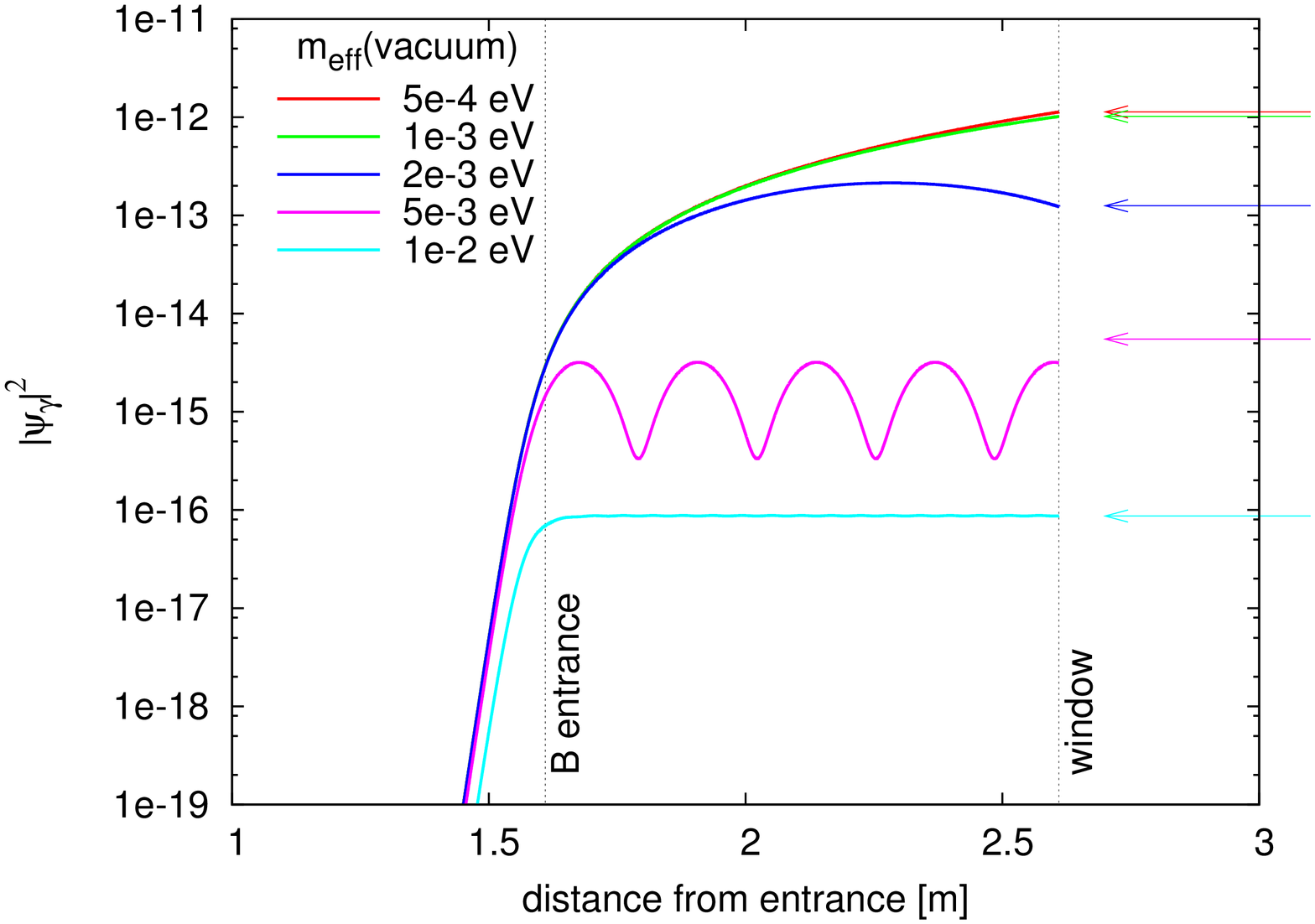}
\includegraphics[angle=270,width=3.3in]{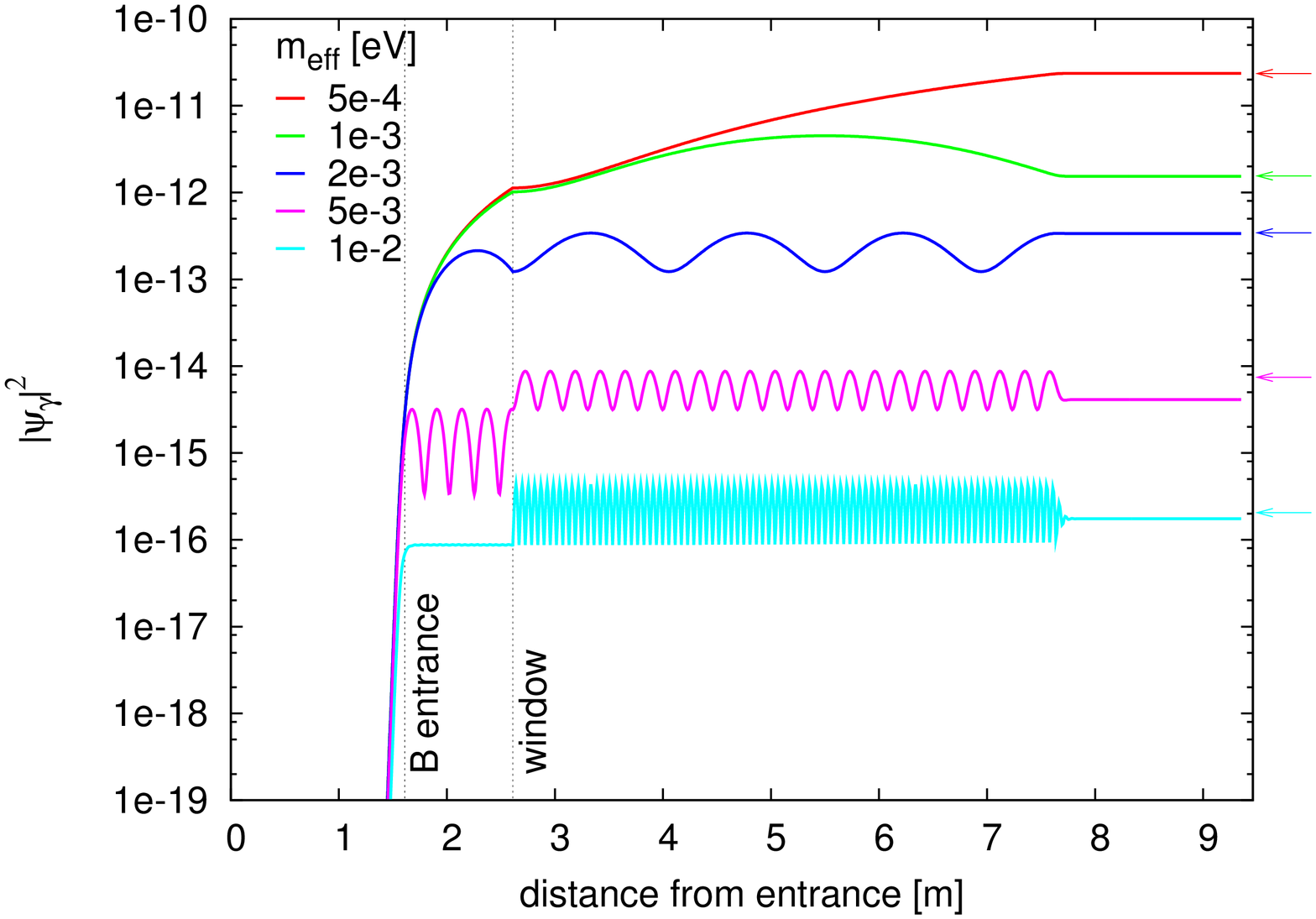}
\caption{Photon amplitude vs.~$z$ for a magnetic field $\Bext = 5$~Tesla with falloff length $\Dzf = 5$~cm, assuming photon coupling $\bgam = 10^{12}$.  Vertical dotted lines show the locations of the window and the beginning of the $\Bext$ region.  Arrows at the right show expected oscillation probabilities for a tophat magnetic field, assuming that a measurement is made at the window. \FigA~Oscillation probability for a particle beginning at the entrance window and measured at the interior window.  \FigB~Oscillation probability for a second particle beginning on the interior window is added to the probability due to the first particle.\label{f:Pgc_realB_1win}}
\end{center}
\end{figure}

Since an increase in the chameleon mass sharply decreases the chameleon-photon oscillation probability~(\ref{e:Pgc}), the two are effectively decoupled near the window. Thus the sudden change in the background chameleon field due to the window does not lead to any sudden changes in the oscillation amplitude.  Figure~\ref{f:Pgc_realB_1win}~\FigA~shows the numerically computed amplitudes for a range of masses up to the location of a window placed one meter inside the $\Bext$ region.  Arrows at the right show the corresponding probabilities for a tophat magnetic field.  The tophat approximation is clearly better here than in Fig.~\ref{f:Pgc_realB_nowin}.  Of course, the smooth transition at the entrance to the $\Bext$ region still has some effect on transition probabilities.  Since it adds a few extra centimeters to the length of the $\Bext$ region, it shifts the locations of the features in the afterglow rate seen in Fig.~\ref{f:Gaft_Gdec_2pt}.  Also, some of the suppression due to the smooth transition remains.  This is best seen in the plot for the $\meff = 5\times 10^{-3}$~eV chameleon in Fig.~\ref{f:Pgc_realB_1win}~\FigA, which has a final photon amplitude about $40\%$ less than the tophat prediction.

Section~\ref{sec:chameleon-photon_oscillation_analytic_calculation} began with a particle in a pure chameleon state immediately after measurement by the entrance window of the vacuum chamber.  In the case of a window in the interior of the $\Bext$ region, a particle can also begin in a pure chameleon state on the surface of the window.  By the homogeneity and isotropy of the chameleon population we expect that for every particle incident on the window from low $z$ with momentum $\vec k$ there will be another leaving the window on the other side with the same momentum.  In order to compute the total oscillation probability due to both partitions of the magnetic field, we must also consider the oscillation due to this second particle.  Fig.~\ref{f:Pgc_realB_1win}~\FigB~shows this combined probability, which once again agrees well (with errors at the $\sim 10\%$ level at high masses) with the tophat magnetic field calculation.

\subsection{Interior windows in \GammeVCHASE}
\label{subsec:interior_windows_in_gammev-chase}

\begin{figure}[tb]
\begin{center}
\includegraphics[angle=270,width=3.3in]{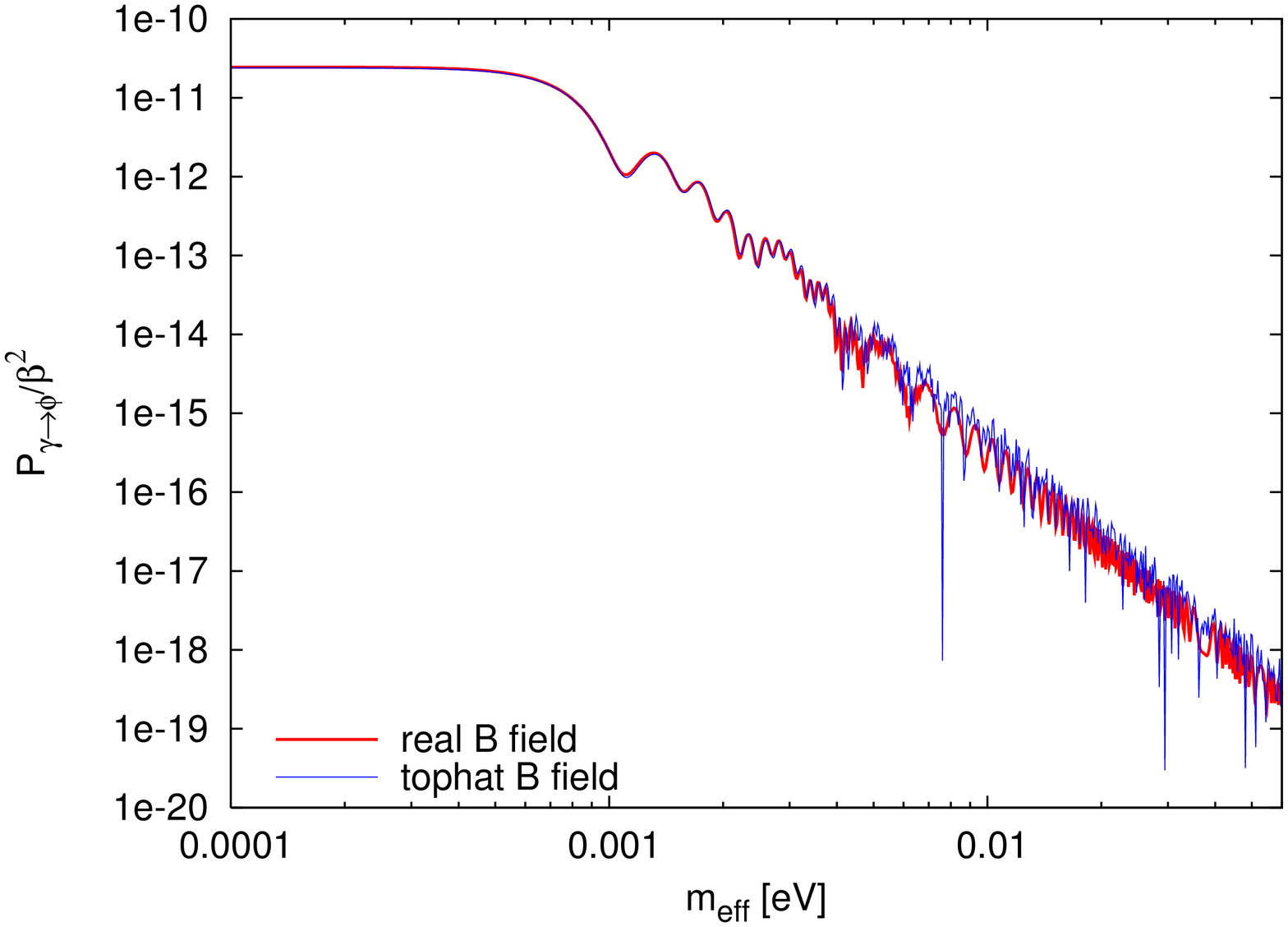}
\includegraphics[angle=270,width=3.3in]{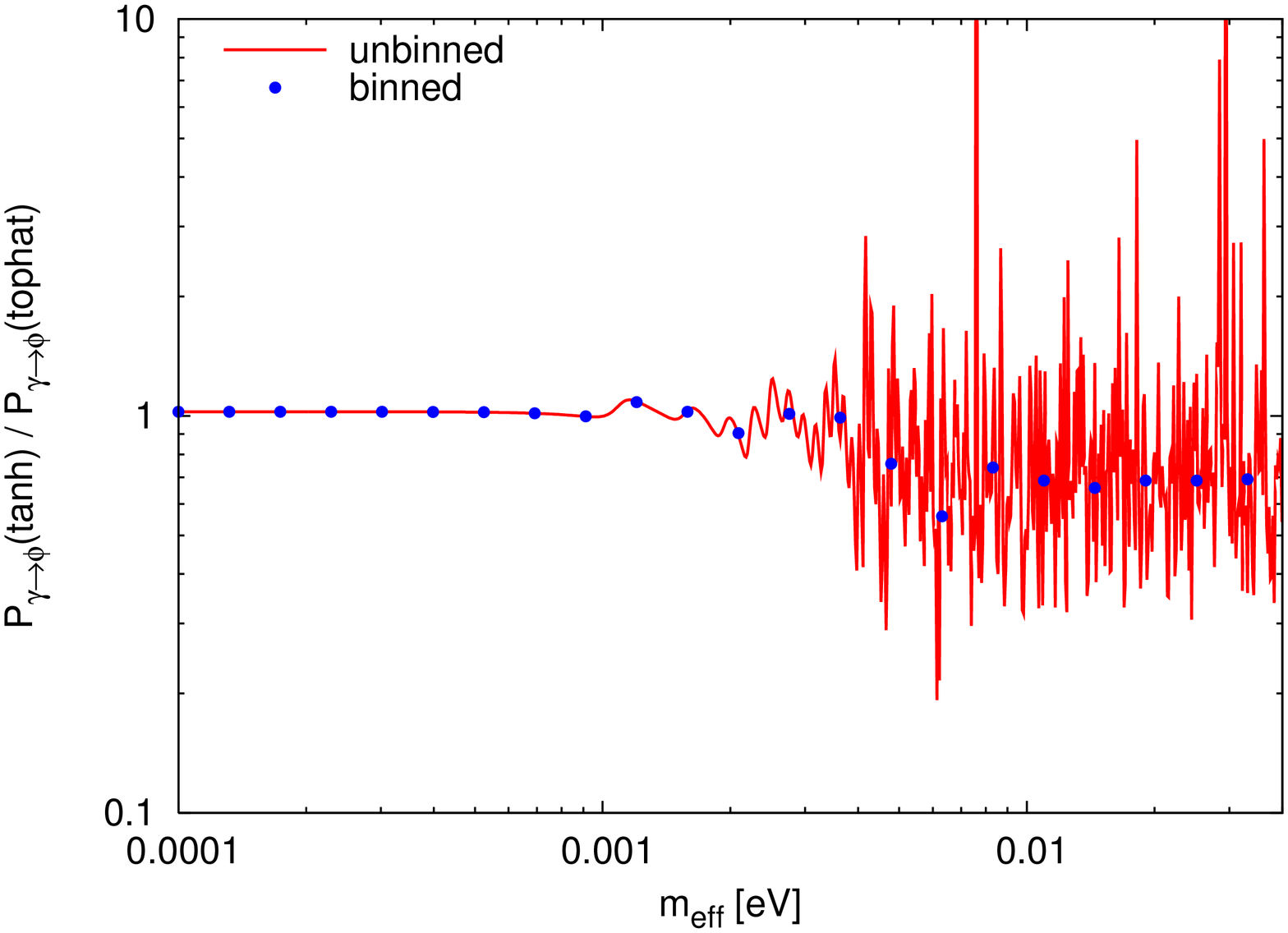}
\caption{\FigA~Oscillation probability vs.~chameleon mass for tophat and $\tanh$ magnetic fields, assuming two interior windows and $\bgam = 10^{12}$.  \FigB~Ratio of $\tanh$ and tophat oscillation probabilities.  \label{f:Pgc12_vs_meff_2win}}
\end{center}
\end{figure}

\GammeVCHASE~uses two interior windows, the first $1.0$~m inside the $\Bext$ region and the second after another $30$~cm.  Each window is $1$~inch in diameter.  Figure~\ref{f:Pgc12_vs_meff_2win}~\FigA~compares the resulting oscillation probability to its tophat approximation.  The adiabatic suppression of oscillation by many orders of magnitude seen in Fig.~\ref{f:Pgc_realB_nowin} is not in evidence here.  However, some suppression remains, as shown in Fig.~\ref{f:Pgc12_vs_meff_2win}~\FigB.  The points in that figure use probabilities averaged in bins of width $10\%$ in $\meff$ in order to average out features due to different effective $\Bext$ region lengths.  This averaging shows that the oscillation probability for the $\tanh$ field is $30\%$ lower than the tophat approximation at high masses.  Since $\bgam \propto \Pgc^{1/2}$, this implies that constraints on $\bgam$ found using the tophat approximation will overestimate actual constraints by $\approx 15\%$ for $\meff \gtrsim 0.01$~eV.  

A more accurate numerical computation would also consider:
\begin{itemize}
\item a range of particle directions $\hat k$, since a particle not propagating parallel to the $\hat z$ axis would see an effective magnetic field falloff length larger than $\Dzf$;
\item a range of particle frequencies within the $\approx 1$~cm linewidth of the laser;
\item reflections from the chamber walls inside the $\Bext$ region, leading to partial measurements of particle content.
\end{itemize}
Each of these corrections will serve to smooth out features in the oscillation probabilities in Fig.~\ref{f:Pgc12_vs_meff_2win}.  Since numerically computing the probability for each path used in a calculation such as that in Section~\ref{sec:chameleon-photon_oscillation_analytic_calculation} would be prohibitive, and since adiabatic suppression is only important in the least interesting region $\meff \gtrsim 0.01$~eV of our parameter space, we have retained the tophat magnetic field approximation in our computations of decay and afterglow rates.  Thus we expect that our constraints will overestimate actual constraints on $\bgam$ at the ten percent level, or approximately the thickness of the line used to plot constraints in Figure~\ref{f:constraints_model-indep} of Section~\ref{subsec:chase_model-independent_constraints}. 

\section{Enhancements in \GammeVCHASE}
\label{sec:enhancements_in_gammev-chase}

The \GammeV~chameleon experiment was similar to the idealized afterglow experiment discussed in Sec.~\ref{subsec:an_idealized_afterglow_experiment}.  \GammeVCHASE~made major improvements to the vacuum system and the detector, as well as adding data runs with lower magnetic field values and partitioning the magnetic field region into sections.  In this section we discuss each of these improvements.

\subsection{Vacuum system}
\label{subsec:pumping_system}

The range of chameleon potentials probed depends on the density ratio between the vacuum and the walls of the vacuum chamber.  A chameleon particle will be trapped inside the vacuum chamber if the containment condition (\ref{e:containment}) is satisfied for $E=2.33$~eV and $\rhom$ equal to the lowest density in the chamber walls.  Meanwhile, an oscillation experiment will only be sensitive to chameleons below some mass $\mmax$ at the density $\rhov$ of the chamber vacuum, since massive chameleons mix very weakly with photons.

Here we study power law and chameleon dark energy potentials (\ref{e:V_powerlaw},\ref{e:V_chameleon_dark_energy}) in order to find the range of exponents $n$ for which some region of the ($\bgam$, $\bmat$, $g$) parameter space is accessible to a given afterglow experiment.   Note that our goal in this section is not to compute actual constraints, but only to provide a simple estimate of the range of potentials which may be probed.  It will be useful for this purpose to neglect $\rhog$, since a greater total density inside the chamber degrades constraints.  This assumption is equivalent to $\bgam \ll \bmat$. 

For potentials (\ref{e:V_powerlaw},\ref{e:V_chameleon_dark_energy}), assuming negligible $\rhog$, eq.~(\ref{e:meffbulk}) for the chameleon mass implies that $\meff \propto \rhom^{(n-2)/(2n-2)}$.  Requiring that $\meff(\phibulk(\rhom)) > E$ (containment) and $\meff(\phibulk(\rhov)) < \mmax$ (mixing), we find
\begin{equation}
\frac{n-2}{2n-2}
>
\frac{\log(E/\mmax)}{\log(\rhom/\rhov)}.
\end{equation}
Now we must consider $n>2$ and $n<0$ chameleons separately.  If $n>2$ and $\rhom/\rhov \leq (E/\mmax)^2$, then the containment and mixing conditions cannot both be satisfied; the experiment is insensitive to these models.  If $n>2$ and $\rhom/\rhov > (E/\mmax)^2$ then 
\begin{equation}
n
>
2
\frac{\log(\rhom/\rhov) - \log(E/\mmax)}
     {\log(\rhom/\rhov) - 2\log(E/\mmax)}.
\end{equation}
That is, for $n$ satisfying this condition, the experiment will be able to probe some region of the ($\bgam$, $\bmat$, $g$) parameter space.  If $n<0$ and $E/\mmax < \rhom/\rhov \leq (E/\mmax)^2$ then
\begin{equation}
n
>
-2
\frac{\log(\rhom/\rhov) - \log(E/\mmax)}
     {2\log(E/\mmax) - \log(\rhom/\rhov)}.
\end{equation}
If $\rhom/\rhov > (E/\mmax)^2$ then all $n<0$ can be probed by the experiment.

\GammeV~used a turbomolecular pump in combination with a roughing pump.  Since any chameleon particle entering the roughing pump would be scooped out of the chamber, \GammeV~was sensitive only to chameleons massive enough to bounce from the $P = 1.9\times 10^{-3}$~torr gas at the intake of that pump.  Thus the lowest-density ``wall'' in \GammeV~had $\rhom = 4\times 10^{-9}$~\gcmc.  The vacuum density was $\rhov = 2\times 10^{-13}$~\gcmc~and $\mmax \approx 10^{-3}$~eV~\cite{Chou_etal_2009}.  Thus \GammeV~was completely insensitive to potentials with $n>2$ and could probe only $-0.77 < n < 0$.  

\GammeVCHASE~maintained a vacuum of $\rhov = 3.7\times 10^{-14}$~\gcmc~using ion pumps and cryo-pumping on the $4$~Kelvin walls of the magnet bore.  This vacuum system did not allow chameleons to escape from the chamber, meaning that the lowest-density regions of the chamber wall were the entrance and exit windows, made of BK7 glass, with density $\rhom = 2.51$~\gcmc.  Although the experimental sensitivity to high-mass chameleons depended on the photon coupling, $\mmax \sim 0.01$~eV is a reasonable approximation.  With this approximation, \GammeVCHASE~can probe $n<0$ and $n > 2.5$.  However, we note that $\rhom \sim \rhog$ in \GammeVCHASE, so the range of potentials probed will be smaller when $\bgam > \bmat$. For example, when $\bgam/\bmat = 10^{3} \rhom/\rhog$, \GammeVCHASE~is sensitive to $n<0$ and $n > 2.8$.

\subsection{Multiple magnetic fields}
\label{subsec:multiple_magnetic_fields}

In order to extend constraints on the photon coupling to lower values, an afterglow experiment would need to use a large magnetic field and collect data over a long time period, as was done by \GammeV.  However, the afterglow signal~(\ref{e:Faft}) falls off exponentially with a rate $\Gdec \propto \bgam^2 \Bext^2$.  Thus increasing $\Bext$ makes the experiment insensitive to chameleon models with large photon couplings.  Since the signal falls off rapidly, collecting data for a longer time does not help.  

\begin{figure*}[tb]
\begin{center}
\includegraphics[angle=270,width=1.7in]{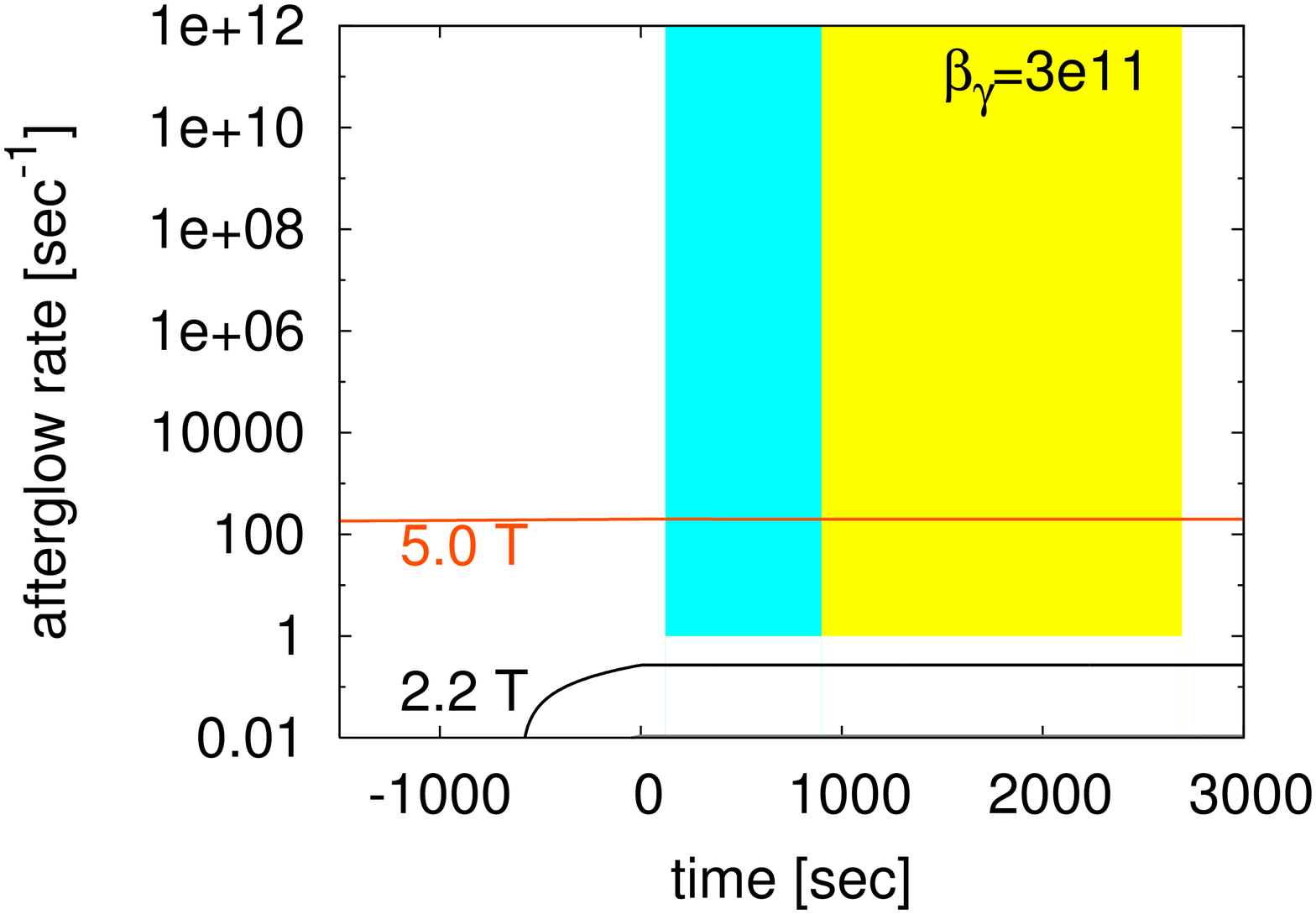}
\includegraphics[angle=270,width=1.7in]{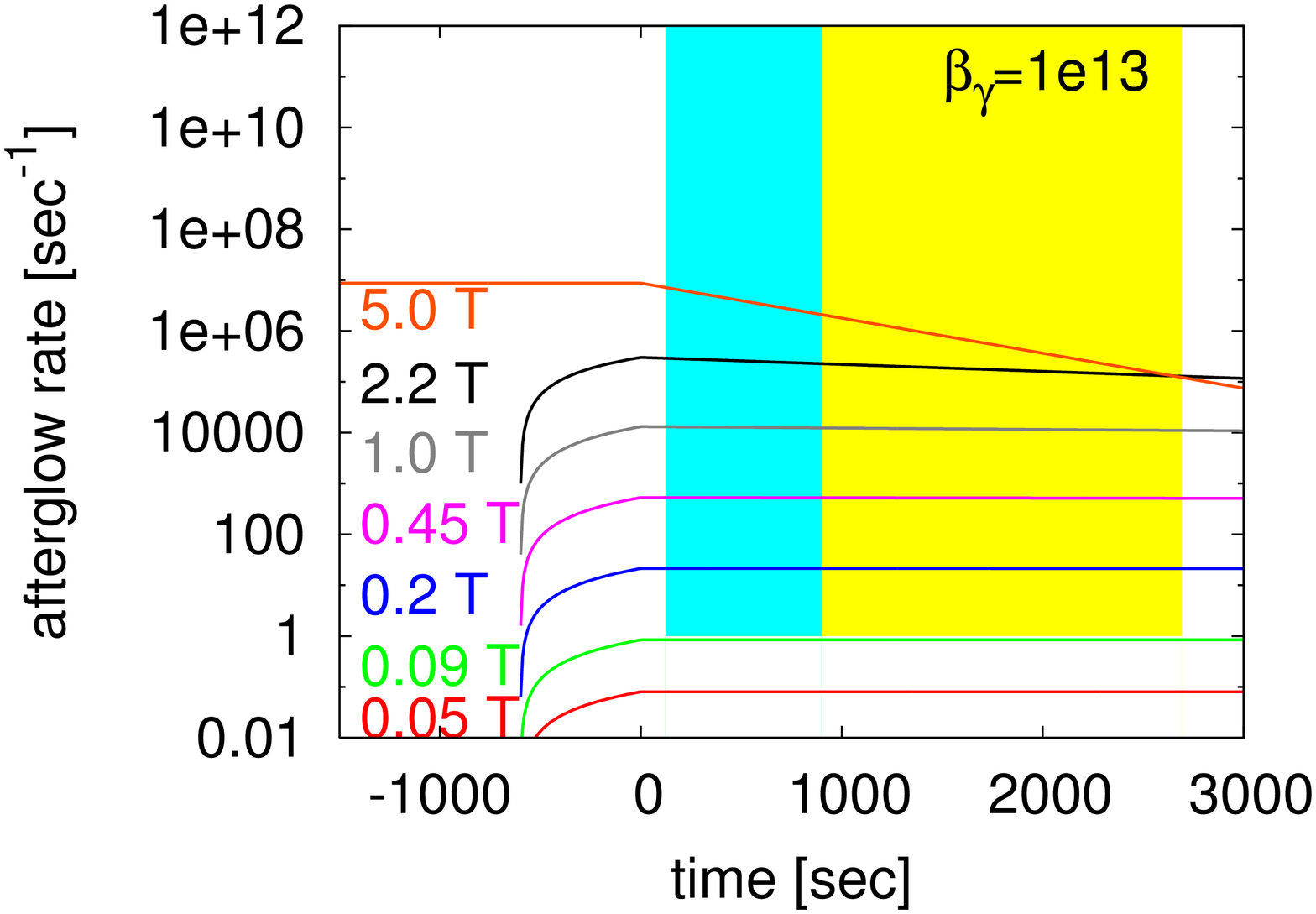}
\includegraphics[angle=270,width=1.7in]{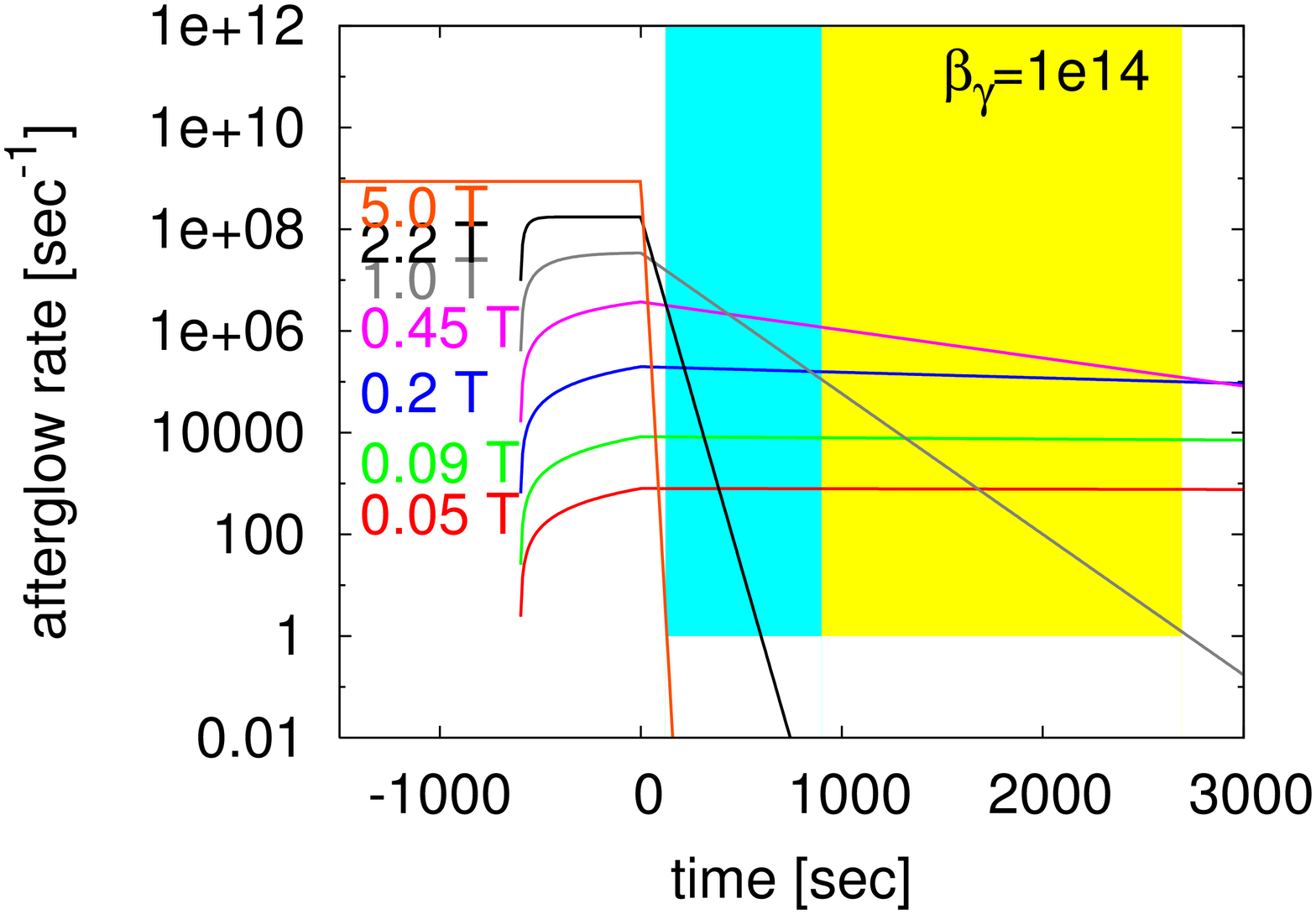}
\includegraphics[angle=270,width=1.7in]{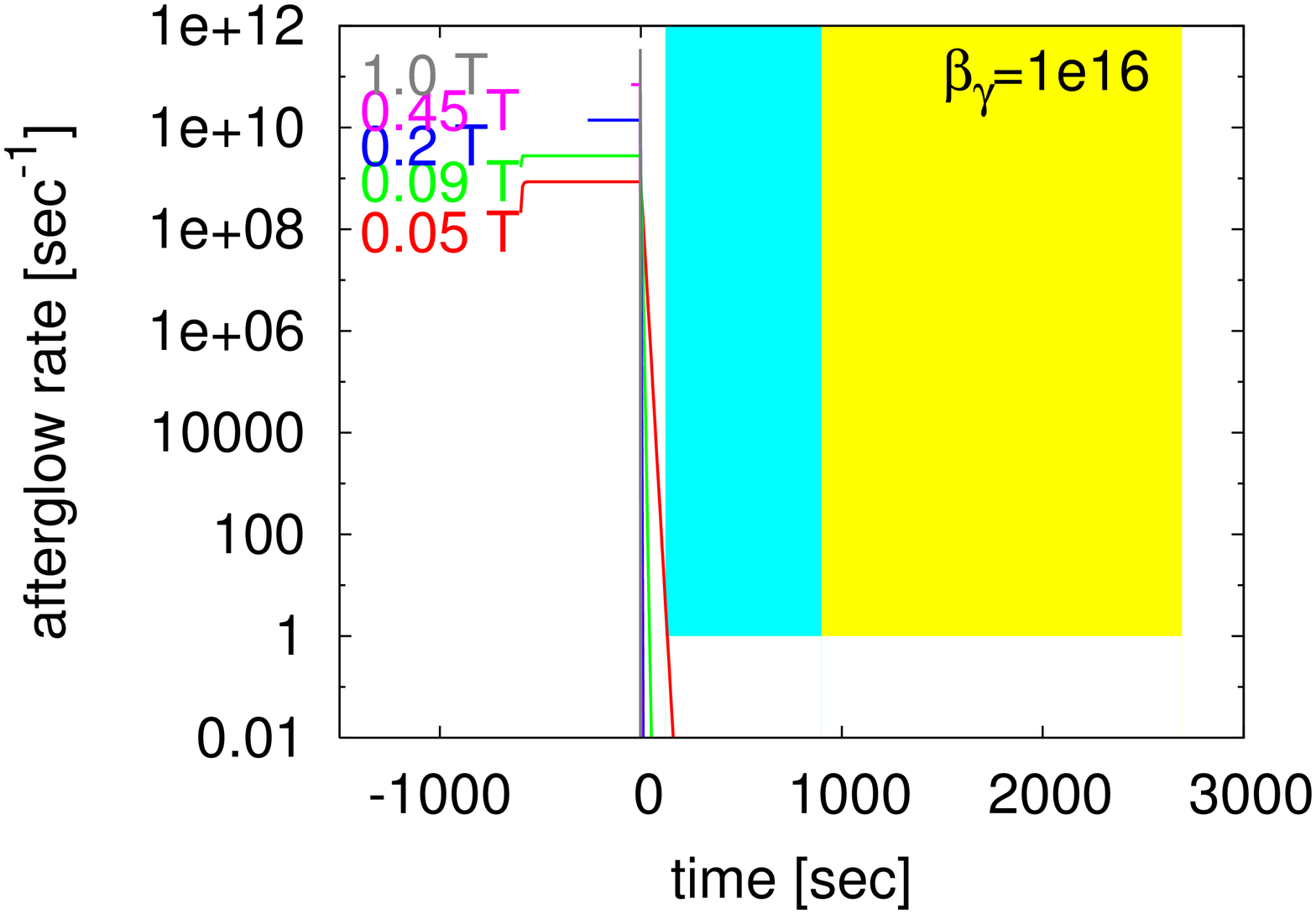}
\caption{Afterglow signal vs. time for various magnetic fields (labelled in plots) and photon couplings (upper right corners of plots).  Approximate observation windows for the short data runs (thin, light blue shaded regions) and long data runs (both shaded regions) are shown in each plot; the long runs are used only for $\Bext = 5$~Tesla.  In all cases, $\meff = 10^{-4}$~eV is assumed.  \label{f:afterglow_signal}}
\end{center}
\end{figure*}

\GammeVCHASE~used short runs with low magnetic fields to supplement the high-$\Bext$, long-duration runs.  For each photon polarization, data were collected in two long runs of $\Bext = 5.0$~Tesla and filling times of $\tprod = 5$~hours.  Short data runs of $\tprod = 10$~minutes each were conducted for magnetic fields of $\Bext = 2.2$~T, $1.0$~T, $0.45$~T, $0.20$~T, $0.089$~T, and $0.050$~T.  Additionally, seven calibration runs were carried out with $\tprod = 10$~minutes and $\Bext = 0$ to allow for further study of backgrounds.  

These low-$\Bext$ runs extended constraints to models with larger $\bgam$.  Figure~\ref{f:afterglow_signal} shows the afterglow signal $\Faft(t)$ from (\ref{e:Faft}) for various magnetic fields and photon couplings.  The narrower, light blue shaded region represents the time window used for the short runs.  The larger, yellow shaded region is the extra observation time in the long runs.  A detailed analysis of the data, properly accounting for statistical and systematic uncertainties, will be provided in Sec.~\ref{sec:analysis_and_constraints}.  Here, Fig.~\ref{f:afterglow_signal} roughly approximates \GammeVCHASE~constraints by assuming that an afterglow signal of at least $1$~Hz two minutes after the laser is turned off will be detected by the experiment.  Thus a curve in Fig.~\ref{f:afterglow_signal} entering the shaded regions will be detected.  

At the lowest $\bgam$ in Fig.~\ref{f:afterglow_signal}, the chameleon is detectable only in the $5$~Tesla run.  By $\bgam=10^{13}$ several other runs can also probe the chameleon.  Meanwhile, the decay time $\Gdec^{-1}$ in the $5$~Tesla run is less than the observation time, so that a decline in $\Faft$ is apparent.  At $\bgam = 10^{14}$ the signal in the $5$~Tesla run falls off so rapidly that it is no longer observable, and \GammeVCHASE~constraints are due entirely to the short runs.  By $\bgam = 10^{16}$, $\Gdec$ is so large, even in the low-$\Bext$ runs, that none of the runs can detect the chameleon signal.  Since collider constraints~\cite{Brax_etal_2009} already exclude $\bgam \gtrsim \Mpl / (1 \mathrm{TeV}) = 2\times 10^{15}$, further \GammeVCHASE~runs at still lower magnetic fields are unnecessary.  

\subsection{Interior windows}
\label{subsec:interior_windows}

As described in Sec.~\ref{subsec:interior_windows_in_gammev-chase}, the  magnetic field region in \GammeVCHASE~is divided into three partitions by interior windows.  In addition to reducing the adiabatic suppression of oscillation rates, these windows fill in some of the gaps in the constraints due to destructive interference during the production phase.  Without the partitions, the afterglow signal~(\ref{e:Faft}) contains a factor of $\Pgc$ due to chameleon production as photons stream through the chamber.  $\Pgc$ vanishes when $\meff^2 = 4n\pi k/L$ for any integer $n$.  This corresponds to maximal destructive interference in oscillation; the chameleon amplitude produced in one part of the chamber exactly cancels that from another.

\begin{figure}[tb]
\begin{center}
\includegraphics[angle=270,width=3.3in]{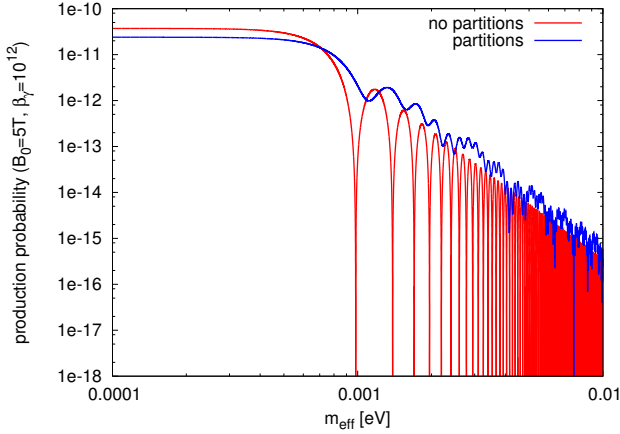}
\caption{Production probability for a chameleon with $\bgam = 10^{12}$ in a magnetic field $\Bext = 5$~Tesla, with and without the \GammeVCHASE~partitions.  \label{f:P_prod_dishrack}}
\end{center}
\end{figure}

Figure~\ref{f:P_prod_dishrack} shows the chameleon production probability per photon, $\Pgc$, as a function of effective mass, with and without partitions in the magnetic field region.  Without partitions, the probability has several zeros near the dark energy mass scale of $2.4\times 10^{-3}$~eV.  The partitions push the first zero of $\Pgc$ from $\meff = 9.8\times 10^{-4}$~eV to $7.6\times 10^{-3}$~eV, eliminating the sharp features in the predicted signal in the region of the dark energy scale. In fact, suppressing destructive interference was the original reason for inclusion of interior windows in the chamber.  The benefits of these windows for mitigating adiabatic suppression of chameleon production were discovered only later.

\subsection{Detector}
\label{subsec:detector}

The dominant source of systematic uncertainty in \GammeV~was instability in the photomultiplier tube used to detect afterglow photons~\cite{Chou_etal_2009}.  The dark rate seen in the PMT had large fluctuations on time scales of $~1$~minute.  As a result, constraints on chameleons had to be obtained by averaging the signal reported by the PMT over long times.  Even the time-averaged rate had a substantial uncertainty; averaging over a large number of hour-long intervals resulted in a mean dark rate of $115$~Hz with a standard deviation of $12$~Hz.  

\GammeVCHASE~used a PMT with a lower mean dark rate, $30$~Hz, which was much more stable with time.  Moreover, the PMT was modulated by a shutter with a period of $30$~sec and a duty cycle of $0.5$, allowing minute-scale variations in the dark rate to be monitored in real time.  Fortunately the dark rate of the \GammeVCHASE~PMT remained stable.  Nevertheless, the experiment was designed to detect any such instability and to subtract its effects from the signal.  

\section{Chameleon-photon oscillation: Monte Carlo simulation}
\label{sec:chameleon-photon_oscillation_monte_carlo_simulation}

Section~\ref{sec:chameleon-photon_oscillation_analytic_calculation} computed the afterglow rate $\Gaft$ and decay rate $\Gdec$ per chameleon particle.  That calculation, which assumed a simplified set of initial conditions, was fast and accurate for a \GammeV-like experiment.  Its extension to the \GammeVCHASE~geometry was found to be accurate to a factor of $\sim 2$ over most of the mass range, makeing it useful when designing broad features of the \GammeVCHASE~geometry.  However, it was inadequate for data analysis for several reasons.
\begin{enumerate}
\item The assumption that the chameleon particle's initial position on the entrance window is irrelevant breaks down when the oscillation length $4\pi k / \meff^2$ of the theory becomes smaller than the chamber radius, corresponding to a mass of $\meff \approx 0.01$~eV.
\item Even at low chameleon masses, the calculation of Sec.~\ref{sec:chameleon-photon_oscillation_analytic_calculation} is inaccurate for nontrivial chamber geometries such as the interior windows in \GammeVCHASE.
\item That calculation cannot include random systematic effects such as diffuse reflection from the chamber walls.
\end{enumerate}
Thus a Monte Carlo simulation of \GammeVCHASE~was used to compute the afterglow signal and to estimate the effects of random systematics in the \GammeVCHASE~analysis~\cite{Steffen_etal_2010}.

\subsection{Particle paths}
\label{subsec:particle_paths}

Relaxing the restrictions on the initial conditions used in Sec.~\ref{sec:chameleon-photon_oscillation_analytic_calculation}, the decay and afterglow rates are given by
\begin{eqnarray}
\Gdec
&=&
\frac{1}{4\pi\Awin}
\int_\mathrm{win}d^2x
\int_\Omega d^2\Omega
\Pdec \frac{\cos\theta}{\ltot}
\\
\Gaft
&=&
\frac{1}{4\pi\Awin}
\int_\mathrm{win}d^2x
\int_\Omega d^2\Omega
\Paft \frac{\cos\theta}{\ltot}
\label{e:Gaft_full_init}
\\
\Pdec 
&=& 
\PabsBexit + \left|\psigBexit\right|^2
\\
\Paft
&=&
\left| (\psigBexit \cdot \hat S)A_S^{N-\nR}\right|^2 \Pdet
+
(S\rightarrow P),
\end{eqnarray}
where the $d^2x$ integrals are carried out over the surfaces of the windows and the $d^2\Omega$ integrals over the unit sphere of $\hat k$ values.
Here $\Awin$ is the total surface area of the window surfaces inside the chamber.  The probabilities $\Paft$ and $\Pdec$ are oscillation probabilities along particle paths which are completely determined by their initial positions and directions.  A Monte Carlo calculation of these rates replaces the integrals with sums over particle paths with randomly chosen initial conditions.  In the limit that the number of particles simulated becomes large, the Monte Carlo rates will approach the actual rates.

A particle in a pure chameleon state can result from any quantum-mechanical measurement of particle content at a glass window, including the entrance and exit windows as well as the interior windows of \GammeVCHASE.  Thus we initialize each particle by randomly choosing a location on one of the window surfaces, with probability  proportional to surface area.  When computing the decay rate, the initial particle direction is chosen randomly from a uniform distribution as appropriate to an isotropic chameleon population.  Although this would also be correct for the afterglow calculation, photons emerging from the chamber at most of these directions would not reach the detector.  This would cause the Monte Carlo calculation of $\Gaft$ to converge very slowly.  Instead we choose particle directions from a distribution that is uniform over the portion of the unit sphere with $\theta < \theta_0 = 0.1$ and zero elsewhere.  Since this probability distribution covers a fraction $f_\theta = (1-\cos\theta_0)/2$ of the unit sphere, we must multiply our result by $f_\theta$ to obtain $\Gaft$.

\begin{figure}[tb]
\begin{center}
\includegraphics[angle=270,width=3.3in]{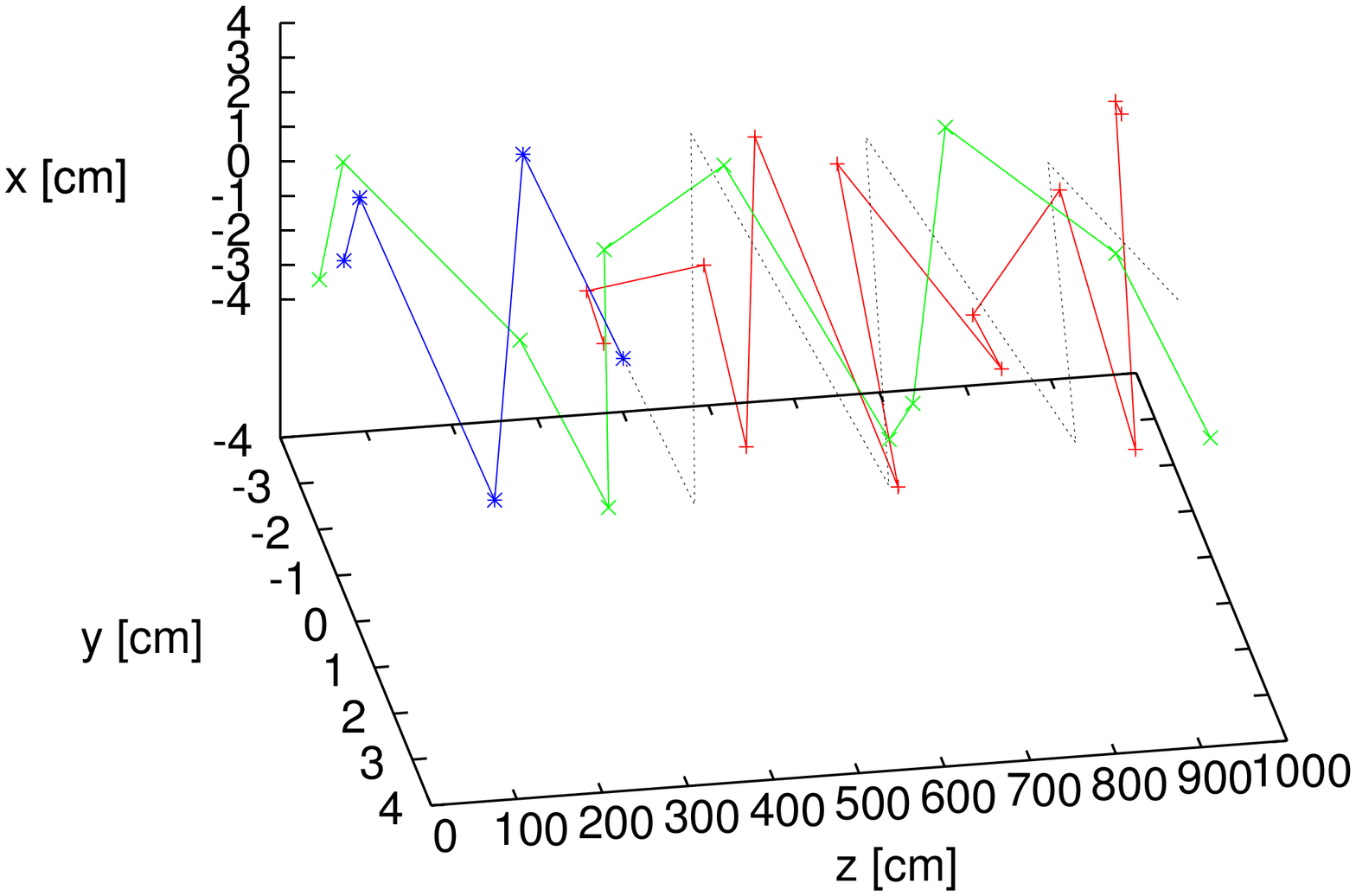}
\includegraphics[angle=270,width=3.3in]{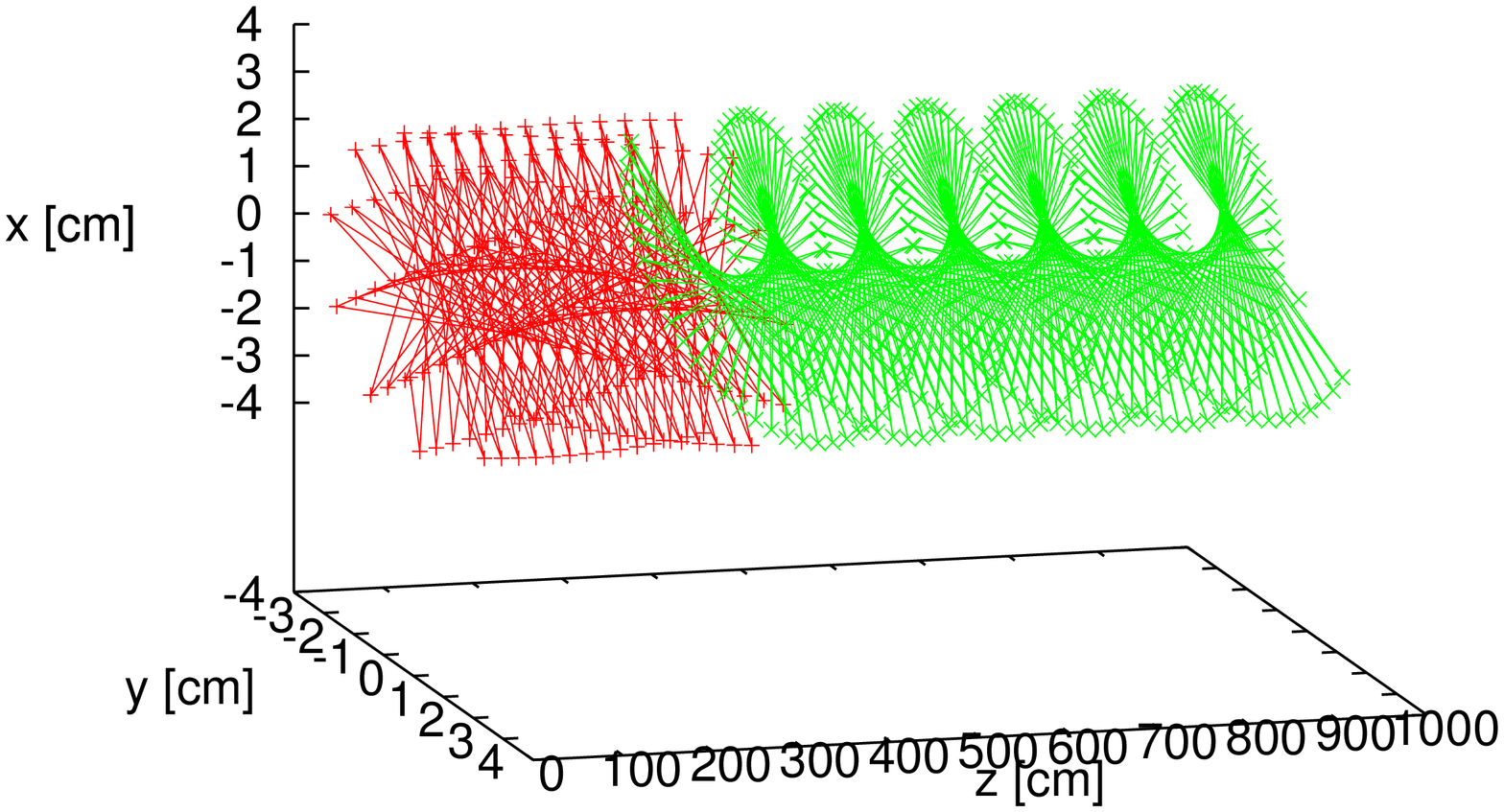}
\caption{Typical particle paths used in the Monte Carlo simulation of \GammeVCHASE.  \FigA~Paths used to compute $\Gaft$.  The red path (with ``$+$'' symbols showing bounces from chamber walls) begins on an interior window and ends on the exit window.  The green path (with ``$\times$'' symbols denoting bounces) begins on the entrance window and ends on the exit window.  The blue path (with ``*'' symbols denoting bounces) begins on the entrance window and ends on an interior window; the subsequent dotted line denotes the photon which may result from the quantum measurement at the interior window.  \FigB~Paths used to compute $\Gdec$.  The red path (``$+$'') begins on an internal window and ends on the entrance window, while the green path (``$\times$'') begins on the exit window and ends on an interior window.  Since a larger range of directions is allowd in the decay rate computation, the typical path bounces from chamber walls much more frequently than those used in the afterglow computation.  \label{f:mc_paths}}
\end{center}
\end{figure}

 Once an initial position and direction have been chosen, the particle is propagated in that direction until it encounters a wall or window.  At a chamber wall, the particle direction is changed and its photon momentum corrected to account for absorption and phase shifting, as described in Sec.~\ref{sec:reflection_from_a_barrier}.  The particle is then propagated forward once again until a window or another wall is encountered.  At a window, the particle content is measured.  Continuing to follow the chameleon particle after this point would constitute a double-counting, since the chameleon position and direction immediately after this bounce from the window is within the allowed set of chameleon initial conditions.  For this chameleon path, $\Pdec$ is the probability that the particle is a photon at the window plus the total probability of absorption during collisions with chamber walls.  In order to compute $\Paft$, we must compute not just the photon amplitude at this window but the probability that the photon will escape from the chamber  and reach the detector.  The factor $|A_S|^{2(N-\nR)}$ in (\ref{e:Gaft_full_init}) is the probability of escape without absorption in the walls, and the factor $\Pdet$ is the probability of detection.  Thus we must also keep track of the photon which may result from quantum measurement at a window. Figure~\ref{f:mc_paths} shows a few sample paths used by the Monte Carlo calculation to compute the afterglow rate (Left) and the decay rate (Right). 

\begin{figure}[tb]
\begin{center}
\includegraphics[angle=270,width=3.3in]{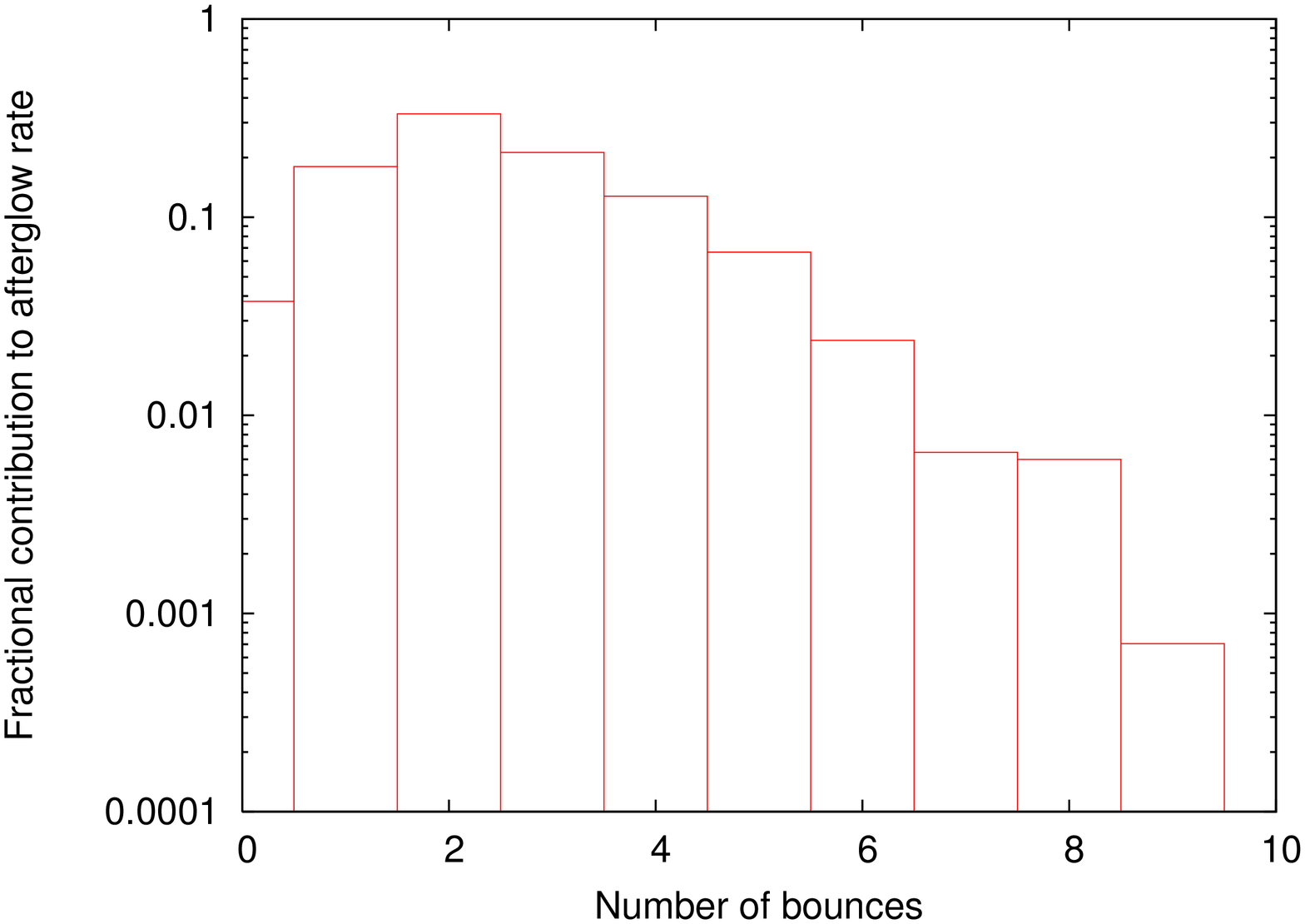}
\includegraphics[angle=270,width=3.3in]{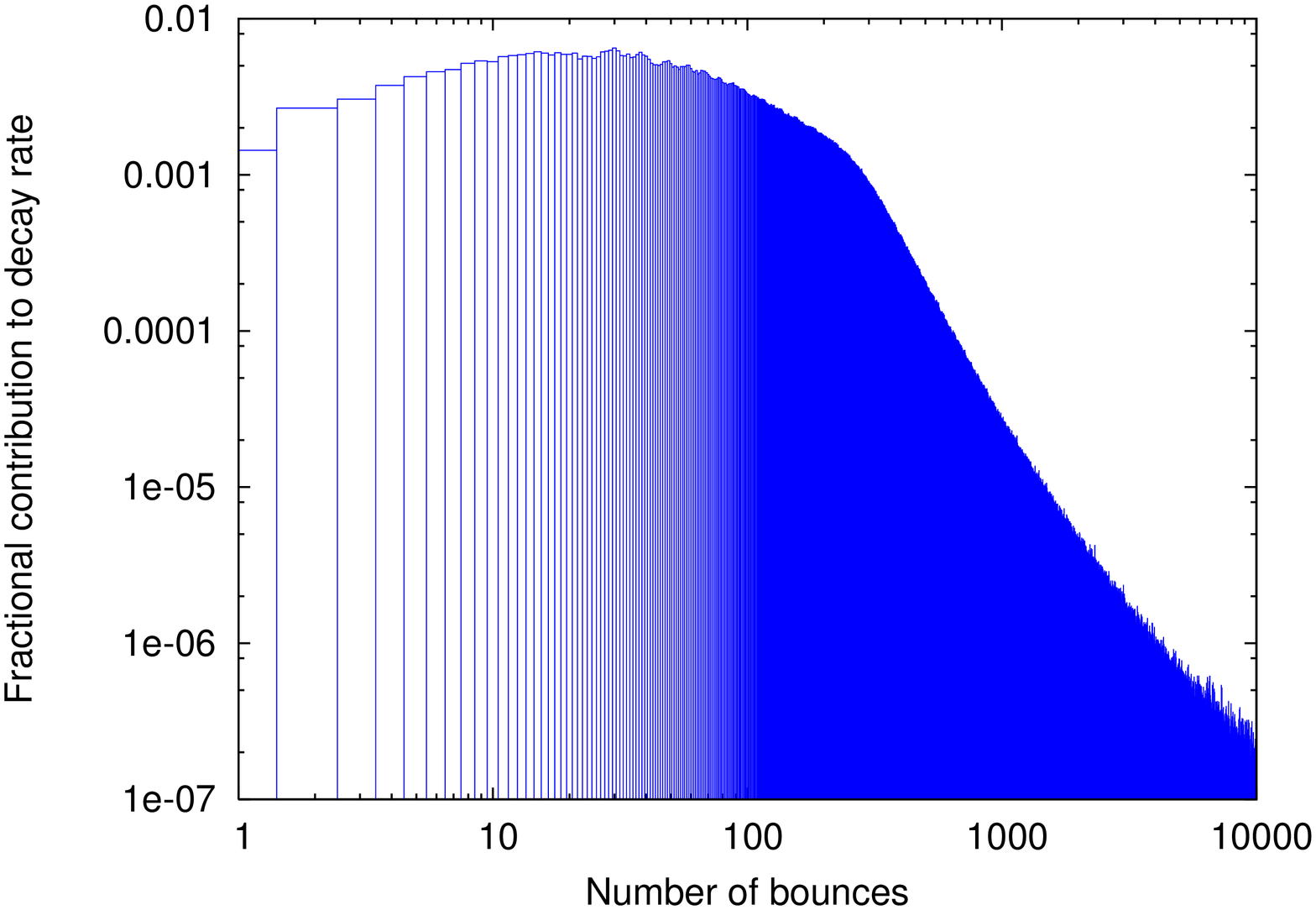}
\caption{Contributions to \FigA~afterglow and to \FigB~decay rates from paths vs. number of wall bounces per path.  The afterglow rate receives no contributions from paths bouncing more than $9$ times, since these emerge from the chamber at too great an angle to reach the \GammeVCHASE~detector.\label{f:mc_Nbounce}}
\end{center}
\end{figure}

From Fig.~\ref{f:mc_paths} it is evident that simulated paths used in the decay rate computation bounce many more times from the chamber walls than paths used to compute the afterglow rate.  This is because paths which bounce many times have large angles $\theta$ with respect to the cylinder axis; thus photons resulting from such paths are less likely to emerge from the \GammeVCHASE~chamber and to reach the detector.  Figure~\ref{f:mc_Nbounce} shows the fractional contributions to the total afterglow and decay rates from paths with different numbers of wall bounces.  As expected, afterglow is dominated by paths with very few bounces; a third of the paths bounce exactly twice, while over $90\%$ bounce between $1$ and $5$ times.  A path with only two bounces must have an angle less than $2R / (\frac{1}{3}\ltot) = 0.020$, or about $1^\circ$.  Thus the afterglow rate is dominated by paths reflecting from the chamber walls at grazing incidence.  In contrast, paths contributing to the decay rate reflect at a large range of angles.  Their contributions to the total rate do not fall off until the number of bounces is around $\ltot / R \approx 300$.

\subsection{Tests of the Monte Carlo simulation}
\label{subsec:tests_of_the_monte_carlo_simulation}

\begin{figure}[tb]
\begin{center}
\includegraphics[angle=270,width=3.3in]{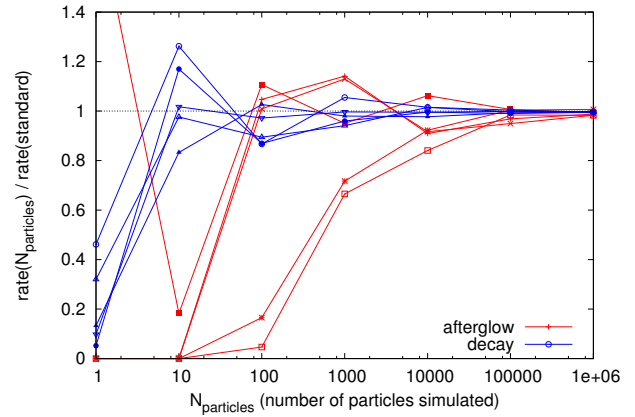}
\caption{Convergence of the Monte Carlo simulated rates as the number of particles simulated is increased.  For each particle number $N_\mathrm{particles}$, each point represents a different random number seed used in the Monte Carlo simulation.  $\bgam=10^{12}$, $\meff = 10^{-4}$~eV, and $\Bext=5$~Tesla are assumed. 
\label{f:MC_convergence}}
\end{center}
\end{figure}

In the limit that the number $N_\mathrm{particles}$ of particles simulated becomes large, the Monte Carlo computations of $\Gaft$ and $\Gdec$ should converge to constant values.  This convergence can be studied by changing the random number seed used by the simulation at fixed $N_\mathrm{particles}$.  The scatter in the resulting rates should be smaller at larger $N_\mathrm{particles}$.  Figure~\ref{f:MC_convergence} shows that the rates do indeed converge, in the sense that the standard deviation of several rate calculations with different random number seeds becomes small.  The decay rate $\Gdec$ has converged to $0.5\%$ by $N_\mathrm{particles}=10^5$ and the afterglow rate has converged to $1\%$ by $N_\mathrm{particles}=10^6$.  Since most of the simulated afterglow paths do not reach the detector, even with the restricted set of initial conditions used, $N_\mathrm{particles}$ must be an order of magnitude larger in the afterglow calculation in order for the rates to converge to the percent level.  Henceforth we use $N_\mathrm{particles}=10^5$ for the $\Gdec$ computation and $N_\mathrm{particles}=10^6$ for the $\Gaft$ computation.

\begin{figure}[tb]
\begin{center}
\includegraphics[angle=270,width=3.3in]{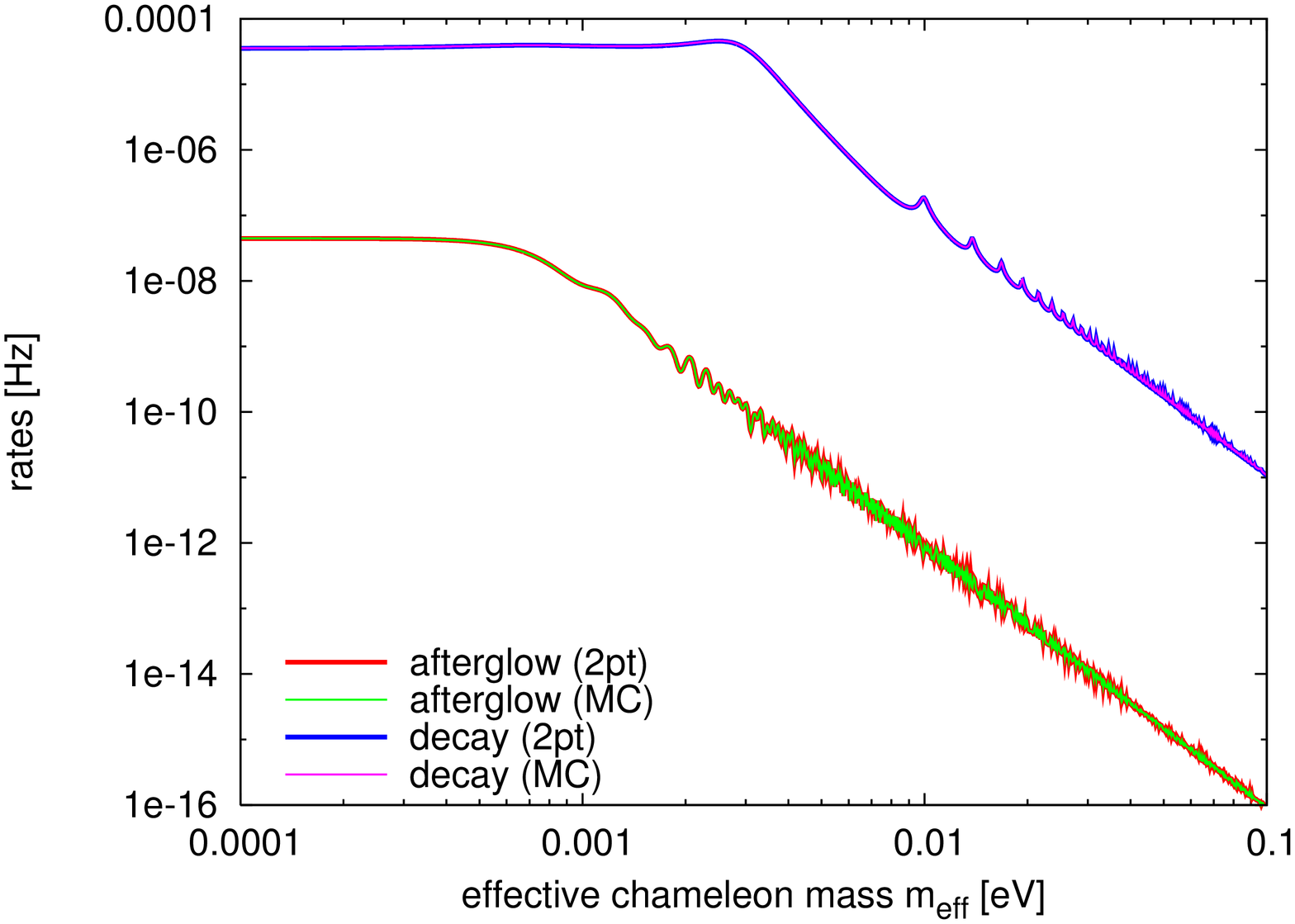}
\includegraphics[angle=270,width=3.3in]{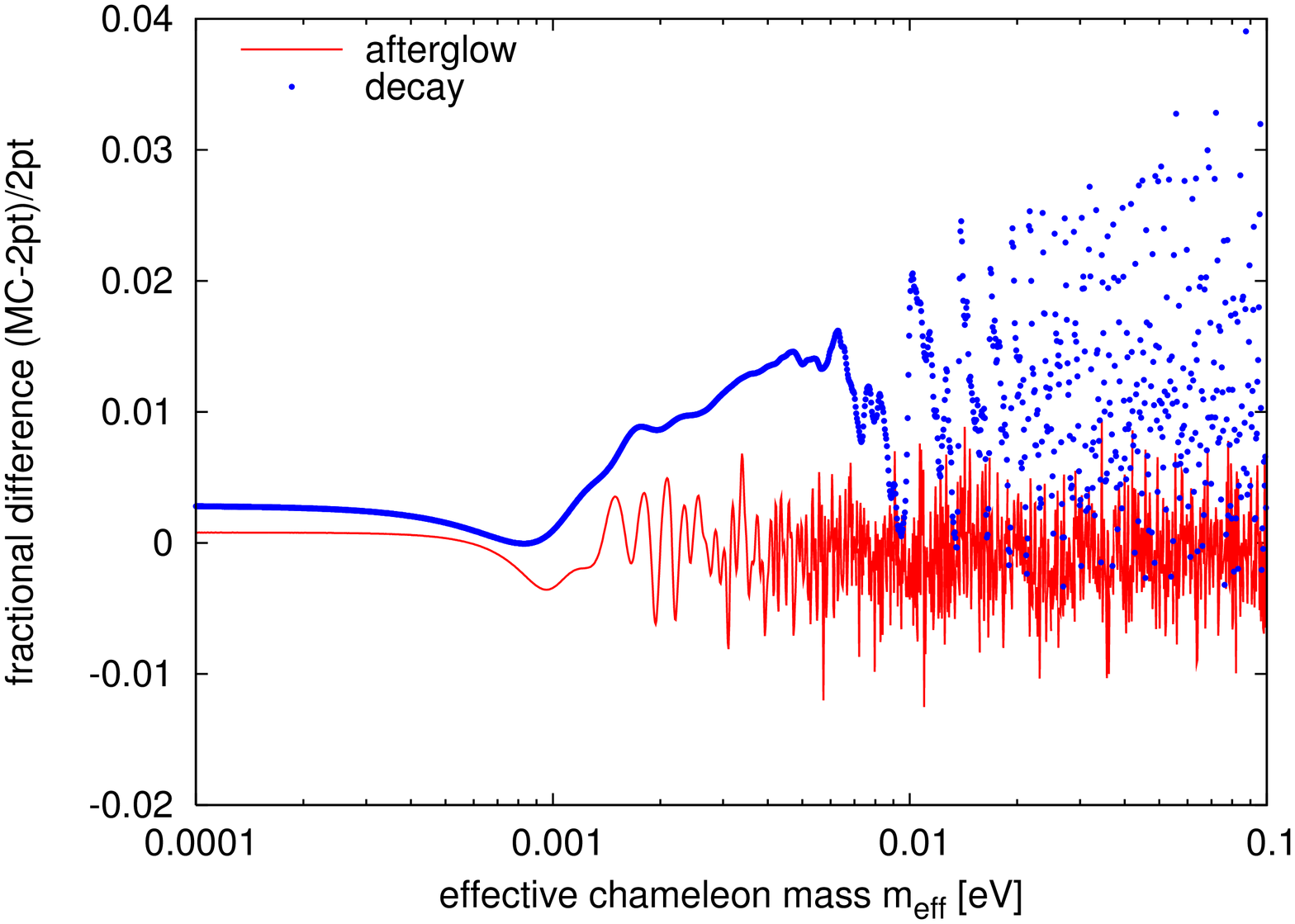}
\caption{\FigA~$\Gaft$ and $\Gdec$ for the 2-point (2pt) calculation of Sec.~\ref{sec:chameleon-photon_oscillation_analytic_calculation} as well as the Monte Carlo (MC) calculation with all particles required to start at the center of the entrance window.  In both calculations, the chamber has no interior windows.  $\bgam=10^{12}$ and $\Bext=5$~Tesla are assumed. \FigB~Fractional difference between the 2-point and Monte Carlo calculations.  \label{f:2pt_vs_mc}}
\end{center}
\end{figure}

We have shown that the Monte Carlo computation of $\Gaft$ and $\Gdec$ is precise.  Next we show that it is accurate.  The 2-point calculation of Sec.~\ref{sec:chameleon-photon_oscillation_analytic_calculation} found these rates exactly for a chamber with no interior windows and a restricted set of initial conditions, namely, that the particles begin at the center of the entrance window.  The same  conditions can be imposed on the Monte Carlo calculation for the sake of comparison.  Fig.~\ref{f:2pt_vs_mc} shows $\Gaft$ and $\Gdec$ computed using the two methods.  Evidently both rates are accurate at the $\approx 1\%$ level across the full range of chameleon masses. 

\begin{figure}[tb]
\begin{center}
\includegraphics[angle=270,width=3.3in]{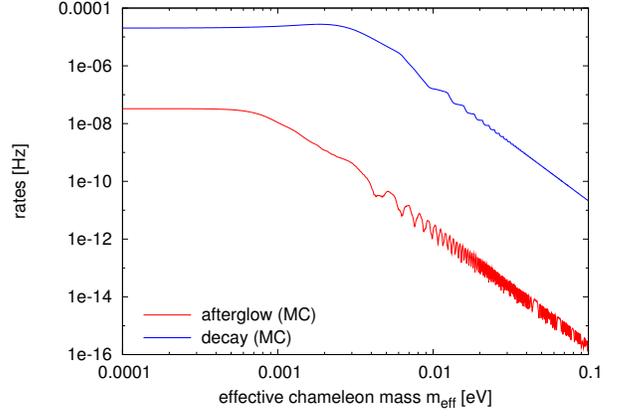}
\caption{Afterglow and decay rates computed using the Monte Carlo simulation.  Particles are allowed to begin at any position on a window surface.  $\xi_V = 0$, $\bgam=10^{12}$, and $\Bext=5$~Tesla are assumed.\label{f:mc_rates}}
\end{center}
\end{figure}

Figure~\ref{f:mc_rates} shows the afterglow and decay rates per chameleon particle, computed with no restrictions on the initial particle position, and including the interior windows.  In Section~\ref{sec:analysis_and_constraints} we will use such Monte Carlo calculations with appropriate values of the potential-dependent phase $\xi_V$ in order to determine \GammeVCHASE~constraints.  

\subsection{Diffuse reflection}
\label{subsec:diffuse_reflection}

Thus far we have assumed perfectly specular reflection of particles from the \GammeVCHASE~chamber walls; that is, the angle of incidence equals the angle of reflection.  Suppose instead that a fraction $\fdiff$ of bounces resulted in perfectly diffuse reflection, in which all directions were equally likely for the reflected particle.  This could result from a rough surface whose local normal vector could differ substantially from that of a perfect cylinder.  Given an incident direction ${\hat k}_I$ and a randomly chosen reflected direction ${\hat k}_R$, the effective local normal vector is proportional to ${\hat k}_R - {\hat k}_I$.

\begin{figure}[tb]
\begin{center}
\includegraphics[angle=270,width=3.3in]{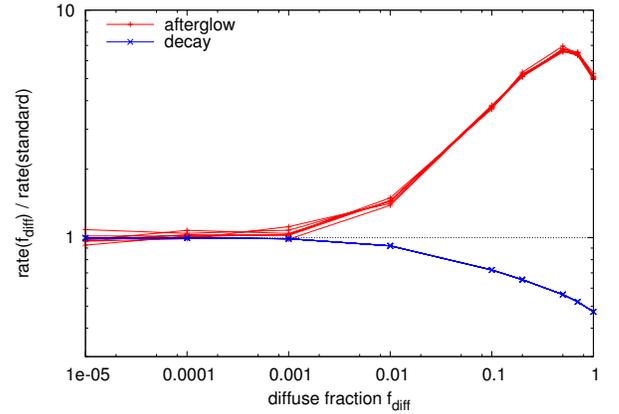}
\caption{Fractional change in afterglow and decay rates vs.~fraction $\fdiff$ of diffuse reflection.  Rates for six different random number seeds are shown.\label{f:spec_vs_diff}}
\end{center}
\end{figure}

Figure~\ref{f:spec_vs_diff} shows the change in afterglow and decay rates when the fraction of diffuse reflections is increased.  For this computation we allow all initial directions for afterglow paths, rather than requiring that $\theta < \theta_0 = 0.1$ as before.  The qualitative effect at $\fdiff \ll 1$ is that diffuse reflection increases the afterglow rate and slightly decreases the decay rate.  This is because such diffuse reflection is most important for particles with large $\theta$, which bounce many times but do not reach the detector.  Diffuse reflection gives such a particle a nonzero chance of reaching the detector, thereby increasing the afterglow rate.  Since increasing $\Gaft$ and decreasing $\Gdec$ both improve \GammeVCHASE~constraints, henceforth we make the conservative assumption that $\fdiff=0$.

\subsection{Sensitivity to chamber properties}
\label{subsec:sensitivity_to_chamber_properties}

\begin{figure}[tb]
\begin{center}
\includegraphics[angle=270,width=3.3in]{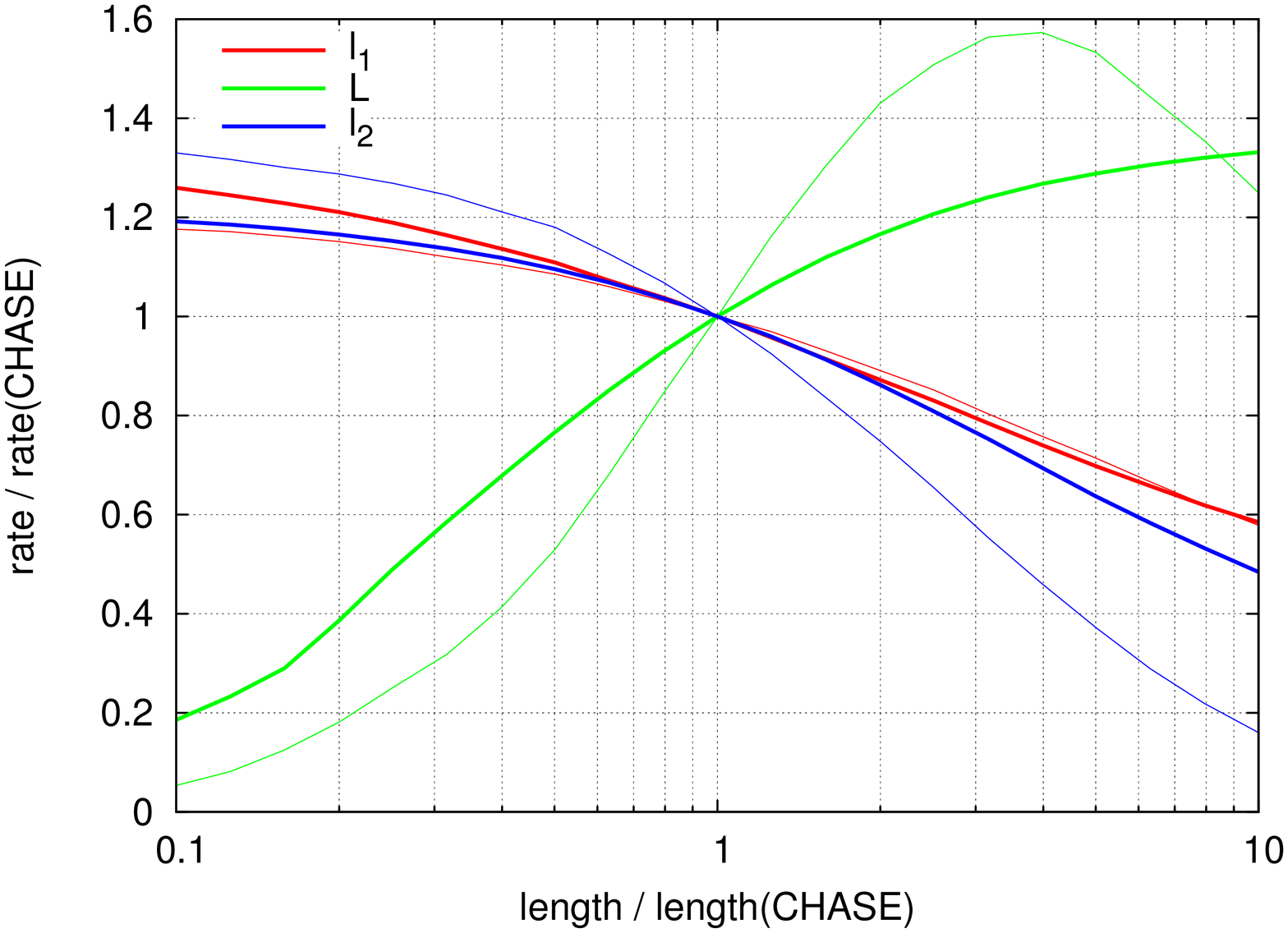}
\includegraphics[angle=270,width=3.3in]{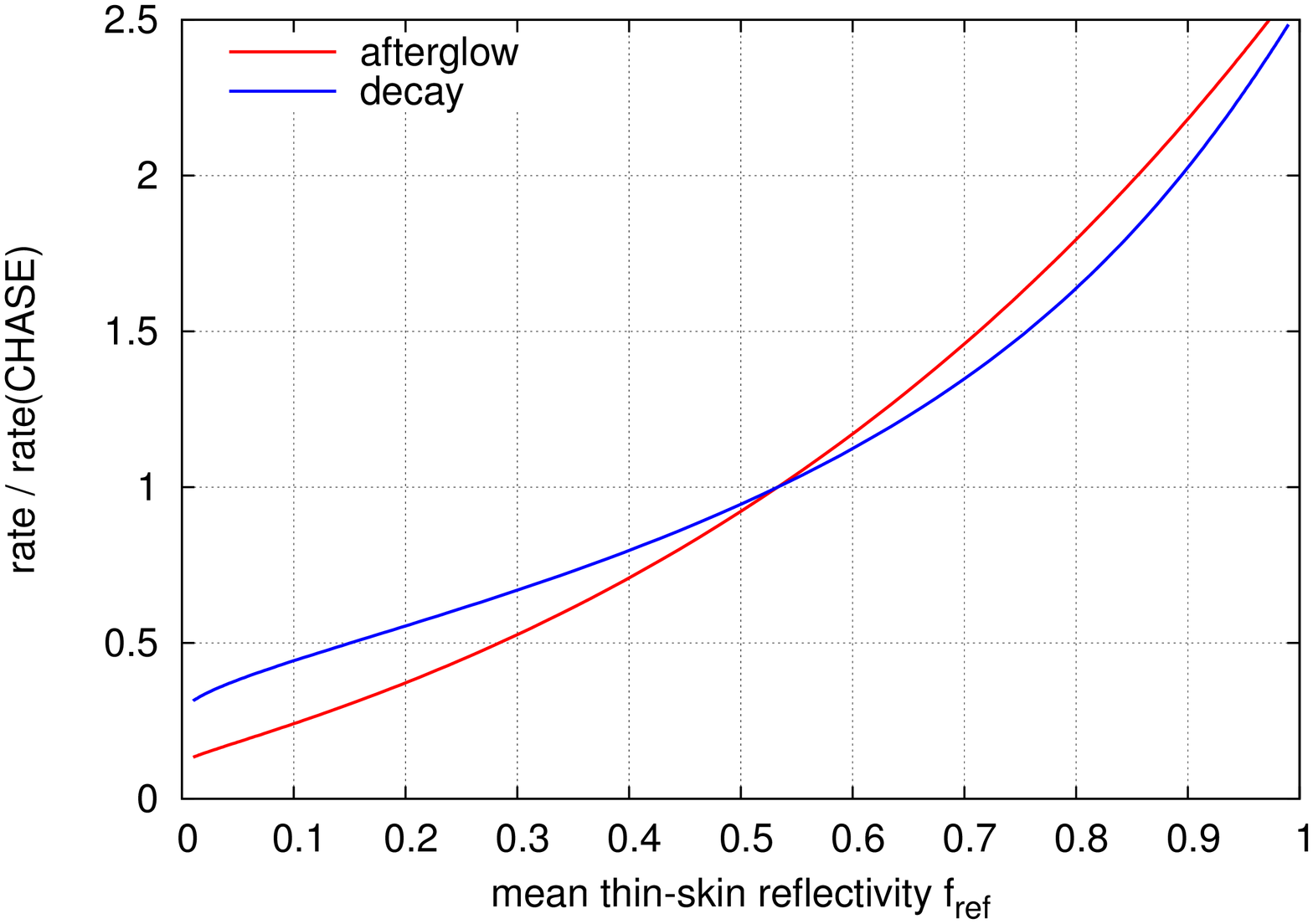}
\caption{\FigA~Afterglow rates (thin lines) and decay rates (thick lines) as the lengths $\ell_1$, $L$, and $\ell_2$ are varied relative to their \GammeVCHASE~values in Table~\ref{t:gammev_apparatus}.  \FigB~Afterglow and decay rates vs. the mean wall reflectivity $\fref$ in the thin-skin limit $\tilde n_1 \rightarrow \infty$.  In both plots, $\meff=10^{-4}$~eV and $\xi_V=0$ have been assumed.  \label{f:vary_parameters}}
\end{center}
\end{figure}

The geometry of the \GammeVCHASE~experiment is illustrated in Figure~\ref{f:gammev_apparatus}, with numerical quantities listed in Table~\ref{t:gammev_apparatus}.  The length $L$ of the magnetic field region, as well as its offsets $\ell_1$ and $\ell_2$ from the entrance and exit windows, respectively, affect the afterglow and decay rates.  The same is true of the reflectivity of the chamber walls. Such properties may be important to the design of future afterglow experiments, so it is instructive to consider their effects on the \GammeVCHASE~rates.

Figure~\ref{f:vary_parameters}~\FigA~shows how $\Gaft$ and $\Gdec$ change when the lengths $\ell_1$, $L$, and $\ell_2$ are varied relative to their \GammeVCHASE~values.  As expected, increasing $L$ from low values causes $\Gaft$ and $\Gdec$ to grow as $L^2$.  Particularly for the afterglow rate, this growth levels off as destructive interference during oscillation becomes important and as each particle must bounce from the walls a greater number of times before reaching the exit window.  Meanwhile, increasing $\ell_1$ and $\ell_2$ cause the chamber volume to increase without adding volume to the magnetic field region.  This means that each particle spends less time in the $\Bext$ region, hence the oscillation rate is smaller.  Increasing $\ell_2$ also has the effect of making particles bounce more between the $\Bext$ region and the exit window, further suppressing $\Gaft$.  However, varying $\ell_1$ and $\ell_2$ by factors of two in either direction only changes $\Gdec$ by about $10\%$ and $\Gaft$ by at most $25\%$.  Thus \GammeVCHASE~constraints will be relatively insensitive to the chamber geometry.  

An additional consideration is the volume occupied by the vacuum system.  Suppose that the cylindrical chamber considered above is only a fraction $\fvol$ of the total volume available to chameleon particles.  For \GammeVCHASE~$\fvol = 0.68$.  Since the distribution of chameleon particles is  homogeneous, only a fraction $\fvol$ of the particles will be in the chamber at any given time, with the remainder inside the vacuum system.  Thus the rates $\Gdec$ and $\Gaft$ computed earlier in this section must be multiplied by $\fvol$.  Constraints presented in Sec.~\ref{sec:analysis_and_constraints} include this factor.

The mean reflectivity $\fref$ is varied in Figure~\ref{f:vary_parameters}~\FigB. Since the photon skin depth and the visibility factor $\fvis$ discussed in Sec.~\ref{subsec:reflection_and_absorption_of_photons} are specific to the material of the chamber wall, Fig.~\ref{f:vary_parameters}~\FigB~makes the simplifying assumption of zero skin depth.  This is equivalent to $\tilde n_1 \rightarrow \infty$ and $\fvis = \fref$ in the notation of Sec.~\ref{subsec:reflection_and_absorption_of_photons}.  Around the \GammeVCHASE~value of $\fref = 0.53$, the afterglow and decay rates scale approximately linearly with $\fref$.  We note that the $7.5\%$ measurement uncertainty in $\fref$ will lead to $\approx 10\%$ uncertainties in $\Gaft$ and $\Gdec$.  Since $\bgam \propto \Faft^{1/4}$ at low $\bgam$, a $10\%$ uncertainty in $\Gaft$ implies a $2.5\%$ uncertainty in the \GammeVCHASE~upper bound on $\bgam$, which is a nontrivial contribution to the total uncertainty.  

In future experiments, highly polished chamber walls would not strengthen constraints by much.  Polished metal, with $\fref \approx 0.85-0.9$, would only double the afterglow signal, improving constraints on $\bgam$ by $2^{1/4}-1 \approx 20\%$.  What is important is measuring $\fref$ at the $10\%$ level so that a precise bound can be placed on $\bgam$.

\section{Analysis and constraints}
\label{sec:analysis_and_constraints}

\subsection{Profile likelihood analysis}
\label{subsec:profile_likelihood_analysis}

Finally, we apply the results of the preceding sections to \GammeVCHASE~data using the profile likelihood method of~\cite{Rolke_Lopez_Conrad_2005}.  The afterglow photon rate $\Faft$ from (\ref{e:Faft}) depends on the afterglow and decay rates per chameleon, $\Gaft$ and $\Gdec$, as well as the production time $\tprod$.  The rates $\Gaft$ and $\Gdec$, as functions of the chameleon parameters $\meff$ $\xi_V$, and $\bgam$, for \GammeVCHASE~as described in Sec.~\ref{sec:enhancements_in_gammev-chase}, are computed in Sec.~\ref{sec:chameleon-photon_oscillation_monte_carlo_simulation}; Fig.~\ref{f:mc_rates} shows these rates for a particular choice of $\xi_V$ and $\bgam$.  Thus for \GammeVCHASE~we know $\Faft(\meff,\xi_V,\bgam,t)$.

A diagram of \GammeVCHASE~is shown in Fig.~\ref{f:gammev_apparatus}~\FigB. In each run, during the afterglow phase of the experiment, \GammeVCHASE~counts photons in $15$~second bins.  A shutter covers the detector, a PMT, in every other bin, allowing the background photon ``dark rate'' to be monitored in real time.  The excess photon rate observed by the PMT must come from the vacuum chamber.  It is some combination of the afterglow rate $\Faft(\meff,\xi_V,\bgam,t)$ and a background systematic rate $\Fsyst(\{{\wp}_i\},t)$ which depends on ``nuisance parameters'' $\{{\wp}_i\}$.

Let the eight data runs and seven calibration runs be labeled by $r$, and the time bins in run $r$ by $b_r$; the mean photon rate observed in the $b_r$th bin of run $r$ is $\Fobs_{r,b_r}$.  The predicted photon rate $\Fpred_{r,b_r}(\meff,\xi_V,\bgam,\{{\wp}_i\})$, for a given choice of the chameleon and nuisance parameters, is found by averaging $\Faft + \Fsyst$ over time $t$ in the appropriate bin.  The uncertainty $\sigma_{r,b_r}$ in that bin is dominated by the Poisson noise in the $28$~Hz dark rate, $\sqrt{28\rm{ Hz}/15\rm{ sec}} = 1.4$~Hz, and contains additional noise from background photons~\cite{Steffen_etal_2010}.  Summing over all runs and all bins in each run, we define
\begin{equation}
\chi^2(\meff,\xi_V,\bgam,\{{\wp}_i\})
=
\sum_{r,b_r}
\frac{\left(\Fobs_{r,b_r}- \Fpred_{r,b_r}\right)^2}
{\sigma_{r,b_r}^2}.
\label{e:chi2nuisance}
\end{equation}
At each point in the chameleon parameter space, the profile likelihood method defines $\chi^2(\meff,\xi_V,\bgam)$ to be $\chi^2(\meff,\xi_V,\bgam,\{{\wp}_i\})$ minimized over the nuisance parameters.  This is compared with the value $\chiSqnull$ for the null model $\bgam = \Faft = 0$, which has no photon-coupled chameleon field.

\subsection{Systematic rate $\Fsyst$}
\label{subsec:systematic_rate_Fsyst}

The total background rate $\Fsyst$ is found to have three important components~\cite{Steffen_etal_2010}:
\begin{enumerate}
\item a dark rate $\Fdark = 28$~Hz in the PMT, which is the same for all data runs;
\item a ``glow'' $\Fpump$ emitted by the ion pump, which varies from run to run with a mean of $\Fpumpmean = 1.2$~Hz and a standard deviation of $\sigma_\mathrm{pump} = 0.4$~Hz;
\item a transient rate $\Ftran(t)$ which is the same function of time for all data runs.
\end{enumerate}
$\Fdark$ is degenerate with individual pump glow rates, so we fix it in the analysis.

At late times, the background is dominated by the first two of these, $\Fsyst \rightarrow \Fdark + {\Fpump}_{,r}$.  Since  $\Fdark$ and ${\Fpump}_{,r}$ are time-independent, neither one depends upon the bin number $b_r$.  Poisson variations in $\Fdark$ of $\sqrt{\Fdark/\Delta t_{r,b_r}}$ are the dominant component of the uncertainty in a bin of width $\Delta t_{r,b_r}$.  The set $\{\wp_i\}$ of nuisance parameters includes the pump glows ${\Fpump}_{,r}$, with the mean and standard deviation above, but not the known dark rate $\Fdark$.  We modify (\ref{e:chi2nuisance}) above to include the term $\sum_r ({\Fpump}_{,r}-\Fpumpmean)^2/\sigma_\mathrm{pump}^2$ in order to account for the uncertainty in ${\Fpump}_{,r}$.

The transient rate $\Ftran(t)$ is studied in~\cite{Steffen_etal_2012}.  It cannot be a chameleon afterglow because it is independent of the magnetic field, and because its amplitude peaks in the orange region of the electromagnetic spectrum rather than the green $2\pi/\omega = 532$~nm of the input photons.  Because of this spectrum, references~\cite{Steffen_etal_2010,Steffen_etal_2012} refer to this transient component as the ``orange glow''.  It is modeled as a run-independent exponentially decaying background photon rate $\Ftran(t) = \Forange \exp(-\Gorange t)$.  

Although $\Ftran$ cannot be a chameleon afterglow signal, it can mimic such a signal in one particular run.  Furthermore, $\Ftran$ will introduce correlations among the errors in the first several data bins.  Therefore we cannot simply measure $\Ftran$ and subtract it from the observed signal.  We must treat it as a systematic to be fit; that is, we must include $\Forange$ and $\Gorange$ in the set $\{\wp_i\}$ of systematics parameters.  Analysis of \GammeVCHASE~calibration data shows $\Forange \approx 7$~Hz and $1/\Gorange \approx 2$~minutes.   In order to avoid the portion of the data most contaminated by this systematic, henceforth we discard the first two minutes of afterglow data in each run.
 
The ``profile $\chi^2$'' minimizes over the nuisance parameters $\{\wp_i\}$.  Since the total systematic rate $\Fsyst(t)$ depends linearly on all of the nuisance parameters except for $\Gorange$, given a value for $\Gorange$ the $\chi^2$-minimizing values for the others can be found by solving a linear system.  Letting $t_{r,b_r}$ and $\Delta t_{r,b_r}$, respectively, be the central time and the width of bin $b_r$ in run $r$, we define
\begin{eqnarray}
{\Faftmean}_{r,b_r}
&=&
\int_{t_{r,b_r}-\Delta t_{r,b_r}/2}^{t_{r,b_r}+\Delta t_{r,b_r}/2} \Faft(t) dt
\\
{\mathcal S}_{r,b_r}
&=&
\frac{1}{2} \, \sinc\left(\frac{1}{2}\Gorange\Delta t_{r,b_r}\right) \exp({-\Gorange t_{r,b_r}}) \quad
\\
{\mathcal A}_{r,m}
&=&
\sum_{b_r} \frac{{\mathcal S}_{r,b_r}^m}{\sigma_{r,b_r}^2}
\\
{\mathcal A}_{r,\mathrm{pump}}
&=&
\frac{1}{\sigma_\mathrm{pump}^2} + \sum_{b_r} \frac{1}{\sigma_{r,b_r}^2}
\\
{\mathcal B}_{r,m}
&=&
\sum_{b_r} \frac{(\Fobs_{r,b_r}-{\Faftmean}_{r,b_r}){\mathcal S}_{r,b_r}^m}{\sigma_{r,b_r}^2}
\end{eqnarray}
for $m = 0,1,2$.  Then $\Forange$ and the ${\Fpump}_{,r}$ are given by
\begin{eqnarray}
\Forange
&=&
\frac{1}{2}
\frac{\sum_r \left({\mathcal B}_{r,1} - {\mathcal B}_{r,0}\frac{{\mathcal A}_{r,1}}{{\mathcal A}_{r,\mathrm{pump}}}\right)}
{\sum_r \left( {\mathcal A}_{r,2} - \frac{{\mathcal A}_{r,1}^2}{{\mathcal A}_{r,\mathrm{pump}}}\right)}
\\
{\Fpump}_{,r}
&=&
\frac{{\mathcal B}_{r,0}}{{\mathcal A}_{r,\mathrm{pump}}} 
- 2 \Forange \frac{{\mathcal A}_{r,1} }{{\mathcal A}_{r,\mathrm{pump}}}.
\end{eqnarray}
$\chi^2$ must be minimized numerically with respect to the remaining nuisance parameter, $\Gorange$, in order to determine the profile $\chi^2$, which depends only upon $\meff$, $\xi_V$, and $\bgam$.
Henceforth we will use $\chi^2$ to refer to the profile $\chi^2$ defined above.

\subsection{What would a chameleon signal look like?}
\label{subsec:what_would_a_chameleon_signal_look_like}

\begin{figure*}[tb]
\begin{center}
\includegraphics[angle=270,width=2.3in]{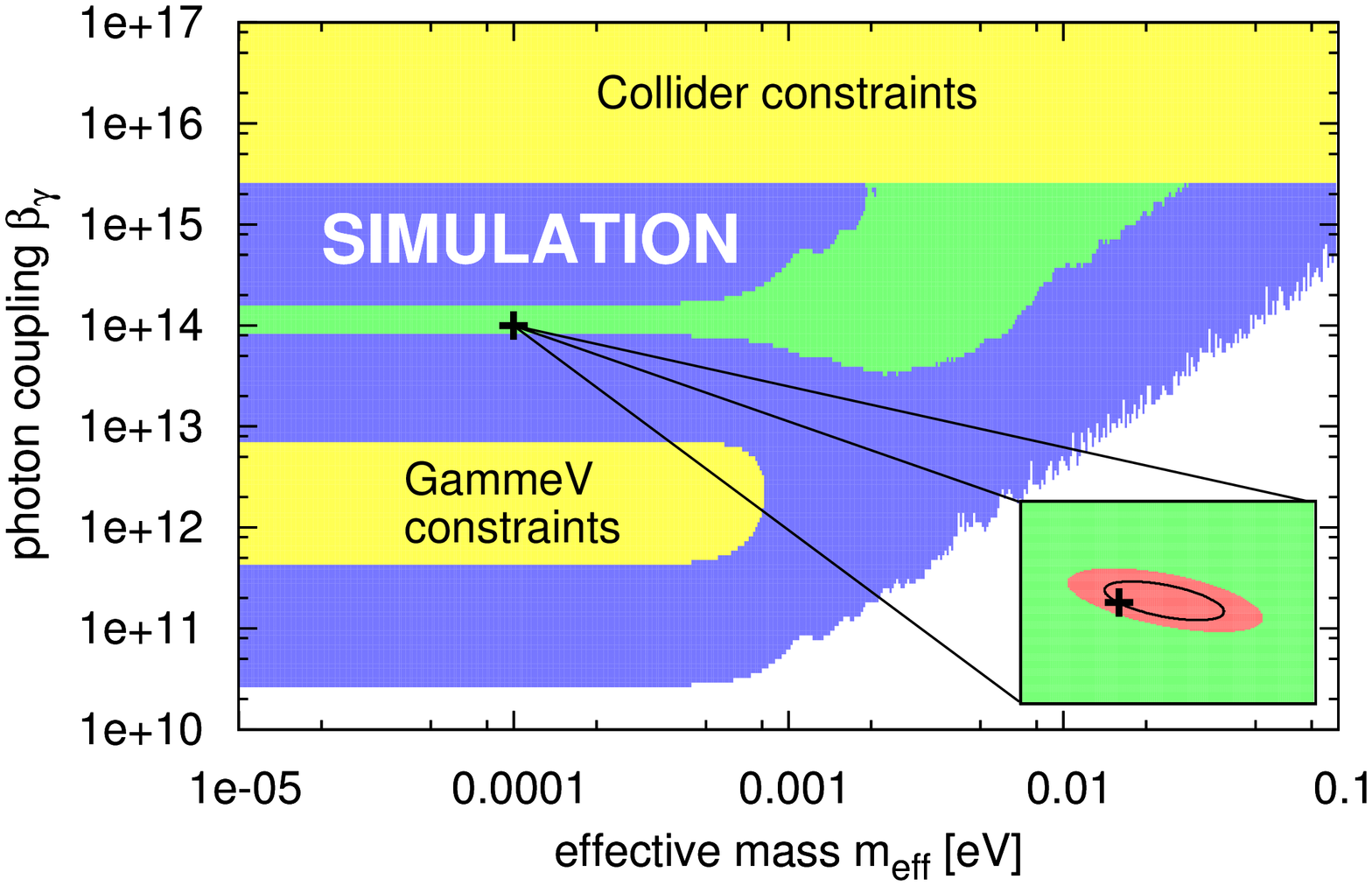}
\includegraphics[angle=270,width=2.3in]{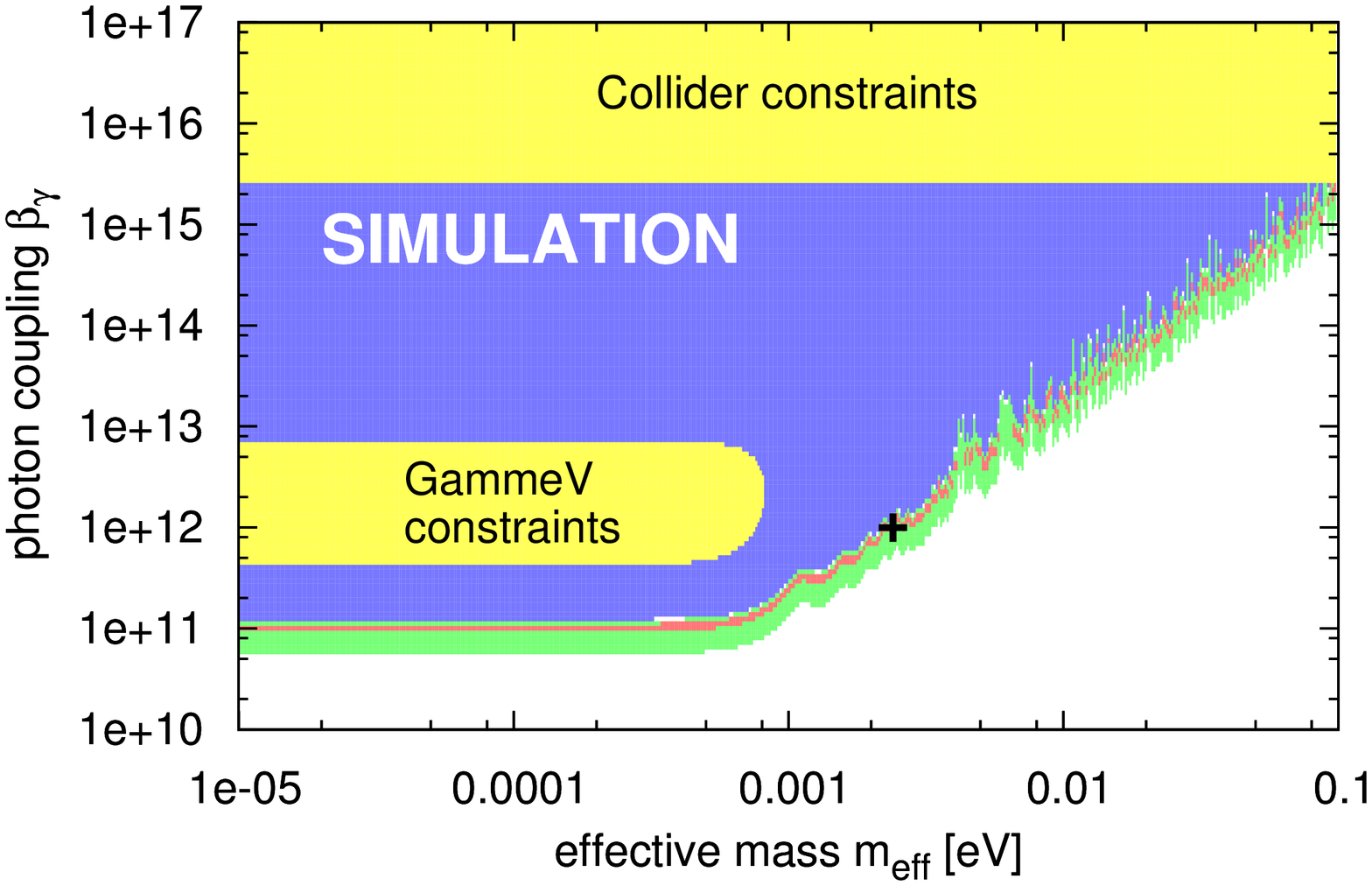}
\includegraphics[angle=270,width=2.3in]{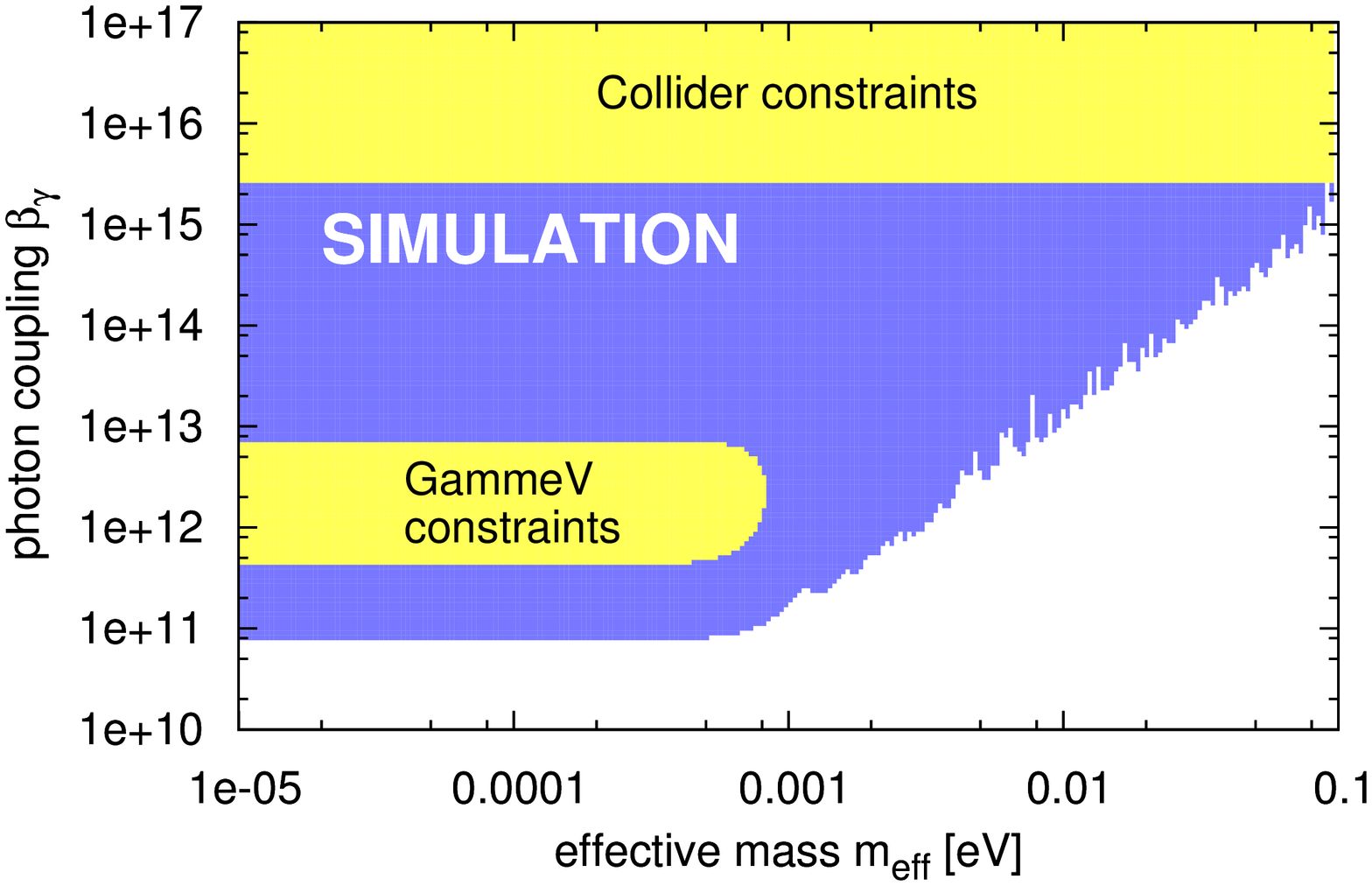}
\caption{Analysis of simulated \GammeVCHASE~data.  Yellow regions show previous constraints; blue regions show models excluded relative to the null (no-chameleon) model by $\Delta\chi^2 = 6.0$, corresponding to a $95\%$ confidence level for a Gaussian probability distribution; green regions show chameleon models preferred relative to the null model by $\Delta\chi^2 = 9.2$, corresponding to a $99\%$ confidence level; pink regions within the green regions show chameleons within $\Delta\chi^2 = 6.0$ of the best-fit chameleon model.  Black ``+'' signs show the fiducial chameleon model.  \FigThreeA~Simulated chameleon near the center of the \GammeVCHASE~sensitivity region, $\meff = 10^{-4}$~eV and $\bgam = 10^{14}$.  The inset shows $0.99\times 10^{-4}$~eV~$<\meff < 1.02\times 10^{-4}$~eV on the horizontal axis and $0.9999\times 10^{14} < \bgam < 1.0001\times 10^{14}$ on the vertical axis; the black ellipse shows the $\Delta\chi^2 = 2.3$ contour, corresponding to the $68\%$ confidence level.  \FigThreeB~Simulated chameleon near the edge of the \GammeVCHASE~sensitivity region, $\meff = 2.4\times 10^{-3}$~eV and $\bgam = 10^{12}$.  \FigThreeC~Simulated data with no chameleon.      In all of the analyses, $\xi_V = 0$ has been assumed. \label{f:simulated_constraints}}
\end{center}
\end{figure*}

Simulated \GammeVCHASE~``data'' show what a chameleon afterglow signal would look like in the experiment.  For a given chameleon model, the predicted rate $\Fpred_{r,b_r}(\meff,\xi_V,\bgam,\{\wp_i\})$ was simulated by setting $\Forange = 7$~Hz and $\Gorange = 1 / 120$~sec, and by choosing the pump glows for each run from a Gaussian distribution of mean $1.2$~Hz and standard deviation $0.4$~Hz.   In bin $b_r$ of run $r$ this implies $\Fpred_{r,b_r} \Delta t_{r,b_r}$ photons.  The simulated ``data'' for this bin are generated by randomly choosing a number from a Poisson distribution of mean $\Fpred_{r,b_r} \Delta t_{r,b_r}$.

``Data'' were simulated for three different scenarios:
\begin{enumerate}
\item a chameleon with $\meff = 10^{-4}$~eV, $\xi_V=0$, and $\bgam = 10^{14}$, near the center of the \GammeVCHASE~sensitivity region;
\item a chameleon with $\meff = 2.4\times 10^{-3}$~eV, $\xi_V=0$, and $\bgam=10^{12}$, near the edge of the sensitivity region; 
\item a null model, with no chameleon afterglow.
\end{enumerate}
Figure~\ref{f:simulated_constraints} shows the constraints resulting from analysis of these three simulations.  For the first scenario, analyzed in Fig.~\ref{f:simulated_constraints}~\FigThreeA, \GammeVCHASE~data would be extremely powerful.  The experiment would be able to constrain the chameleon mass to $\sim 1\%$ and the photon coupling to better than $0.01\%$.  In the second scenario, \GammeVCHASE~would detect the presence of a chameleon to high significance ($\Delta\chi^2 > 50$), but severe parameter degeneracies would make a determination of $\meff$ and $\bgam$ difficult.  Further study of this chameleon would require a new or redesigned experiment.  Finally, in the third scenario, \GammeVCHASE~would exclude chameleon models over a large range of parameters.  The profile likelihood analysis prevents a spurious identification of the orange glow systematic as a chameleon afterglow.

\subsection{\GammeVCHASE~model-independent constraints}
\label{subsec:chase_model-independent_constraints}

Now that the \GammeVCHASE~analysis has been tested on simulations, we proceed to the actual data.  Here we calculate constraints which are model-independent, in the sense that the chameleon parameters are the effective mass $\meff$ inside the oscillation chamber, the phase $\xi_V$, and the photon coupling $\bgam$, rather than the potential and the matter coupling.  In order for these constraints to be applicable, the chameleon mass in the chamber walls must be large enough that chameleon particles are contained inside the chamber.

\begin{figure}[tb]
\begin{center}
\includegraphics[angle=270,width=3.3in]{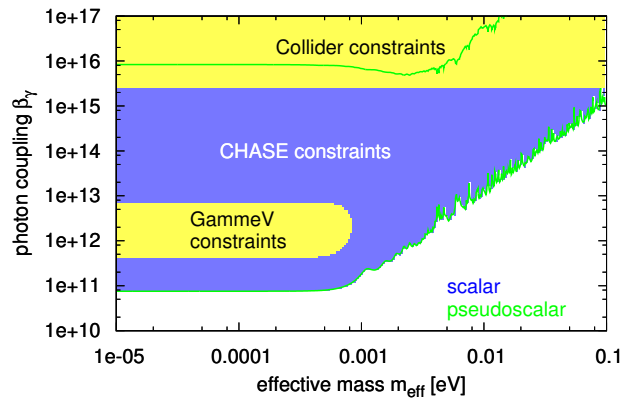}
\caption{Model-independent chameleon constraints for $\xi_V=0$, from~\cite{Steffen_etal_2010}.  The solid blue region is exluded at the $\Delta\chi^2 = 6.0$ level ($95\%$~CL for Gaussian probability) for scalar chameleons, and the interior of the green curve is excluded at that level for pseudoscalar chameleons.  \label{f:constraints_model-indep}}
\end{center}
\end{figure}

\begin{figure}[tb]
\begin{center}
\includegraphics[angle=270,width=3.3in]{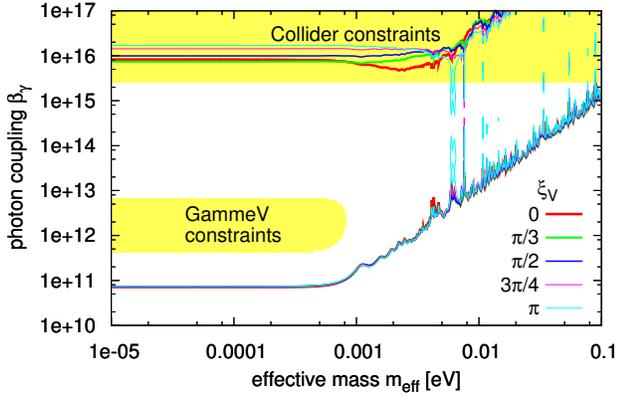}
\caption{Model-independent scalar chameleon constraints for several reflection phases $\xi_V$.  \label{f:constraints_model-indep_vary_xi}}
\end{center}
\end{figure}

Figure~\ref{f:constraints_model-indep} shows \GammeVCHASE~constraints~\cite{Steffen_etal_2010} on scalar and pseudoscalar chameleons with $\xi_V=0$.  Yellow regions are previous constraints in this parameter space; \GammeV~constraints are taken from~\cite{Chou_etal_2009}, and collider constraints were found by~\cite{Kleban_Rabadan_2005,Brax_etal_2009,Brax_etal_2010}.  In the blue region, the null model is preferred relative to the chameleon by $\Delta\chi^2 = 6.0$, which corresponds to exclusion at the $95\%$ confidence level for a Gaussian probability density function; henceforth we use ``$95\%$ CL'' to refer to this $\Delta\chi^2=6.0$ contour.  \GammeVCHASE~constraints improve upon those of \GammeV~by: extending to higher $\meff$, well beyond the dark energy scale of $2.4\times 10^{-3}$~eV; bridging the gap between \GammeV~and collider constraints; improving the low-$\meff$ upper bound on $\bgam$ through a tighter control of systematic uncertainties.  

As shown in Figure~\ref{f:constraints_model-indep_vary_xi}, these results are not strongly dependent on $\xi_V$.  This is because the afterglow rate $\Gaft$ is dominated by particles bouncing from the walls at grazing incidence, for which photon reflection itself contributes the large polarization-dependent phase shifts shown in Fig.~\ref{f:reflectivity}.  Averaging over polarizations weakens the dependence of $\Gaft$ on $\xi_V$.

\begin{figure}[tb]
\begin{center}
\includegraphics[angle=270,width=3.3in]{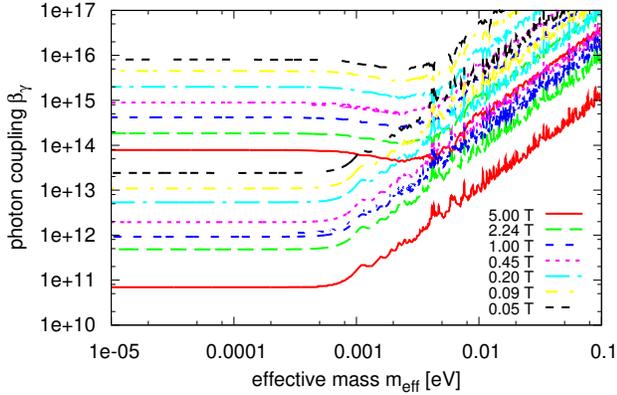}
\caption{Model-independent scalar chameleon constraints for each of the magnetic fields used in \GammeVCHASE, assuming $\xi_V=0$. \label{f:constraints_model-indep_indivB}}
\end{center}
\end{figure}

The model-independent plots in this section and in Ref.~\cite{Steffen_etal_2010} assume that $\meff$ is the same in all runs.  However, in some models $\meff$ can depend significantly on the magnetic field.  This is particularly true at the largest values of $\Bext$, where the electromagnetic energy density is of the same order of magnitude as the gas density inside the vacuum chamber.

Thus, for completeness, we show in Figure~\ref{f:constraints_model-indep_indivB} the constraints resulting from each of the seven different magnetic fields individually.  A joint analysis of data and calibration runs is necessary to distinguish between the ``orange glow'' systematic and an actual chameleon afterglow, as described in Sec.~\ref{subsec:systematic_rate_Fsyst}.  In order to avoid double-counting the calibration data, we analyze a different one of the seven calibration runs along with each magnetic field value in Fig.~\ref{f:constraints_model-indep_indivB}.


\subsection{\GammeVCHASE~constraints on $\phi^4$ theory}
\label{subsec:chase_constraints_on_phi4_theory}

\begin{figure}[tb]
\begin{center}
\includegraphics[angle=270,width=3.3in]{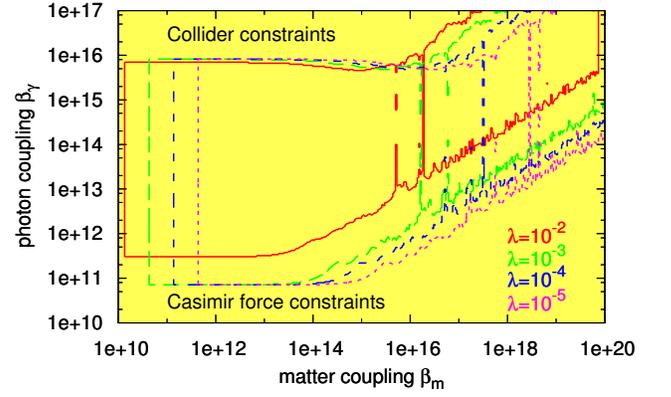}
\caption{\GammeVCHASE~constraints on $\phi^4$ chameleons.  Chameleon fragmentation becomes important for $\lambda \gtrsim 0.01$ and low $\bgam$.  Note that the entire parameter space shown has already been excluded by the Casimir force constraints of~\cite{Brax_etal_2007c}. \label{f:constraints_phi4}}
\end{center}
\end{figure}

Chameleon fragmentation in $\phi^4$ theory, $V(\phi) = \frac{\lambda}{4!} \phi^4$, was studied in Sec.~\ref{subsec:chameleon_fragmentation}.  Figure~\ref{f:constraints_phi4} shows \GammeVCHASE~constraints on $\phi^4$ theory.  The effects of fragmentation are evident at large $\lambda$ and low $\bgam$, where fragmentation competes with oscillation as a means of reducing the detectable chameleon population in the experiment.  Below $\lambda \approx 0.01$, fragmentation has only a negligible effect on afterglow constraints.  Meanwhile, at low $\bmat$ the constraining power of \GammeVCHASE~is limited by the chameleon containment requirement; for low $\bmat$ and $\lambda$, chameleons can escape though the walls of the chamber.  

As a result, \GammeVCHASE~constraints on $\phi^4$ chameleons are considerably weaker than fifth force constraints from measurements of the Casimir force~\cite{Brax_etal_2007c}.  The entire parameter space region shown in Fig.~\ref{f:constraints_phi4} is already excluded by Casimir force constraints.  

\subsection{\GammeVCHASE~constraints on dark energy}
\label{subsec:chase_constraints_on_dark_energy}

Next, we apply \GammeVCHASE~data to chameleon dark energy models (\ref{e:V_chameleon_dark_energy}).  Since the constraints of the previous section apply also to $n=4$ chameleon dark energy, we focus here on inverse power laws, $n<0$.  We also look at the exponential potential (\ref{e:V_exp}), which we write in a form
\begin{equation}
V(\phi) 
=
M_\Lambda^4 + M_\Lambda^4 \exp\left(-\frac{\kappa \phi}{M_\Lambda}\right)
\label{e:V_exp_de}
\end{equation}
suitable to a dark energy model.  As discussed in Sec.~\ref{subsec:phase_change_due_to_reflection}, these potentials are the easiest to study because their phase changes $\xi_V$ are independent of incident angle; more complicated potentials would require a numerical computation of $\xi_V$ as a function of $\theta$.  

\begin{figure}[tb]
\begin{center}
\includegraphics[angle=270,width=3.3in]{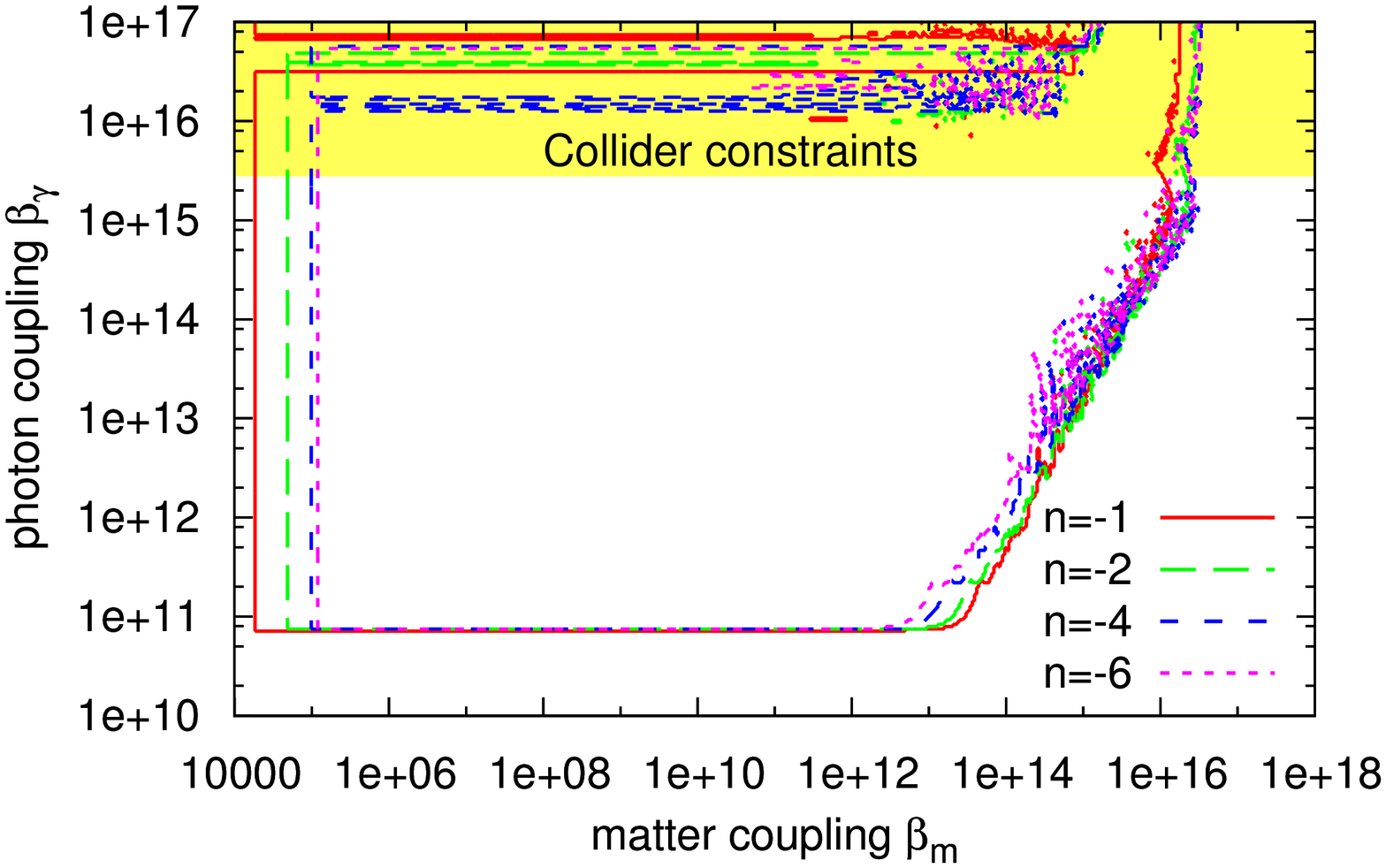}
\includegraphics[angle=270,width=3.3in]{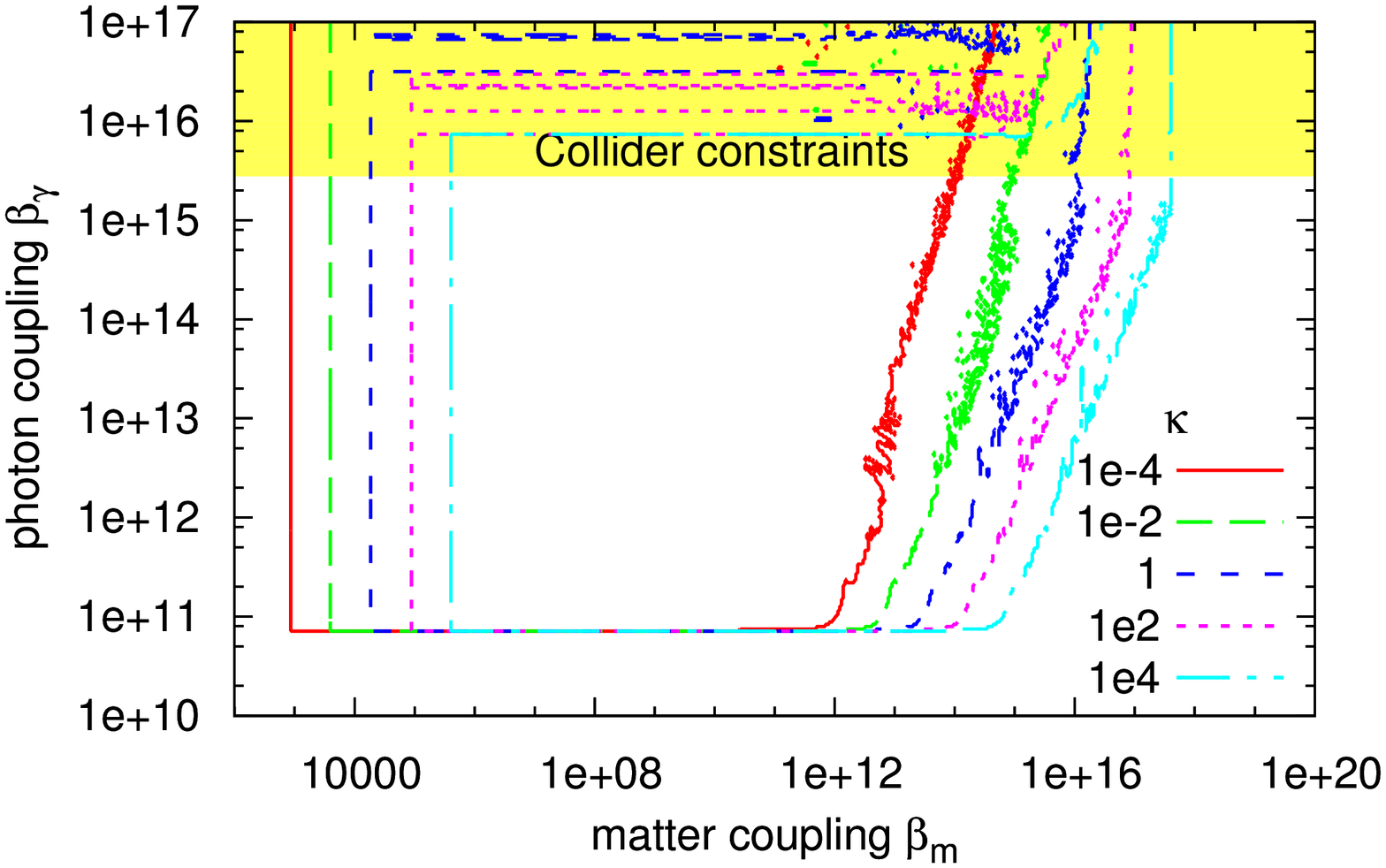}
\caption{Constraints on chameleon dark energy models~(\ref{e:V_chameleon_dark_energy}).  \FigA~assumes $\kappa=1$ and varies $n$; \FigB~fixes $n=-1$ and varies $\kappa$.  \label{f:constraints_powerlaw}}
\end{center}
\end{figure}

\begin{figure}[tb]
\begin{center}
\includegraphics[angle=270,width=3.3in]{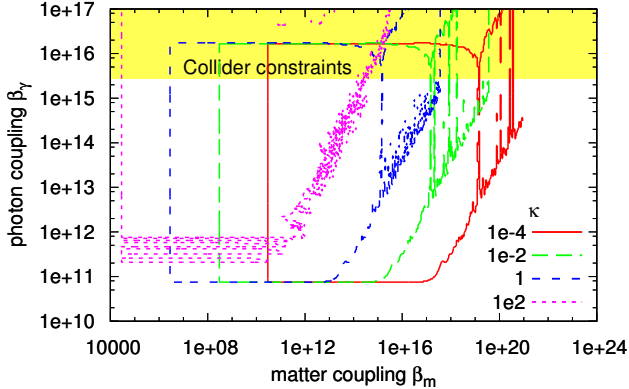}
\caption{Constraints on exponential dark energy models~(\ref{e:V_exp_de}).  Models with $\kappa \gtrsim 10^4$ have particles too massive to be probed by \GammeVCHASE.  \label{f:constraints_exp}}
\end{center}
\end{figure}

\GammeVCHASE~constraints on inverse power law chameleon dark energy are shown in Figure~\ref{f:constraints_powerlaw}.  Twelve orders of magnitude in $\bmat$ are excluded for a range of $\bgam$ in each of the models shown in Fig.~\ref{f:constraints_powerlaw}~\FigA, with the excluded region limited at low $\bmat$ by the containment requirement and at  high $\bmat$ by destructive interference in oscillation due to a large effective mass inside the chamber.  Fig.~\ref{f:constraints_powerlaw}~\FigB~shows that changing $\kappa$ has the effect of translating the excluded region in the $\bmat$ direction; the same behavior is seen for other $n$.  Figure~\ref{f:constraints_exp} constrains exponential dark energy~(\ref{e:V_exp_de}).  In these models $\meff = \sqrt{\kappa (\bmat\rhom + \bgam\rhog) / ( M_\Lambda \Mpl )}$ grows so rapidly with $\kappa$ that $\kappa \gtrsim 10^4$ is inaccessible to \GammeVCHASE~\cite{Upadhye_Steffen_Weltman_2010}.

\subsection{Combined constraints on $n=-1$ dark energy}
\label{combined_constraints_on_n=-1_dark_energy}

Chameleon dark energy is described by a complicated parameter space, with several experimental constraints which depend on the parameters in different ways.  While the previous discussion attempted to compare \GammeVCHASE~to existing constraints, such constraints do not exist (or have not been published) for all of the potentials considered here.  Furthermore, new experiments have constrained chameleon models since the publication of \GammeVCHASE~results in~\cite{Steffen_etal_2010}, and forecasts of planned experiments have been made.

\begin{figure}[tb]
\begin{center}
\includegraphics[angle=270,width=3.3in]{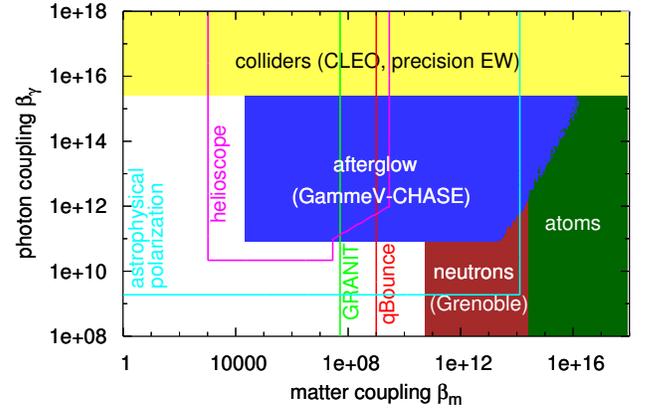}
\caption{Combined constraints on $n=-1$, $\kappa=1$ chameleon dark energy~(\ref{e:V_chameleon_dark_energy}). Solid regions show current constraints from analyses of experimental data, while lines represent forecasts and preliminary analyses. \label{f:constraints_combined_n-1}}
\end{center}
\end{figure}

It is instructive to include as many chameleon constraints as possible on a single plot.  For that we choose a specific potential, chameleon dark energy~(\ref{e:V_chameleon_dark_energy}) with $n=-1$ and $\kappa=1$.  Figure~\ref{f:constraints_combined_n-1} shows current constraints and forecasts on this model.  Collider constraints have been studied extensively~\cite{Brax_etal_2009,Brax_etal_2010,Kleban_Rabadan_2005,Balest_etal_1995} and are weakly dependent on $\bmat$ and $V(\phi)$, but only exclude extremely strong photon couplings.  Afterglow constraints are due to \GammeVCHASE.  At large matter couplings, the chameleon-mediated fifth force would affect electronic energy levels in atoms, which are inconsistent with the data~\cite{Brax_Burrage_2011}.  More recently, fifth force contributions to the quantized energy levels of neutrons in a classical gravitational field have been used to constrain chameleons~\cite{Nesvizhevsky_etal_2002,Brax_Pignol_2011}.  Since the neutron wavefunction in such experiments is de-localized to a cloud of several microns, chameleon screening is ineffective for suppressing fifth forces, resulting in powerful constraints at large $\bmat$.

The upcoming neutron experiments qBounce and GRANIT are expected to improve constraints on the matter coupling by several orders of magnitude~\cite{Kreuz_etal_2009,Brax_Pignol_2011,Bassler_etal_2012}.  ``Helioscopes'' designed to convert solar axions into photons can also be used to constrain chameleons~\cite{Brax_Zioutas_2010,Brax_Lindner_Zioutas_2011,Baker_etal_2012}.  Refs.~\cite{Burrage_Davis_Shaw_2009,Burrage_Davis_Shaw_2009b} study chameleon-photon oscillation using more distant astrophysical sources; however, since they describe their analysis as ``preliminary'' and systematic effects due to astrophysical uncertainties remain to be analyzed, we include this constraint as a forecast.  Not shown on the plot are Casimir force tests~\cite{Brax_etal_2007} and the original \GammeV~Experiment~\cite{Chou_etal_2009}, which do not constrain the potential chosen here.  The E\"ot-Wash torsion pendulum experiment~\cite{Kapner_etal_2007} likely excludes a range of models around $\bmat  = 1$, but constraints for this specific model have yet to be computed.

\section{Conclusion}
\label{sec:conclusion}

We conducted a thorough investigation of the physics of chameleon particles in afterglow experiments, focusing on questions which arose during the design and analysis of \GammeVCHASE.  Afterglow experiments rely upon two assumptions about chameleon particles: reflection from dense matter, which is necessary for trapping chameleon particles; and oscillation, which allows chameleon production as well as photon regeneration.  By studying the interaction of a chameleon particle with atoms, in isolation as well as in a lattice, we showed that the matter density can be treated as homogeneous inside the chamber as well as in its walls.  Chameleon-atom scattering is negligible in the vacuum; Fig.~\ref{f:beta_min} showed that the chameleon particle ignores the electron cloud, so that the cross section is dominated by hard-sphere scattering from the nucleus alone.  Fig.~\ref{f:m_lattice} showed that a chameleon particle ``sees'' the chamber wall as a homogeneous solid until its mass becomes much larger than the $\sim 1$~eV needed for reflection, so trapping is unaffected by the fact that real matter is a lattice of atoms.

On the other hand, the smoothly varying magnetic field found in an afterglow experiment could potentially suppress chameleon production significantly.  If the length scale on which the magnetic field drops from its maximum value to zero, at the edge of the magnetic field region, is much larger than the chameleon oscillation length, then the transition will be adiabatic, and an incoming photon will emerge from the magnetic field in a pure photon state.  This adiabatic suppression can be mitigated substantially by placing glass windows inside the magnetic field region, as is evident from a comparison of Figs.~\ref{f:Pgc_realB_nowin} and~\ref{f:Pgc12_vs_meff_2win}~\FigA.  Such interior windows were included in \GammeVCHASE~and will be essential to any future afterglow experiment which seeks to improve constraints at greater chameleon masses.

After studying these general effects, we proceeded to calculate the signal expected in an afterglow experiment.  The analytic approximation of~\cite{Upadhye_Steffen_Weltman_2010} was improved upon in 
Sec.~\ref{sec:chameleon-photon_oscillation_analytic_calculation}.  This approximation, in the limit where it becomes exact, was used to verify that the Monte Carlo simulation of Sec.~\ref{sec:chameleon-photon_oscillation_monte_carlo_simulation} was correct.  The Monte Carlo simulation, in turn, was used to show that the predicted signal will not be decreased by diffuse reflection in the chamber walls.  The dependence of the signal on the wall reflectivity and the chamber geometry was shown in Fig.~\ref{f:vary_parameters}.  Afterglow and decay rates per chameleon particle for \GammeVCHASE~were shown in Fig.~\ref{f:mc_rates}.

Finally, in Sec.~\ref{sec:analysis_and_constraints}, we explained in greater detail the analysis underlying the constraints of~\cite{Steffen_etal_2010}.  The model-independent constraints of that reference were shown for different chameleon phase shifts and for each of the magnetic field runs used in the experiment.  \GammeVCHASE~data were then used to constrain a wide range of models, including dark energy candidates.  Since additional constraints on chameleons have been released since the publication of~\cite{Steffen_etal_2010}, and since further experiments are planned, we compared several different constraints and forecasts in Fig.~\ref{f:constraints_combined_n-1}.  For the model shown, \GammeVCHASE~excluded five orders of magnitude in the photon coupling over a range of ten orders of magnitude in the matter coupling.  \GammeVCHASE has made a substantial contribution to the study of photon-coupled chameleon field theories.

\vspace{12pt}
\subsection*{Acknowledgments}

We are grateful to A. Baumbaugh, R. Cowsik, P. O. Mazur, R. Tomlin, A. Weltman, and W. Wester for many informative discussions.  JS thanks the Brinson Foundation for its generous support.  
This work is supported by the U.S. Department of Energy, 
Basic Energy Sciences, Office of
Science, under contract No. DE-AC02-06CH11357.
This work was supported by Fermi National Accelerator Laboratory,
 operated by the U.S. Department of Energy under contract No. DE-AC02-07CH11359.

The submitted manuscript has been created by UChicago Argonne, LLC, Operator
of Argonne National Laboratory (“Argonne”). Argonne, a U.S. Department of
Energy Office of Science laboratory, is operated under Contract No.
DE-AC02-06CH11357. The U.S. Government retains for itself, and others acting
on its behalf, a paid-up nonexclusive, irrevocable worldwide license in said
article to reproduce, prepare derivative works, distribute copies to the
public, and perform publicly and display publicly, by or on behalf of the
Government.


\bibliographystyle{unsrt}
\bibliography{chameleon}

\end{document}